\documentclass[letterpaper]{article} %, draft 
\usepackage{amsmath,amsthm,amsfonts,amssymb}
\usepackage{url,algorithm}
{\normalfont}{\normalfont}

\usepackage{enumitem}

\usepackage{natbib}
\newcommand{\ba}[1]{\begin{array}{#1}}
\newcommand{\ea}{\end{array}}

\usepackage{graphicx}
\usepackage{epstopdf}
\graphicspath{{images/}}
\usepackage{subfigure}
\usepackage{listings} %Per inserire codice
\usepackage[usenames]{color} %Per permettere la colorazione dei caratteri 

\begin{document}
\title{Dynamic asymptotic homogenization for periodic viscoelastic materials}
\author{Rosaria Del Toro$^{1}$, Andrea Bacigalupo$^{1}$, Marco Paggi$^{1}$}
\date{\small $^{1}$ IMT School for Advanced Studis Lucca, Piazza S. Francesco $19$, $55100$ Lucca, Italy}
\maketitle
\begin{abstract}
A non-local dynamic homogenization technique for the analysis of a viscoelastic heterogeneous material which displays a periodic microstructure is herein proposed. The asymptotic expansion of the micro-displacement field in the transformed Laplace domain allows obtaining, from the expression of the micro-scale field equations, a set of recursive differential problems defined over the periodic unit cell. Consequently, the cell problems are derived in terms of perturbation functions depending on the geometrical and physical-mechanical properties of the material and its microstructural heterogeneities.\\ 
A down-scaling relation is formulated in a consistent form, which correlates the microscopic to the macroscopic transformed displacement field and its gradients through the perturbation functions. \\
Average field equations of infinite order are determined by substituting the down-scale relation into the micro-field equation. Based on a variational approach, the macroscopic field equations of a non-local continuum is delivered and the local and non-local overall constitutive and inertial tensors of the homogenized continuum are determined.
The problem of wave propagation in case of a bi-phase layered material with orthotropic phases and axis of orthotropy parallel to the direction of layers is investigated as an example. In such a case, the local and non-local overall constitutive and inertial tensors are determined analytically and the dispersion curves obtained from the non-local homogenized model are analysed.
\end{abstract}
\section{Introduction}
Over the last few years, the technological progress led to a fast development of composite materials. Such materials play a crucial role for various applications in civil, naval, aerospace and mechanical engineering, since they boast remarkable mechanical and physico-chemical properties such as high strength, corrosion and thermal resistance, enhanced durability, light weight and ease handling, \cite{gibson2011principles}. Among them, the class of polymer matrix composites is very promising since they can achieve performances superior to metals with a reduced weight, \cite{wang2011polymer}. The matrix is usually a resin (epoxy or polyester) with high toughness, reinforced by fibers (glass, aramid, boron, etc.), which have very high strength. The combination of the two materials is very effective: the matrix diffuses the load among the fibers and protects them from abrasion, fracture, and damage. At the same time, the reinforcing fibres increase the overall strength and stiffness of the composite. 
Similarly, in laminates, the polymeric matrix is used to bond other materials together and increase the toughness of the composite, see e.g. photovoltaic modules, \cite{paggi2016global}.
To reduce the cost of synthetic fiber-reinforced composites and produce
environmentally sustainable materials, bio-fibre-reinforced polymer composites are very promising and are becoming increasing popular in emergent countries. The variant of hybrid composites, where synthetic and natural fiber reinforcements are mixed together offer also a possible trade-off solution, \cite{dhakal2018abrasive}.\\
If the heterogeneities of the reinforcement are sufficiently regular and their size is much smaller than the dimension of the component, then the material is said to be a composite with periodic or quasi-periodic microstructure. The constitutive response of polymeric composites and their variants is that of viscoelastic materials, which exhibit creep and stress relaxation phenomena.\\
 An intense knowledge of the behaviour of viscoelastic materials allows manufacturing devices, which can be applied to a wide range of fields, including biomedical, industrial, defence and construction. \cite{krushynska2016visco} studied the wave dispersion properties and attenuation capability of dissipative solid acoustic metamaterials with local resonators having subwavelength band gaps. The characterization and simulation of mechanical problems involving such materials is very expensive, due to the presence of heterogeneities.\\
 Therefore, based on the premises above, the theory of homogenization may represent an excellent methodology to recognize and model the effects of the microscopic behaviour on the overall properties of materials. Such a theory allows replacing a heterogeneous material with an equivalent homogenous one, which can be modelled through either a first order (Cauchy) or a non-local continuum. Generally, three main classes of homogenization techniques are possible: the asymptotic techniques (\citealp{Bensoussan1978}; \citealp{Bakhvalov1984}; \citealp{GambinKroner1989}; \citealp{Allaire1992};   \citealp{meguid1994asymptotic}; \citealp{Boutin1996}; \citealp{fish2001higher}; \citealp{AndrianovBolshakov2008}; \citealp{panasenko2009boundary}; \citealp{Tran2012}; \citealp{Bacigalupo2014};), the variational-asymptotic techniques (\citealp{Smyshlyaev2000}; \citealp{Smyshlyaev2009};   \citealp{BacigalupoGambarotta2014b}; \citealp{bacigalupo2014effective}) and many identification approaches, involving the analytical (\citealp{bigoni2007analytical}; \citealp{bacca2013amindlin}; \citealp{bacca2013anisotropic}; \citealp{bacca2013mindlin2}; \citealp{bacigalupo2013multi}; \citealp{bacigalupo2017identification}) and the computational techniques (\citealp{Forest1998};   \citealp{ostoja1999couple}; \citealp{KouznetsovaGeersBrekelmans2002}; \citealp{forest2002homogenization};  \citealp{feyel2003multilevel}; \citealp{KouznetsovaGeers2004};  \citealp{kaczmarczyk2008scale}; \citealp{yuan2008micromechanical}; \citealp{BacigalupoGambarotta2010b};\citealp{de2011cosserat}; \citealp{forest2011generalized};   \citealp{addessi2013micromechanical}; \citealp{zah2013computational}; \citealp{trovalusci2015scale}). Such techniques have been expanded to the multi-field case, such as termomechanics \citealp{aboudi2001linear}; \citealp{kanoute2009multiscale}; \citealp{zhang2007thermo} and thermo-diffusive, \citealp{BacigalupoMoriniPiccolroaz2016b}; \citealp{bacigalupo2016multiscale} and thermo-piezoelectricity phenomena, \citealp{fantoni2017multi}.\\
In case of viscoelastic materials with periodic microstructure,
which are the object of the present article, there are still a few contributions devoted to homogenization techniques applied to this paricular class of composites. Specifically, the computational techniques are proposed in the works of \citealp{ohono2000}; \citealp{Haasemann2010}; \citealp{tran2011simple} and \citealp{chen2017finite} and the asymptotic techniques are analysed in \citealp{yi1998asymptotic} and \citealp{hui2013nonlocal}.\\
In the context of the computational homogenization techniques, \cite{ohono2000} dealt with an homogenization model for elastic-viscoplastic periodic materials, but without taking into account an asymptotic expansion of the field variables. Their method allows determining the macroscopic and the microscopic stress and strain states in nonlinear time-dependent periodic materials and it encompasses any problem where the history of the macro-strain and the macro-stress depends upon the time. \cite{Haasemann2010} considered a microstructure where all the constituents are linear viscoelastic. The constitutive laws at the microscale were converted into a Laplace-Carson domain, where the constitutive equations have a quite similar form to those of a linear elastic material and then a homogenization approach based on the Hill-Mandel condition was exploited. The Laplace-Carson transform associated with the application of a FE-method enables the computation of the relaxation tensor in the Laplace domain and the inverse Laplace-Carson transformation provides the material properties in the time domain. \cite{tran2011simple} presented a computational homogenization method to determine the  response of a linear viscoelastic heterogeneous material. The components of the relaxation tensor, which appear in the constitutive law at the macro-scale, are numerically determined in the time domain, without involving the Laplace transform.\\
Although the employment of computational approaches is more and more extensive thanks to up-to-date computer facilities, they have a high computational cost, they cannot challenge dynamic problems and they are not able to provide higher-order approximations of the homogenized constitutive tensors.\\
Concerning with the asymptotic techniques applied to viscoelastic materials, \cite{hui2013nonlocal} proposed a non-local homogenization method with multiple length scales for detecting wave propagation in viscoelastic composite materials, by proceeding with an asymptotic expansion of the governing system of equations defined in the time domain, then recast into the Laplace domain. The higher-order terms, derived from this approach, identify the micro heterogeneities producing the wave dispersions and predicting the creation of bandgaps. Neverthless, such a method cannot provide an average field equation of infinite order, containing local and non-local higher-order tensor components.\\
Motivated by the state-of-the-art literature on homogenization, the present study proposes a dynamic variational-asymptotic homogenization technique for the analysis of a viscoelastic material with periodic microstructure modelled with a non-local continuum, based on the asymptotic and variational methods (\cite{Smyshlyaev2000} and \cite{Bacigalupo2014}) and on the works related to the variational principles of linear viscoelasticity (\cite{leitman1966variational}, \cite{fabrizio1992mathematical} and  ).\\
The field equation at the micro-scale, which describes the heterogeneous viscoelastic domain, is determined in the time domain and it is converted into the Laplace domain, with the help of the two-sided Laplace transform. The micro-displacement field is expressed as an asymptotic expansion in the transformed Laplace space and its replacement into the field equation at the micro-scale enables to produce a sequence of recursive differential problems defined over the periodic unit cell. 
Then solvability conditions are imposed to such nonhomogeneous recursive cell problems to determine the down-scaling relation, linking the microscopic transformed displacement field to the macroscopic one and its gradients through the perturbation functions. Such functions rely on the geometrical and physical-mechanical properties of the material and measure the microstructural heterogeneities. Average field equations of infinite order are determined by substituting the down-scale relation into the micro-field equations. Its formal solution is provided with the help of an asymptotic expansion of the transformed macro-displacement and, by considering only the terms at the zeroth order, the field equations related to the equivalent viscoelastic Cauchy continuum are retrieved.\\
Section 2 deals with the description of the field equations in the time domain and in the Laplace domain at the microscale. Section 3 shows the recursive differential problems and their solutions and  Section 4 presents the cell problems and the related perturbation functions. Section 5 defines the down-scaling relation, the up-scaling relation and the average field equations of infinite order.  
In Section 6, by means of a variational approach, the overall constitutive tensors and the overall inertial tensor related to the homogenized continuum are derived in the Laplace domain for the class of periodic viscoelastic materials, after introducing the energy-like functional in the Laplace domain (\cite{fabrizio1992mathematical}). Moreover, the Euler-Lagrangian differential equation at the  macro-scale is determined, expressed in terms of the transformed macro-displacement and its gradients up to the fourth order. In Section 7, the variational-asymptotic homogenization technique is applied to a bi-phase layered material with isotropic phases subject to periodic body forces. To verify the reliability of the proposed homogenization procedure, the solution of the homogenized problem is compared with the one obtained from the heterogeneous problem and a good agreement between the models is obtained. Finally, the problem of wave propagation and the related dispersion curves is studied.
Concluding remarks complete the article.

\section{Problem setting and field equation in the Laplace domain}
\begin{figure}[b!]
	\centering
	\includegraphics[height=0.34\textheight]{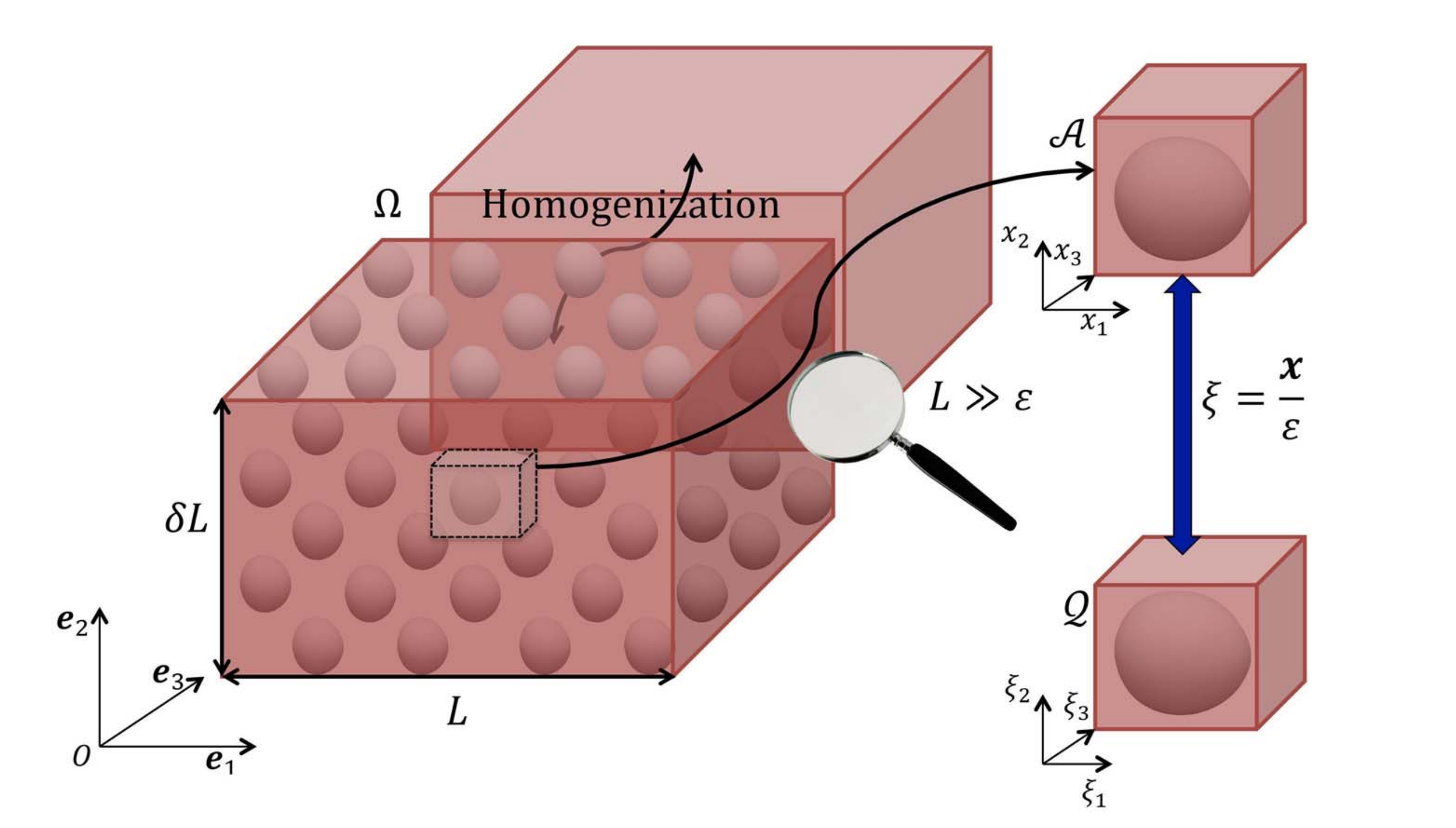}
	\caption{Heterogeneous and homogeneous $3D$ domain $\Omega$ with periodic cell $\mathcal{A}$ and the corresponding nondimensional cell $\mathcal{Q}$.}
	\label{permistd}
\end{figure}  
Let $\Omega$ be a three-dimensional viscoelastic heterogeneous material which displays a periodic microstructure. A generic point of the material is identified by the position vector $\boldsymbol{x} = x_{1}\boldsymbol{e}_{1}+x_{2}\boldsymbol{e}_{2}+x_{3}\boldsymbol{e}_{3}$ related to a system of coordinates with origin at point $O$ and orthogonal base $\{{} \boldsymbol{e}_{1},\boldsymbol{e}_{2}, \boldsymbol{e}_{3}\}$. Let $\mathcal{A}=[0,\varepsilon]\times [0,\delta \varepsilon]\times [0,\varepsilon]$ be a periodic cell with characteristic size $\varepsilon$. $\mathcal{A}$ is described by three orthogonal periodicity vectors $\boldsymbol{v}_{1}$, $\boldsymbol{v}_{2}$ and $\boldsymbol{v}_{3}$ defined as $\boldsymbol{v}_{1} =  d_{1}\boldsymbol{e}_{1} = \varepsilon \boldsymbol{e}_{1} $,  $\boldsymbol{v}_{2}= d_{2}\boldsymbol{e}_{2} = \delta \varepsilon \boldsymbol{e}_{2}$ and $\boldsymbol{v}_{3} =  d_{3}\boldsymbol{e}_{3} = \varepsilon \boldsymbol{e}_{3}$. The material domain is set up by the repetition of the cell $\mathcal{A}$ in accordance with the directions of $\boldsymbol{v}_{1}$, $\boldsymbol{v}_{2}$ and $\boldsymbol{v}_{3}$, see Fig. \ref{permistd}. Since the material is $\mathcal{A}$-periodic, the micro relaxation tensor $\mathbb{G}^{m}(\boldsymbol{x},t)=G^{m}_{ijhk}\boldsymbol{e}_{i}\otimes \boldsymbol{e}_{j}\otimes \boldsymbol{e}_{h}\otimes \boldsymbol{e}_{k}$, which depends on time and accounts for the viscoelastic effects, and the material density $\rho^{m}(\boldsymbol{x})$ comply with the following conditions:
\begin{equation}
\label{GG}
\mathbb{G}^{m}(\boldsymbol{x}+\boldsymbol{v}_{i},t)=\mathbb{G}^{m}(\boldsymbol{x},t), \quad i=1,2,3 \quad \forall \boldsymbol{x} \in \mathcal{A}, 
\end{equation} 
\begin{equation}
\label{roro}
\rho^{m}(\boldsymbol{x}+\boldsymbol{v}_{i},t)=\rho^{m}(\boldsymbol{x},t), \quad i=1,2,3 \quad \forall \boldsymbol{x} \in \mathcal{A}. 
\end{equation}               
The micro stress $\boldsymbol{\sigma}(\boldsymbol{x},t)$ constitutive relation, which models the viscoelastic elements of the heterogeneous material, is expressed in terms of the hereditary integral, \cite{christensen2012theory}:
\begin{equation}
\label{eqn:vis}
\boldsymbol{\sigma}(\boldsymbol{x},t) = \int_{-\infty}^{t} \mathbb{G}^{m} (\boldsymbol{x},t-\tau) \dot{\boldsymbol{\varepsilon}}(\boldsymbol{x},\tau) d\tau,
\end{equation}
where the superscript \emph{m} refers to the microscale and $\boldsymbol{\varepsilon}(\boldsymbol{x},t)=\varepsilon_{ij}\boldsymbol{e}_{i}\otimes \boldsymbol{e}_{j}$ is the micro-strain tensor. Moreover $t$ denotes the time coordinate and the superimposed dot indicates time derivative. The material undergoes 
small displacements and so the micro-strain tensor is defined as
$\boldsymbol{\varepsilon}(\boldsymbol{x},t) = \frac{1}{2} (\nabla \boldsymbol{u} (\boldsymbol{x},t)+\nabla^{T} \boldsymbol{u}(\boldsymbol{x},t))$, where $\nabla \boldsymbol{u}$ is the gradient of the micro-displacement $\boldsymbol{u}(\boldsymbol{x},t)$.  
In the time domain, the deformation response of the material under dynamic loading is expressed by the momentum balance equation:
\begin{equation}
\label{eqn:din}
\nabla \cdot \boldsymbol{\sigma}(\boldsymbol{x},t)+\boldsymbol{b}(\boldsymbol{x},t) = \rho^{m}(\boldsymbol{x}) {\ddot{\boldsymbol{u}}}(\boldsymbol{x},t),
\end{equation}  
where $\boldsymbol{u}(\boldsymbol{x},t)$ is the micro-displacement field and $\boldsymbol{b}(\boldsymbol{x},t)$ are the body forces.
In the derivation of the theory, the heterogeneous material is supposed to be subjected to a system of $\mathcal{L}$-periodic  body forces $\boldsymbol{b}(\boldsymbol{x},t)$, with zero mean values over $\mathcal{L} = [0, L] \times [0, \delta L]$. The structural (or macroscopic)  length L is assumed to be much greater than the  microstructural length $\varepsilon$, i.e. L$>$$>$ $\varepsilon$, to allow the scales separation condition and so $\mathcal{L}$ is considered as an actual representative portion of the material. Let $\mathcal{Q} = [0,1]\times [0,\delta]\times [0,1]$ be the nondimensional cell reproducing the periodic microstructure. $\mathcal{Q}$ is determined by rescaling the size of the periodic cell $\mathcal{A}$ for the characteristic length $\varepsilon$. Accordingly, two variables are introduced to differentiate the two scales, namely the macroscopic (or slow) one, $\boldsymbol{x} \in \mathcal{A}$, which measures the slow fluctuations, and the microscopic (or fast) variable, $\boldsymbol{\xi} = \frac{\boldsymbol{x}}{\varepsilon} \in \mathcal{Q}$, which measures the fast propagation of the signal. Thanks to cell $\mathcal{Q}$, the properties ~\eqref{GG} and $~\eqref{roro}$ may be rewritten in terms of the microscopic variable $\boldsymbol{\xi}$ and so $\mathbb{G}^{m}$ and $\rho^{m}$ are assumed to be $\mathcal{Q}$-periodic and defined on $\mathcal{Q}$ as 
\begin{equation}
\mathbb{G}^{m}(\boldsymbol{x},t)=\mathbb{G}^{m}(\boldsymbol{\xi}=\boldsymbol{x}/\varepsilon,t), \quad \rho^{m}(\boldsymbol{x},t)=\rho^{m}(\boldsymbol{x}/\varepsilon,t).
\end{equation} 
Bearing in mind the definition of the strain tensor $\boldsymbol{\varepsilon}$, the minor simmetry of the relaxation tensor $\mathbb{G}^{m}$ is applied to $\boldsymbol{\dot{\varepsilon}}$ in the integral $~\eqref{eqn:vis}$ and the substitution of Eq.$~\eqref{eqn:vis}$ into Eq.$~\eqref{eqn:din}$ yields to
\begin{equation}
\label{eqn:sm}
\nabla \cdot \Big[\int_{-\infty}^{t} \mathbb{G}^{m}\Big (\frac{\boldsymbol{x}}{\varepsilon}     ,t-\tau\Big )\nabla \dot{\boldsymbol{u}}(\boldsymbol{x},\tau) d\tau\Big] +\boldsymbol{b}(\boldsymbol{x},t) = \rho^{m}(\boldsymbol{x}) {\ddot{\boldsymbol{u}}}(\boldsymbol{x},t).
\end{equation} 
Denoting with $[[f]]=f^{i}(\Sigma)-f^{j}(\Sigma)$ the jump of the function values $f$ at the interface $\Sigma$ between two different phases $i$ and $j$ in the periodic cell $\mathcal{A}$, the following fully-bonded interface conditions hold
\begin{equation}
\label{eqn:conint}
[[\boldsymbol{u}(\boldsymbol{x})]]\vert_{\boldsymbol{x} \in \Sigma}= \boldsymbol{0}, \quad
\Big [\Big [ \int_{-\infty}^{t} \mathbb{G}^{m}\Big (\frac{\boldsymbol{x}}{\varepsilon}     ,t-\tau\Big )\nabla \dot{\boldsymbol{u}}(\boldsymbol{x},\tau) d\tau \cdot \boldsymbol{n} \Big] \Big ]\Big \vert_{\boldsymbol{x} \in \Sigma}= \boldsymbol{0},
\end{equation}
where $\boldsymbol{n}$ represents the outward normal to the interface $\Sigma$. 
Since $\mathbb{G}^{m}$ and $\rho^{m}$ are $\mathcal{Q}$-periodic and the body forces are $\mathcal{L}$-periodic, the micro-displacement depends on both the slow variable $\boldsymbol{x}$ and the fast one $\boldsymbol{\xi}$ and can be expressed as
\[
\boldsymbol{u} = \boldsymbol{u}\Big(\boldsymbol{x},\frac{\boldsymbol{x}}{\varepsilon},t\Big).
\]
The two-sided Laplace transform of an arbitrary, real valued, time varying function, $\emph{f} \in \mathbb{R}$, is defined as, \cite{paley1934fourier},
\begin{equation}
\label{eqn:lap}
\mathcal{L}(f(t))= \hat{f}(s) = \int_{-\infty}^{+\infty} f(t) e^{-st} dt, \quad s\in \mathbb{C}, 
\end{equation} 
where, the Laplace argument, $s$, and the Laplace transform, $\hat{f}$, are complex valued (i.e. $\hat{f}:\mathbb{C} \to \mathbb{C}$ ). The derivative rule for the Laplace transform is provided by
\begin{equation}
\label{eqn:der}
\mathcal{L}\Big (\frac{\partial^{n}f(t)}{\partial t^{n}}\Big) = s^{n}\hat{f}(s),
\end{equation}
and the convolution rule of $f_{1}$ and $f_{2}$ is given as
\begin{equation}
\label{eqn:conv}
\mathcal{L}(f_{1}(t) \ast f_{2}(t)) = \mathcal{L}(f_{1}(t)) \mathcal{L}(f_{2}(t)), 
\end{equation}
Equation $~\eqref{eqn:sm}$ governing the periodic viscoelastic material in the time domain will be recast in the Laplace domain employing the Laplace transform ~\eqref{eqn:lap}, the convolution rule ~\eqref{eqn:conv} and the derivative rule ~\eqref{eqn:der}. Therefore, in the Laplace domain, it results
\begin{equation}
\nabla \cdot \Big[\mathcal{L}\Big(\mathbb{G}^{m}\Big(\frac{\boldsymbol{x}}{\varepsilon},t\Big)\Big) \mathcal{L}\Big(\nabla{\dot{u}}\Big(\frac{\boldsymbol{x}}{\varepsilon},\boldsymbol{x},t\Big)\Big)\Big] + \mathcal{L}(\boldsymbol{b}(\boldsymbol{x},t)) = \rho^{m}\Big(\frac{\boldsymbol{x}}{\varepsilon}\Big) \mathcal{L}\Big(\boldsymbol{\ddot{u}}\Big(\frac{\boldsymbol{x}}{\varepsilon},\boldsymbol{x},t\Big)\Big),
\end{equation} 
or, in other terms:
\begin{equation}
\label{eqn:tl}
\nabla \cdot \Big [\mathbb{\hat{G}}^{m}\Big(\frac{\boldsymbol{x}}{\varepsilon},s\Big) s \nabla \hat{u}\Big(\frac{\boldsymbol{x}}{\varepsilon},\boldsymbol{x},s\Big)\Big ] + \hat{\boldsymbol{b}}(\boldsymbol{x},s) = \rho^{m}\Big(\frac{\boldsymbol{x}}{\varepsilon}\Big) s^{2} \boldsymbol{\hat{u}} \Big(\frac{\boldsymbol{x}}{\varepsilon},\boldsymbol{x},s \Big),
\end{equation}
where $\hat{u}$ and $\nabla \hat{u}$ represent the micro-displacement field and the gradient of the micro-displacement field converted in the Laplace domain. Moreover, $\mathbb{\hat{G}}^{m}$ is the micro-relaxation tensor and $\hat{\boldsymbol{b}}(\boldsymbol{x},s)$ are the body forces transformed in the Laplace domain. In addition, it is convenient to consider $\mathbb{\hat{C}}^{m}\Big(\frac{\boldsymbol{x}}{\varepsilon},s \Big) =s\mathbb{\hat{G}}^{m}\Big(\frac{\boldsymbol{x}}{\varepsilon},s \Big)$.
The governing equation of the periodic viscoelastic material defined in the Laplace domain is 
\begin{eqnarray}
\label{eqn:fe}
\nabla \cdot (\hat{\mathbb{C}}^{m}\nabla \hat{\boldsymbol{u}})+ \hat{\boldsymbol{b}}=\rho^{m}s^{2}\hat{\boldsymbol{u}}.
\end{eqnarray}
Denoting with $[[f]]=f^{i}(\Sigma)-f^{j}(\Sigma)$ the jump of the function values $f$ at the interface $\Sigma$ between two phases $i$ and $j$ in the periodic cell $\mathcal{A}$, the following continuity conditions hold for a perfectly bonded interface
\begin{equation}
\label{icld}
[[\hat{\boldsymbol{u}}(\boldsymbol{x},s)]]\vert_{\boldsymbol{x} \in \Sigma}= \boldsymbol{0}, \quad
\Big [ \Big [ \Big ( \hat{\mathbb{C}}^{m}\Big (\frac{\boldsymbol{x}}{\varepsilon},s\Big) \nabla \hat{\boldsymbol{u}}\Big(\boldsymbol{x},\frac{\boldsymbol{x}}{\varepsilon},s \Big)\Big ) \cdot \boldsymbol{n} \Big] \Big] \Big \vert_{\boldsymbol{x} \in \Sigma}= \boldsymbol{0},
\end{equation}
where $\boldsymbol{n}$ represents the outward normal to the interface $\Sigma$. 
The solution of Eq. \eqref{eqn:fe} is too expensive from both a numerical and an analytical point of view, because the coefficients are $\mathcal{Q}$-periodic. In order to cope with such a drawback, it is convenient to employ a non-local asymptotic homogeneization technique to turn the heterogeneous material into an equivalent homogeneous one. Such a procedure generates equations, equivalent to \eqref{eqn:fe}, whose coefficients are not affected by oscillations and their solutions are close to those of the original equation. Moreover, the computational cost to solve \eqref{eqn:fe}  significantly reduces.\\
 In the equivalent homogenized material, by considering a reference system $\{O,\boldsymbol{e}_{1},\boldsymbol{e}_{2}\}$, the macro-displacement transformed in the Laplace domain is denoted as $\hat{\boldsymbol{U}}(\boldsymbol{x})=\hat{U}_{i}\boldsymbol{e}_{i}$, with respect to a point $\boldsymbol{x}$, and the transformed displacement gradiend is defined as $\nabla \hat{\boldsymbol{U}}(\boldsymbol{x})= \frac{\partial \hat{U}_{i}}{\partial x_{j}}\boldsymbol{e}_{1}\otimes \boldsymbol{e}_{2}$.

%\begin{equation}
%\label{eqn:conint}
%[[\hat{u}_{h}]]\vert_{\boldsymbol{x} \in \Sigma}=0,\quad %[[\hat{\sigma_{ij}}n_{j}]]\vert_{\boldsymbol{x} \in \Sigma}=0,
%%\end{equation}
%where $\boldsymbol{n}$ represents the outward normal to the %interface $\Sigma$.  
%Interface conditions $eq.~\eqref{eqn:conint}$ can be %expressed in terms of the fast variable $\boldsymbol{\xi}$
\section{Asymptotic expansion of the microscopic displacement}
\label{sec:2}
Based on the asymptotic approach developed in \cite{Bakhvalov1984}, \cite{Smyshlyaev2000}, \cite{BacigalupoGambarotta2014}, the micro-displacement $\boldsymbol{u}$ is expressed as an asymptotic expansion in terms of the parameter $\varepsilon$ that separates the slow $\boldsymbol{x}$ variable from the fast one $\boldsymbol{\xi} = \frac{\boldsymbol{x}}{\varepsilon}$,
\begin{flalign}
\label{eqn:at}
&u_{h}\Big(\boldsymbol{x},\frac{\boldsymbol{x}}{\varepsilon},t\Big) = \sum_{l=0}^{+\infty} \varepsilon^{l} u^{(l)}_{h} = u^{(0)}_{h}\Big(\boldsymbol{x},\frac{\boldsymbol{x}}{\varepsilon},t\Big) + \varepsilon u^{(1)}_{h}\Big(\boldsymbol{x},\frac{\boldsymbol{x}}{\varepsilon},t\Big) +\varepsilon^{2}u^{(2)}_{h}\Big(\boldsymbol{x},\frac{\boldsymbol{x}}{\varepsilon},t\Big)+\textrm{O}(\varepsilon^{3}),&
\end{flalign}
The Laplace transform $\eqref{eqn:lap}$ is applied to Eq.$~\eqref{eqn:at}$ and leads to
\begin{flalign}
\label{eqr:tralapex}
&\mathcal{L}\Big(u_{h}\Big(\boldsymbol{x},\frac{\boldsymbol{x}}{\varepsilon},t\Big)\Big)  =  \sum_{l=0}^{+\infty} \varepsilon^{l} \hat{u}^{(l)}_{h} = \hat{u}^{(0)}_{h}\Big(\boldsymbol{x},\frac{\boldsymbol{x}}{\varepsilon},s\Big) + \varepsilon \hat{u}^{(1)}_{h}\Big(\boldsymbol{x},\frac{\boldsymbol{x}}{\varepsilon},s\Big)+\varepsilon^{2}\hat{u}^{(2)}_{h}\Big(\boldsymbol{x},\frac{\boldsymbol{x}}{\varepsilon},s\Big)+ \textrm{O}(\varepsilon^{3}),&
\end{flalign} 
which is equivalent to the asymptotic expansion of the micro-displacement performed in the time domain. 
Let us consider the formula
\begin{flalign}
\label{eqn:cr}
&\frac{D}{Dx_{k}} \hat{\boldsymbol{u}}\Big (\boldsymbol{x},\boldsymbol{\xi} =\frac{\boldsymbol{x}}{\varepsilon}\Big )= \Big [\frac{\partial \hat{u}_{h}(\boldsymbol{x},\boldsymbol{\xi})}{\partial x_{k}} + \frac{\partial \hat{u}_{h}(\boldsymbol{x},\boldsymbol{\xi})}{\partial \xi_{k}} \frac{\partial \xi_{k}}{\partial x_{k}}\Big]\Big \vert _{\boldsymbol{\xi}= \frac{\boldsymbol{x}}{\varepsilon}}=\Big [\frac{\partial}{\partial x_{k}} \hat{u}_{h}(\boldsymbol{x},\boldsymbol{\xi})+\frac{1}{\varepsilon}\hat{u}_{h,k}\Big]\Big \vert _{\boldsymbol{\xi}= \frac{\boldsymbol{x}}{\varepsilon}},&
\end{flalign}
which introduces the macroscopic derivative  $\frac{\partial}{\partial x_{k}} \hat{u}_{h}$  and the microscopic derivative $\hat{u}_{h,k}$ in the transformed Laplace domain,
and let us apply it to the asymptotic expansion$~\eqref{eqr:tralapex}$, leading to:
\begin{flalign}
\label{eqn:ep}
&\frac{D}{Dx_{k}} \hat{\boldsymbol{u}}\Big (\boldsymbol{x},\boldsymbol{\xi} = \frac{\boldsymbol{x}}{\varepsilon}\Big )=\Big [ \frac{\partial \hat{u}^{(0)}_{h}}{\partial x_{k}} + \varepsilon \frac{\partial \hat{u}^{(1)}_{h}}{\partial x_{k}} + \varepsilon^{2} \frac{\partial \hat{u}^{(2)}_{h}}{\partial x_{k}}+...\Big ]+\frac{1}{\varepsilon} \Big [\hat{u}^{0}_{h,k}+ \varepsilon \hat{u}^{(1)}_{h,k} + \varepsilon^{2} \hat{u}^{(2)}_{h,k}+...\Big ]\Big \vert _{\boldsymbol{\xi}= \frac{\boldsymbol{x}}{\varepsilon}}.&
\end{flalign}
The asymptotic technique searches for the solution of Eq. \eqref{eqn:fe} as a decomposition in increasing powers of the microscopic lenght $\varepsilon$. To this purpose, the replacement of the asymptotic expansion $~\eqref{eqr:tralapex}$ into the microscopic field equation $~\eqref{eqn:fe}$ in the Laplace domain and the rearrangement of the terms with equal power $\varepsilon$  yield to the asymptotic field equation
\begin{flalign}
\label{eqn:ce}
&\varepsilon^{-2}\Big (\hat{C}^{m}_{ijhk}\hat{u}^{(0)}_{h,k} \Big )_{,j}+\varepsilon^{-1} \Big [ \Big( \hat{C}^{m}_{ijhk}\Big ( \frac{\partial \hat{u}^{(0)}_{h}}{\partial x_{k}}+\hat{u}^{(1)}_{h,k} \Big) \Big)_{,j}+\frac{\partial}{\partial x_{j}} \Big (\hat{C}^{m}_{ijhk}\hat{u}^{(0)}_{h,k} \Big )\Big ]+&
\end{flalign}
\begin{flalign*}
&+ \varepsilon^{0} \Big [ \Big( \hat{C}^{m}_{ijhk}\Big ( \frac{\partial \hat{u}^{(1)}_{h}}{\partial x_{k}}+\hat{u}^{(2)}_{h,k} \Big) \Big)_{,j}+\frac{\partial}{\partial x_{j}} \Big(\hat{C}^{m}_{ijhk}\Big(\frac{\partial\hat{u}^{(0)}_{h}}{\partial x_{k}}+\hat{u}^{(1)}_{h,k}\Big)\Big)+\hat{b}_{i}-\rho^{m}s^{2}u^{(0)}_{h}\Big ]+&
\end{flalign*}
\begin{flalign*}
&\varepsilon \Big [ \Big( \hat{C}^{m}_{ijhk}\Big (\frac{\partial \hat{u}^{(2)}_{h}}{\partial x_{k}}+\hat{u}^{(3)}_{h,k} \Big) \Big)_{,j}+\frac{\partial}{\partial x_{j}} \Big (\hat{C}^{m}_{ijhk}\Big (\frac{\partial\hat{u}^{(1)}_{h}}{\partial x_{k}}+\hat{u}^{(2)}_{h,k}\Big )\Big)-\rho^{m}s^{2}\hat{u}^{(1)}_{h}\Big ]+\textrm{O}(\varepsilon^{2})\Big ] \Big \vert_{\boldsymbol{\xi}= \frac{\boldsymbol{x}}{\varepsilon}}=0.&
\end{flalign*}
Interface conditions $~\eqref{icld}$ are rephrased with respect to the fast variable $\boldsymbol{\xi}$ since the micro-displacement $\hat{u}_{h}(\boldsymbol{x},\boldsymbol{\xi})$ is supposed to be $\mathcal{Q}-$periodic with respect to $\boldsymbol{\xi}$ and smooth in the slow variable $\boldsymbol{x}$. Indicating with $\Sigma_{1}$ the interface between two different phases in the unit cell $\mathcal{Q}$ and considering the asymptotic expansion $~\eqref{eqr:tralapex}$ of the micro-displacement, interface conditions read
\begin{flalign}
\label{eqn:condinteq}
&\Big [\Big[\hat{u}^{(0)}_{h}\Big]\Big]\Big \vert_{\boldsymbol{\xi} \in \Sigma_{1}}+\varepsilon \Big[\Big[\hat{u}^{(1)}_{h}\Big ]\Big]\Big \vert_{\boldsymbol{\xi} \in \Sigma_{1}}+\varepsilon^{2}\Big [\Big [\hat{u}^{(2)}_{h}\Big]\Big]\Big \vert_{\boldsymbol{\xi} \in \Sigma_{1}}+...=0&
\end{flalign}
\begin{flalign*}
&\frac{1}{\varepsilon}\Big [\Big[\Big(\hat{C}^{m}_{ijhk}\hat{u}^{(0)}_{h,k}\Big)n_{j}\Big]\Big ]\Big \vert_{\boldsymbol{\xi} \in \Sigma_{1}}+\varepsilon^{0}\Big [\Big[ \Big ( \hat{C}^{m}_{ijhk}\Big (\frac{\partial\hat{u}^{(0)}_{h}}{\partial \hat{x}_{k}}+\hat{u}^{(1)}_{h,k}\Big) \Big )n_{j}\Big ]\Big]\Big \vert_{\boldsymbol{\xi} \in \Sigma_{1}}+&
\end{flalign*}
\begin{flalign*}
&+\varepsilon\Big [\Big[ \Big ( \hat{C}^{m}_{ijhk}\Big (\frac{\partial\hat{u}^{(1)}_{h}}{\partial \hat{x}_{k}}+\hat{u}^{(2)}_{h,k}\Big) \Big )n_{j}\Big ]\Big]\Big \vert_{\boldsymbol{\xi} \in \Sigma_{1}}++\varepsilon^{2}\Big [\Big[ \Big ( \hat{C}^{m}_{ijhk}\Big (\frac{\partial\hat{u}^{(2)}_{h}}{\partial \hat{x}_{k}}+\hat{u}^{(3)}_{h,k}\Big) \Big )n_{j}\Big ]\Big]\Big \vert_{\boldsymbol{\xi} \in \Sigma_{1}}+...+=0.&
\end{flalign*} 
\subsection*{Recursive differential problems and their solutions}
The asymptotic field equation $~\eqref{eqr:tralapex}$ produces a set of recursive differential problems that determine sequentially the solutions $\hat{u}^{0}$, $\hat{u}^{1}$...\quad In particular, at the order $\varepsilon^{-2}$, the differential problem, which stems from problem $~\eqref{eqn:ce}$, is
\begin{equation}
\label{eqn:e-2}
\Big(\hat{C}^{m}_{ijhk}\hat{u}^{(0)}_{h,k}\Big)_{,j}=f^{(0)}_{i}(\boldsymbol{x}),
\end{equation} 
with interface conditions
\[
\Big[\Big[\hat{u}^{(0)}_{h}\Big]\Big]\Big\vert_{\boldsymbol{\xi} \in \Sigma_{1}}=0 \quad\Big[\Big[\Big(\hat{C}^{m}_{ijhk}\hat{u}^{(0)}_{h,k}\Big)n_{j}\Big]\Big]\Big\vert_{\boldsymbol{\xi} \in \Sigma_{1}}=0.
\]
The solvabiliy condition of this differential problem, in the class of $\mathcal{Q}-$periodic solutions $\hat{u}^{(0)}_{h}$, implies that $f^{(0)}_{i}(\boldsymbol{x})=0$ and so the differential problem $~\eqref{eqn:e-2}$ develops in the form
\begin{equation}
\label{eqn:sol}
\Big(\hat{C}^{m}_{ijhk}\hat{u}^{(0)}_{h,k}\Big )_{,j}=0.
\end{equation}
The solution results to be
\begin{equation}
\label{eqn:so_l}
\hat{u}^{(0)}_{h}(\boldsymbol{x},\boldsymbol{\xi},s) = \hat{U}^{M}_{h}(\boldsymbol{x},s),
\end{equation}
where $\hat{U}^{M}_{h}(\boldsymbol{x},s)$ is the transformed macroscopic displacement that does not depend on the microstructure.\\
Bearing in mind the solution $~\eqref{eqn:so_l}$, the differential problem from $\eqref{eqn:ce}$ at the order $\varepsilon^{-1}$ is
\begin{equation}
\label{eqn:e1n}
\Big (\hat{C}^{m}_{ijhk}\hat{u}^{(1)}_{h,k}\Big)_{,j}+\hat{C}^{m}_{ijhk,j}\frac{\partial \hat{U}^{M}_{h}}{\partial x_{k}}=f^{(1)}_{i}(\boldsymbol{x}),
\end{equation} 
since $\hat{U}^{M}_{h,k}=0$. Its interface conditions are
\[
\Big [\Big[\hat{u}^{(1)}_{h}\Big]\Big]\Big\vert_{\boldsymbol{\xi}\in \Sigma_{1}}=0 \quad \Big [\Big[\Big(\hat{C}^{m}_{ijhk}\Big (\hat{u}^{(1)}_{h,k}+\frac{\partial \hat{U}^{M}_{h}}{\partial x_{k}}\Big )\Big)n_{j}\Big ]\Big ]\Big \vert_{\boldsymbol{\xi} \in \Sigma_{1}} =0.
\]
Similarly, the solvability condition in the class of $\mathcal{Q}-$periodic functions ensures that  
\begin{equation}
\label{eq:e1cs}
f^{(1)}_{i}(\boldsymbol{x})=\langle \hat{C}^{m}_{ijhk,j} \rangle \frac{\partial \hat{U}^{(M)}_{h}}{\partial x_{k}}, 
\end{equation}  
where $\langle (\cdot) \rangle=\frac{1}{|\mathcal{Q}|} \int_{\mathcal{Q}}^{} (\cdot) d\boldsymbol{\xi}$ and  $|\mathcal{Q}|= \delta$. Moreover the $\mathcal{Q}$-periodicity of the components $\hat{C}^{m}_{ijhk}$ and the divergence theorem entail $f^{(1)}_{i}(\boldsymbol{x})=0$ and the differential problem
\begin{equation}
\label{eqn:solsp}
\Big(\hat{C}^{m}_{ijhk}\hat{u}^{(1)}_{h,k}\Big)_{,j}+\hat{C}^{m}_{ijhk,j}\frac{\partial \hat{U}^{M}_{h}}{\partial x_{k}}=0, \quad \forall \frac{\partial \hat{U}^{M}_{h}}{\partial x_{k}} 
\end{equation}
has the following solution
\begin{equation}
\label{eqn:solp1}
\hat{u}^{(1)}_{h}(\boldsymbol{x},\boldsymbol{\xi},s)=N^{(1,0)}_{hpq_{1}}(\boldsymbol{\xi})\frac{\partial \hat{U}^{M}_{p}}{\partial x_{q_{1}}},
\end{equation}
where $N^{(1,0)}_{hpq_{1}}$ is the perturbation function, which depends on the fast variable $\boldsymbol{\xi}$. The perturbation functions are supposed to have zero mean over the unit cell $\mathcal{Q}$  and so $N^{(1,0)}_{hpq_{1}}$ complies with the normalization condition
\begin{equation}
\langle N^{(1,0)}_{hpq_{1}} \rangle = \frac{1}{|\mathcal{Q}|} \int_{\mathcal{Q}} N^{(1,0)}_{hpq_{1}}(\boldsymbol{\xi}) d\boldsymbol{\xi}=0.
\end{equation}  
Moreover, the perturbation functions exclusively depend on the geometry and on the mechanical properties of the microstructure.
The differential problem at order $\varepsilon^{0}$ is 
\begin{equation}
\label{eqn:e-1}
\Big( \hat{C}^{m}_{ijhk}\Big ( \frac{\partial \hat{u}^{(1)}_{h}}{\partial x_{k}}+\hat{u}^{(2)}_{h,k} \Big) \Big)_{,j}+\frac{\partial}{\partial x_{j}} \Big( \hat{C}^{m}_{ijhk}\Big (\frac{\partial \hat{u}^{(0)}_{h}}{\partial x_{k}}+\hat{u}^{(1)}_{h,k}\Big )\Big)-\rho^{m}s^{2}\hat{u}^{(0)}_{i} = f^{(2)}_{i}(\boldsymbol{x})
\end{equation}
with interface conditions
\[
\Big[\Big[\hat{u}^{(2)}_{h}\Big]\Big]\Big\vert_{\boldsymbol{\xi} \in \Sigma_{1}}=0 \quad \Big [\Big[ \Big ( \hat{C}^{m}_{ijhk}\Big (\frac{\partial\hat{u}^{(1)}_{h}}{\partial \hat{x}_{k}}+\hat{u}^{(2)}_{h,k}\Big) \Big )n_{j}\Big ]\Big]\Big \vert_{\boldsymbol{\xi} \in \Sigma_{1}}=0.
\]
Considering the solutions $~\eqref{eqn:so_l}$ and $~\eqref{eqn:solp1}$ of the differential problems at order $\varepsilon^{-2}$ and $\varepsilon^{-1}$, respectively, the differential problem $~\eqref{eqn:e-1}$ is turned into
\begin{flalign}
\label{eqn:prob0}
&\Big (\hat{C}^{m}_{ijhk}\hat{u}^{(2)}_{h,k}\Big)_{,j}+\Big(\Big(\hat{C}^{m}_{ijhk}N^{(1,0)}_{hpq_{1}}\Big)_{,j}+\hat{C}^{m}_{ijhq_{1}}+\Big (\hat{C}^{m}_{ijhk}N^{(1,0)}_{hpq_{1},k}\Big)\Big)\frac{\partial^{2}\hat{U}^{M}_{p}}{\partial x_{q_{1}}\partial x_{j}}-\rho^{m}s^{2}\hat{U}^{M}_{i} = f^{(2)}_{i}(\boldsymbol{x}),&
\end{flalign}
with interface conditions
\[
\Big[\Big[\hat{u}^{(2)}_{h}\Big]\Big]\Big\vert_{\boldsymbol{\xi} \in \Sigma_{1}}=0, \quad \Big [\Big[ \Big ( \hat{C}^{m}_{ijhk}\Big (\hat{u}^{(2)}_{h,k}+N^{(1,0)}_{hpq_{1}}\frac{\partial^{2}\hat{U}^{M}_{p}}{\partial x_{q_{1}}\partial x_{k}}\Big) \Big )n_{j}\Big ]\Big]\Big \vert_{\boldsymbol{\xi} \in \Sigma_{1}}=0.
\]
Again, solvability condition of differential problem $~\eqref{eqn:prob0}$ in the class of $\mathcal{Q}-$periodic functions and the divergence theorem lead to 
\begin{equation}
f^{(2)}_{i}(\boldsymbol{x})= \langle \hat{C}^{m}_{ijhq_{1}}+\hat{C}^{m}_{ijhk}N^{(1,0)}_{hpq_{1},k}\rangle\frac{\partial^{2}\hat{U}^{M}_{p}}{\partial x_{q_{1}}\partial x_{j}}-\langle\rho^{m}\rangle s^{2}\hat{U}^{M}_{i} 
\end{equation} 
and consequentely the solution of the differential problem at the order $\varepsilon^{0}$ is
\begin{equation}
\label{eqn:sol0}
\hat{u}^{(2)}_{h}(\boldsymbol{x},\boldsymbol{\xi},s) = N^{(2,0)}_{hpq_{1}q_{2}}\frac{\partial^{2}\hat{U}^{M}_{p}}{\partial x_{q_{1}}\partial x_{q_{2}}}+N^{(2,2)}_{hp} s^{2}U^{M}_{p},
\end{equation}  
where $N^{(2,2)}_{hp}$ is the perturbation function depending on the parameter $s$.

\section{Cell problems and perturbation functions}
\label{sec:four} 
In the Section \ref{sec:2}, the solutions $\hat{u}^{(0)}_{h}$, $\hat{u}^{(1)}_{h}$, $\hat{u}^{(2)}_{h}$, $...$\quad have been established. Such solutions are employed to formulate the cell problems, which are classified according to the even power of the parameter $s$. 

\subsection*{Cell problems related to $s^{0}$}
The substitution of solution $\eqref{eqn:solp1}$ into problem $\eqref{eqn:solsp}$ leads to the following cell problem at the order $\varepsilon{^{-1}}$
\begin{equation}
\label{cps1}
\Big (\hat{C}^{m}_{ijhk}N^{(1,0)}_{hpq_{1},k}\Big)_{,j}+\hat{C}^{m}_{ijpq_{1},j}=0
\end{equation}
with interface conditions derived in terms of the perturbation function $N^{(1,0)}_{hpq_{1},k}$ 
\begin{equation}
\Big[\Big[N^{(1,0)}_{ipq_{1}}\Big]\Big]\Big\vert_{\boldsymbol{\xi}\in \Sigma_{1}}=0 \quad \Big [\Big [\Big ( \hat{C}^{m}_{ijhk}\Big(N^{(1,0)}_{hpq_{1},k}+\delta_{hp}\delta_{kq_{1}}\Big )\Big )n_{j}\Big ]\Big ]\Big \vert_{\boldsymbol{\xi} \in \Sigma_{1}}=0,
\end{equation}
where $\delta_{hp}$ and $\delta_{kq_{1}}$ are the Kronecker delta functions.
Once the perturbation function $N^{(1,0)}_{hpq_{1},k}$ has been determined, and thanks to equation $\eqref{eqn:prob0}$ and its solution $\eqref{eqn:sol0}$, the cell problem at the order $\varepsilon^{0}$ is derived and the symmetrized version with respect to indices $q_{1}$ and $q_{2}$ is
\begin{flalign*}
\label{eqn:sym}
&\Big (\hat{C}^{m}_{ijhk}N^{(2,0)}_{hpq_{1}q_{2},k}\Big )_{,j}+\frac{1}{2}\Big [\Big(\hat{C}^{m}_{ikhq_{2}}N^{(1,0)}_{hpq_{1}}\Big)_{,k}+\hat{C}^{m}_{iq_{2}pq_{1}}+\Big (\hat{C}^{m}_{iq_{2}hk}N^{(1,0)}
_{hpq_{1},k}\Big)+&
\end{flalign*} 
\begin{flalign*}
&+\Big (\hat{C}^{m}_{ikhq_{1}}N^{(1,0)}_{hpq_{2}}\Big )_{,k}+\hat{C}^{m}_{iq_{1}pq_{2}}+\Big(\hat{C}^{m}_{iq_{1}hk}N^{(1,0)}_{hpq_{2},k}\Big)\Big ]=&
\end{flalign*} 
\begin{flalign}
&=\frac{1}{2}\langle \hat{C}^{m}_{iq_{2}hq_{1}}+\Big (\hat{C}^{m}_{iq_{2}hk}N^{(1,0)}
_{hpq_{1},k}\Big)+\hat{C}^{m}_{iq_{1}hq_{2}}+\Big (\hat{C}^{m}_{iq_{1}hk}N^{(1,0)}_{hpq_{2},k}\Big) \rangle,&
\end{flalign}
with interface conditions
\begin{equation}
\label{eqn:icsym}
\Big[\Big[N^{(2,0)}_{ipq_{1}q_{2}}\Big]\Big]\Big\vert_{\boldsymbol{\xi} \in \Sigma_{1}}=0,  
\end{equation}
\[
\Big [\Big[\Big (\hat{C}^{m}_{ijhk}N^{(2,0)}_{hpq_{1}q_{2},k}+\frac{1}{2}\Big(\hat{C}^{m}_{ijhq_{2}}N^{(1,0)}_{hpq_{1}}+\hat{C}^{m}_{ijhq_{1}}N^{(1,0)}_{hpq_{2}}\Big ) \Big ) n_{j}\Big ] \Big ]\Big \vert_{\boldsymbol{\xi} \in \Sigma_{1}}=0.
\]
The solution of the cell problem \eqref{eqn:sym} and \eqref{eqn:icsym} is the perturbation function  $N^{(2,0)}_{ipq_{1}q_{2}}$. 
The cell problem at the order $\varepsilon^{w}$ with $w\in \mathbb{Z}$ and $w \geq 1$ is
\begin{flalign*}
\label{meq}
&\Big(\hat{C}^{m}_{ijhk}N^{(w+2,0)}_{hpq_{1}...q_{w+2},k}\Big)_{,j}+\frac{1}{w+2}\sum_{\mathcal{P}^{*}(q)}^{}\Big [(\hat{C}^{m}_{ijhq_{w+2}}N^{(w+1,0)}_{hpq_{1}...q_{w+1}})_{,j}+&
\end{flalign*}
\begin{flalign*}
&+\hat{C}^{m}_{iq_{w+2}hj}N^{(w+1,0)}_{hpq_{1}...q_{w+1},j}+\hat{C}^{m}_{iq_{w+2}hq_{w+1}}N^{(w,0)}_{hpq_{1}...q_{w}}\Big ]=&
\end{flalign*}
\begin{flalign}
&=\frac{1}{w+2}\sum_{\mathcal{P}^{*}(q)}\langle \hat{C}^{m}_{iq_{w+2}hj}N^{(w+1,0)}_{hpq_{1}...q_{w+1},j}+
\hat{C}^{m}_{iq_{w+2}hq_{w+1}}N^{(w,0)}_{hpq_{1}...q_{w}}\rangle,&
\end{flalign}
and the corresponding interface conditions are
\begin{equation}
\Big[\Big[N^{(w+2,0)}_{ipq_{1}...q_{w+2}}\Big]\Big]\Big\vert_{\boldsymbol{\xi} \in \Sigma_{1}}=0,
\end{equation}
\[
\Big [\Big [ \Big ( \hat{C}^{m}_{ijhk} N^{(w+2,0)}_{hpq_{1}...q_{w+2},k}+\frac{1}{w+2}\sum_{P^{*}(q)}\hat{C}^{m}_{ijhq_{w+2}}N^{(w+1,0)}_{hpq_{1}...q_{w+1}} \Big ) n_{j} \Big ] \Big ] \Big \vert_{\boldsymbol{\xi} \in \Sigma_{1}}=0,                    
\]
where symbol $\mathcal{P}^{*}(q)$ denotes all the possible permutations of the multi-index $q=q_{1},q_{2},...,q_{l}$ that does not exhibit fixed indices (see Appendix B). The resolution of cell problem \eqref{meq}  allows to determining the form of the perturbation function   $N^{(w+2,0)}_{ipq_{1}...q_{w+2}}$.
\subsection*{Cell problems related to $s^{2}$}
The substitution of solution $\eqref{eqn:sol0}$ into Eq. $\eqref{eqn:prob0}$ generates the cell problem at the order $\varepsilon^{0}$
\begin{equation}
\label{22N}
\Big (\hat{C}^{m}_{ijhk}N^{(2,2)}_{hp,k}\Big )_{,j}-\rho^{m}\delta_{ip}=-\delta_{ip}	\Big \langle \rho^{m}\Big \rangle.
\end{equation}
with interface conditions:
\begin{equation}
\label{22icN}
\Big[\Big[N^{(2,2)}_{ip}\Big]\Big]\Big\vert_{\boldsymbol{\xi} \in \Sigma_{1}}=0, \quad \Big[\Big[\hat{C}^{m}_{ijhk}N^{(2,2)}_{hp,k}\Big]\Big]\Big\vert_{\boldsymbol{\xi} \in \Sigma_{1}}=0,
\end{equation}
as well as the cell problem $~\eqref{eqn:sym}$ related to the case $s^{0}$. From the resolution of problem \eqref{22N} and \eqref{22icN}, the perturbation function $N^{(2,2)}_{ip}$ is derived. \\  
The perturbation function $N^{(3,2)}_{ipq_{1}}$ is the solution of the cell problem obtained at the order $\varepsilon^{1}$
\begin{flalign}
\label{32N}
&\Big(\hat{C}^{m}_{ijhk}N^{(3,2)}_{hpq_{1},k}\Big)_{j}+\Big[(\hat{C}^{m}_{ijhq_{1}}N^{(2,2)}_{hp})_{,j}+\hat{C}^{m}_{iq_{1}hj}N^{(2,2)}_{hp,j}-\rho^{m}N^{(1,0)}_{ipq_{1}}\Big ]=\Big \langle \hat{C}^{m}_{iq_{1}hj}N^{(2,2)}_{hp,j}-\rho^{m}N^{(1,0)}_{ipq_{1}}\Big\rangle&
\end{flalign}
with interface conditions
\begin{equation}
\label{intconN32}
\Big[\Big[N^{(3,2)}_{ipq_{1}}\Big]\Big]\Big\vert_{\boldsymbol{\xi} \in \Sigma_{1}}=0, \quad \Big[\Big[\hat{C}^{m}_{ijhk}N^{(3,2)}_{hpq_{1},k}+\hat{C}^{m}_{ijhq_{1}}N^{(2,2)}_{hp})n_{j}\Big]\Big]\Big\vert_{\boldsymbol{\xi} \in \Sigma_{1}}=0.
\end{equation}
Meanwhile, at the order $\varepsilon^{2}$, the perturbation function $N^{(4,2)}_{ipq_{1}q_{2}}$ is solution of the cell problem 
\begin{flalign*}
&\Big(\hat{C}^{m}_{ijhk}N^{(4,2)}_{hpq_{1}q_{2},k}\Big)_{,j}+\frac{1}{2}\Big[ ( \hat{C}^{m}_{ijhq_{2}}N^{(3,2)}_{hpq_{1}})_{,j}+\hat{C}^{m}_{iq_{2}hq_{1}}N^{(2,2)}_{hp}+\hat{C}^{m}_{iq_{2}hk}N^{(3,2)}_{hpq_{1},k}+&
\end{flalign*}  
\begin{flalign*} 
&-\rho^{m}N^{(2,0)
}_{hpq_{1}q_{2}}+(\hat{C}^{m}_{ijhq_{1}}N^{(3,2)}_{hpq_{2}})_{,j}+\hat{C}^{m}_{iq_{1}hq_{2}}N^{(2,2)}_{hp}+\hat{C}^{m}_{iq_{1}hk}N^{(3,2)}_{hpq_{2},k}+&
\end{flalign*}
\begin{flalign*}
&-\rho^{m}N^{(2,0)
}_{hpq_{2}q_{1}} \Big ] = \frac{1}{2} \langle \hat{C}^{m}_{iq_{2}hq_{1}}N^{(2,2)}_{hp}+\hat{C}^{m}_{iq_{2}hk}N^{(3,2)}_{hpq_{1},k}-\rho^{m}N^{(2,0)
}_{hpq_{1}q_{2}}+&
\end{flalign*}
\begin{flalign}
&+\hat{C}^{m}_{iq_{1}hq_{2}}N^{(2,2)}_{hp}+\hat{C}^{m}_{iq_{1}hk}N^{(3,2)}_{hpq_{2},k}-\rho^{m}N^{(2,0)
}_{hpq_{2}q_{1}} \rangle,&
\end{flalign}
whose interface conditions are
\begin{flalign}
\Big[\Big[N^{(4,2)}_{ipq_{1}q_{2}}\Big]\Big]\Big\vert_{\boldsymbol{\xi} \in \Sigma_{1}}=0,
\end{flalign}
\begin{flalign*}
&\Big [ \Big [ \Big( \hat{C}^{m}_{ijhk}N^{(4,2)}_{hpq_{1}q_{2},k}+\frac{1}{2}\Big ( \hat{C}^{m}_{ijhq_{2}}N^{(3,2)}_{hpq_{1}}+\hat{C}^{m}_{ijhq_{1}}N^{(3,2)}_{hpq_{2}}\Big ) \Big )n_{j}\Big ]\Big ]\Big \vert_{\boldsymbol{\xi} \in \Sigma_{1}}=0.&
\end{flalign*}
Finally at the order $\varepsilon^{w+2}$, with $w\in \mathbb{Z}$ and  $w\ge 1$, the perturbation function $N^{(w+4,2)}_{ipq_{1}...q_{w+2}}$ is derived from the cell problem 
\begin{flalign*}
&\Big (\hat{C}^{m}_{ijhk}N^{(w+4,2)}_{hpq_{1}....q_{w+2},k}\Big )_{,j}+\frac{1}{w+2}\sum_{P^{*}(q)}\Big[ \Big (\hat{C}^{m}_{ijhq_{w+2}}N^{(w+3,2)}_{hpq_{1}...q_{w+1}}\Big )_{,j}+&
\end{flalign*}
\begin{flalign*}
&+\hat{C}^{m}_{iq_{w+2}hq_{w+1}}N^{(w+2,2)}_{hpq_{1}....q_{w}}+\hat{C}^{m}_{iq_{w+2}hj}N^{(w+3,2)}_{hpq_{1}...q_{w+1},j}-\rho^{m}N^{(w+2)}_{ipq_{1}...q_{w+2}} \Big ] =&
\end{flalign*}
\begin{flalign}
&=\frac{1}{w+2}\sum_{P^{*}(q)}\langle \hat{C}^{m}_{iq_{w+2}hq_{w+1}}N^{(w+2,2)}_{hpq_{1}....q_{w}}+\hat{C}^{m}_{iq_{w+2}hj}N^{(w+3,2)}_{hpq_{1}...q_{w+1},j}-\rho^{m}N^{(w+2,0)}_{ipq_{1}...q_{w+2}} \rangle,&
\end{flalign}
equipped with the interface conditions
\begin{flalign}
\Big[\Big[N^{(w+4,2)}_{ipq_{1}...q_{w+2}}\Big]\Big]\Big \vert_{\boldsymbol{\xi} \in \Sigma_{1}}=0,
\end{flalign}
\begin{flalign*}
\Big [ \Big [ \Big( \hat{C}^{m}_{ijhq_{w+2}}N^{(w+4,2)}_{hpq_{1}...q_{w+2},k}+\frac{1}{w+2}\sum_{P^{*}(q)}\Big ( \hat{C}^{m}_{ijhq_{w+2}}N^{(w+3,2)}_{hpq_{1}...q_{w+1}}\Big ) \Big )n_{j}\Big ]\Big ]\Big \vert_{\boldsymbol{\xi} \in \Sigma_{1}}=0.
\end{flalign*}
\subsection*{Cell problems related to $s^{2n}$}
In the present subsection, the cell problems related to power $s^{2n}$ are devised and their corresponding perturbation functions are established. At the order $\varepsilon^{(2n-2)}$, with $n \in \mathbb{Z}$ and $n \geq 2$, the cell problem is    

\begin{equation}
\Big (\hat{C}^{m}_{ijhk}N^{(2n,2n)}_{hp,k}\Big )_{,j}-\rho^{m}N^{(2n-2,2n-2)}_{ip} = -\langle \rho^{m} N^{(2n-2,2n-2)}_{ip}\rangle,
\end{equation}
with interface conditions
\begin{equation}
\Big [\Big[N^{(2n,2n)}_{ip}\Big]\Big]\Big\vert_{\boldsymbol{\xi} \in \Sigma_{1}}=0, \quad
\Big[\Big[\hat{C}^{m}_{ijhk}N^{(2n,2n)}_{hp,k}\Big]\Big] \Big\vert_{\boldsymbol{\xi} \in \Sigma_{1}}=0,
\end{equation}
and its solution is the perturbation function $N^{(2n,2n)}_{ip}$.\\
Whereas at the order $\varepsilon^{(2n-1)}$, the perturbation function $N^{(2n+1,2n)}_{ipq_{1}}$ is the solution of the cell problem
\begin{flalign*}
&\Big(\hat{C}^{m}_{ijhk}N^{(2n+1,2n)}_{hpq_{1},k}\Big)_{,j}+\Big [\Big(\hat{C}^{m}_{ijhq_{1}}N^{(2n,2n)}_{hp}\Big)_{,j}+\hat{C}^{m}_{iq_{1}hk}N^{(2n,2n)}_{hp,k}-\rho^{m}N^{(2n-1,2n-2)}_{ipq_{1}}\Big ]=&
\end{flalign*}
\begin{flalign}
&= \Big \langle \hat{C}^{m}_{iq_{1}hk}N^{(2n,2n)}_{hp,k}-\rho^{m}N^{(2n-1,2n-2)}_{ipq_{1}} \Big \rangle,&
\end{flalign}
with interface conditions:
\begin{flalign*}
\Big[\Big[N^{(2n+1,2n)}_{ipq_{1}}\Big]\Big]\Big\vert_{\boldsymbol{\xi} \in \Sigma_{1}}=0,
\end{flalign*}
\begin{flalign}
\Big[\Big[(\hat{C}^{m}_{ijhk}N^{(2n+1,2n)}_{hpq_{1},k}+\hat{C}^{m}_{ijhq_{1}}N^{(2n,2n)}_{hp})n_{j}\Big]\Big]\Big\vert_{\boldsymbol{\xi} \in \Sigma_{1}}=0.
\end{flalign}
The cell problem evaluated for $\varepsilon^{(2n)}$ is
\begin{flalign*}
&\Big(\hat{C}^{m}_{ijhk}N^{(2n+2,2n)}_{hpq_{1}q_{2},k}\Big)_{,j}+\frac{1}{2}\Big [ \Big(\hat{C}^{m}_{ijhq_{2}}N^{(2n+1,2n)}_{hpq_{1}}\Big)_{,j}+\hat{C}^{m}_{iq_{2}hq_{1}}N^{(2n,2n)}_{hp}+&
\end{flalign*}
\begin{flalign*}
&+\hat{C}^{m}_{iq_{2}hj}N^{(2n+1,2n)}_{hpq_{1},j}-\rho^{m}N^{(2n,2n-2)}_{ipq_{1}q_{2}}+ \Big(\hat{C}^{m}_{ijhq_{1}}N^{(2n+1,2n)}_{hpq_{2}}\Big)_{,j}+&
\end{flalign*}
\begin{flalign*}
&+\hat{C}^{m}_{iq_{1}hq_{2}}N^{(2n,2n)}_{hp}+\hat{C}^{m}_{iq_{1}hj}N^{(2n+1,2n)}_{hpq_{2},j}-\rho^{m}N^{(2n,2n-2)}_{ipq_{2}q_{1}}\Big ]=&
\end{flalign*}
\begin{flalign*}
& \Big \langle \frac{1}{2} \hat{C}^{m}_{iq_{2}hq_{1}}N^{(2n,2n)}_{hp}+\hat{C}^{m}_{iq_{2}hj}N^{(2n+1,2n)}_{hpq_{1},j}-\rho^{m}N^{(2n,2n-2)}_{ipq_{1}q_{2}}+&
\end{flalign*}
\begin{flalign}
&+\hat{C}^{m}_{iq_{1}hq_{2}}N^{(2n,2n)}_{hp}+\hat{C}^{m}_{iq_{1}hj}N^{(2n+1,2n)}_{hpq_{2},j}-\rho^{m}N^{(2n,2n-2)}_{ipq_{2}q_{1}} \rangle,&
\end{flalign}
with interface conditions
\begin{equation}
\Big[\Big[N^{(2n+2,2n)}_{ipq_{1}q_{2}}\Big]\Big]\Big\vert_{\boldsymbol{\xi} \in \Sigma_{1}}=0,
\end{equation}
\begin{flalign*}
&\Big [ \Big [ \Big( \hat{C}^{m}_{ijhk}N^{(2n+2,2n)}_{hpq_{1}q_{2},k}+\frac{1}{2}\Big ( \hat{C}^{m}_{ijhq_{2}}N^{(2n+1,2n)}_{hpq_{1}}+\hat{C}^{m}_{ijhq_{1}}N^{(2n+1,2n)}_{hpq_{2}}\Big ) \Big )n_{j}\Big ]\Big ]\Big \vert_{\boldsymbol{\xi} \in \Sigma_{1}}=0,&
\end{flalign*}
and its solution is the perturbation function $N^{(2n+2,2n)}_{ipq_{1}q_{2}}$.\\
Finally, the perturbation function $N^{(w+2n+2,2n)}_{ipq_{1}...q_{w+2}}$ is the solution of the cell problem for $\varepsilon^{w+2n}$
\begin{flalign*}
&\Big (\hat{C}^{m}_{ijhk}N^{(w+2n+2,2n)}_{hpq_{1}...q_{w+2},k}\Big)_{,j}+\frac{1}{w+2}\sum_{P^{*}(q)}\Big [\Big(\hat{C}^{m}_{ijhq_{w+2}}N^{(w+2n+1,2n)}_{hpq_{1}...q_{w+1}}\Big)_{,j}+&
\end{flalign*}
\begin{flalign*}
&+\hat{C}^{m}_{iq_{w+2}hq_{w+1}}N^{(w+2n,2n)}_{hpq_{1}...q_{w}}+\hat{C}^{m}_{iq_{w+2}hj}N^{(w+2n+1,2n)}_{hpq_{1}...q_{w+1},j}-\rho^{m}N^{(w+2n,2n-2)}_{ipq_{1}...q_{w+2}}\Big ] =&
\end{flalign*}
\begin{flalign*}
&= \frac{1}{w+2}\sum_{P(q)}^{}\langle \hat{C}^{m}_{iq_{w+2}hq_{w+1}}N^{(w+2n,2n)}_{hpq_{1}...q_{w}}+\hat{C}^{m}_{iq_{w+2}hj}N^{(w+2n+1,2n)}_{hpq_{1}...q_{w+1},j}-\rho^{m}N^{(w+2n,2n-2)}_{ipq_{1}...q_{w+2}}\rangle,&
\end{flalign*}
whose interface conditions are
\begin{flalign}
\label{cps2n}
\Big[\Big[N^{(w+2n+2,2n)}_{ipq_{1}...q_{w+2}}\Big]\Big]\Big\vert_{\boldsymbol{\xi} \in \Sigma_{1}}=0,
\end{flalign}
\begin{flalign*}
\Big [ \Big [ \Big( \hat{C}^{m}_{ijhk}N^{(w+2n+2,2n)}_{hpq_{1}...q_{w+2},k}+\frac{1}{w+2}\sum_{P^{*}(q)}\Big ( \hat{C}^{m}_{ijhq_{w+2}}N^{(w+2n+1,2n)}_{hpq_{1}...q_{w+1}}\Big ) \Big )n_{j}\Big ]\Big ]\Big \vert_{\boldsymbol{\xi} \in \Sigma_{1}}=0.
\end{flalign*}
In \cite{Bakhvalov1984}, it is emphasized that the uniqueness of the perturbation functions $N^{(i,2r)}_{hpq_{1}...q_{i-2r}}$, derived from the cell problems $\eqref{cps1}$-$\eqref{cps2n}$, is guaranteed by imposing the normalization condition $\langle N^{(i,2r)}_{hpq_{1}...q_{i-2r}} \rangle =0$.
\section{Down-scaling relation, average field equation of infinite order and macroscopic problems}
The down-scaling relation referred to the transformed micro-displacement is expressed as an asymptotic expansion of powers of the microscopic length $\varepsilon$ relying on the transformed macro-displacement $\hat{U}^{M}_{h}(\boldsymbol{x},s)$, its gradients and the $\mathcal{Q}$-periodic perturbation functions. Such functions are delivered by solving the cell problems that are listed in the Section \eqref{sec:four}. Therefore, the replacement of the solutions of the recursive differential problems $~\eqref{eqn:so_l}$,$~\eqref{eqn:solp1}$,$~\eqref{eqn:sol0}$,$~\eqref{eqn:eqsol}$ and $\eqref{eqn:soleq}$ into the asymptotic expansion $\eqref{eqr:tralapex}$ enables establishing the transformed micro-displacement $\hat{u}_{h}(\boldsymbol{x},\boldsymbol{\xi},s)$ as 
\begin{flalign}
\label{eqn:dsr}
\hat{u}_{h}\Big(\boldsymbol{x},\frac{\boldsymbol{x}}{\varepsilon},s\Big)& = \Big ( \sum_{l,j=0}^{+\infty}\varepsilon^{j+l}\sum_{|q|=l}^{}N^{(2j+l,2j)}_{hpq}(\boldsymbol{\xi})\frac{\partial^{l}\hat{U}^{M}_{p}}{\partial x_{q}}s^{2j}\Big)\Big \vert _{\boldsymbol{\xi}= \frac{\boldsymbol{x}}{\varepsilon}}=&
\end{flalign}
\begin{flalign*}
&=\Big (\hat{U}^{M}_{h}(\boldsymbol{x},s)+\varepsilon N^{(1,0)}_{hpq_{1}}(\boldsymbol{\xi})\frac{\partial \hat{U}^{M}_{p}}{\partial x_{q_{1}}}+\varepsilon^{2}\Big ( N^{(2,0)}_{hpq_{1}q_{2}}(\boldsymbol{\xi})\frac{\partial^{2}\hat{U}^{M}_{p}}{\partial x_{q1}\partial x_{q2}}++N^{(2,2)}_{hp}(\boldsymbol{\xi})s^{2}\hat{U}^{M}_{p}\Big )+&
\end{flalign*}
\begin{flalign*}
&+\varepsilon^{3}\Big ( N^{(3,0)}_{hpq_{1}q_{2}q_{3}}(\boldsymbol{\xi})\frac{\partial^{3}\hat{U}^{M}_{p}}{\partial x_{q1}\partial x_{q2}\partial x_{q3}}+N^{(3,2)}_{hpq_{1}}(\boldsymbol{\xi})s^{2}\frac{\partial\hat{U}^{M}_{p}}{\partial x_{q1}}\Big )+&
\end{flalign*}
\begin{flalign*}
&\varepsilon^{4}\Big( N^{(4,0)}_{hpq_{1}q_{2}q_{3}q_{4}}(\boldsymbol{\xi})\frac{\partial^{4}\hat{U}^{M}_{p}}{\partial x_{q1}\partial x_{q2}\partial x_{q3}\partial x_{q4}}+N^{(4,2)}_{hpq_{1}q_{2}}(\boldsymbol{\xi})s^{2}\frac{\partial^{2}\hat{U}^{M}_{p}}{\partial x_{q1}\partial x_{q2}}+N^{(4,4)}_{hp}(\boldsymbol{\xi})s^{4}\hat{U}^{M}_{p}\Big )+O(\varepsilon^{5})\Big) \Big \vert _{\boldsymbol{\xi}= \frac{\boldsymbol{x}}{\varepsilon}}.&
\end{flalign*}
In Eq. $\eqref{eqn:dsr}$, $|q|$ describes the lenght of the multi-index and the derivative with respect to $q$ is written as $\frac{\partial^{l}(\cdot)}{\partial x_{q}} = \frac{\partial^{l}(\cdot)}{\partial x_{q_{1}}...x_{q_{l}}}$. Moreover, the perturbation function $N^{(0,0)}_{hp}$ stands for the Kronecker delta $\delta_{hp}$. There is point in observing that the $\mathcal{Q}$-periodic perturbation functions $N^{(2j+l,2j)}_{hpq}$ are affected by the microstructural inhomogeneities of the material and this is emphasized by their dependency on the fast variable $\boldsymbol{\xi}=\frac{\boldsymbol{x}}{\varepsilon}$.\\
 On the other hand, the transformed macro-displacement $\hat{U}^{M}_{h}(\boldsymbol{x},s)$ is $\mathcal{L}$-periodic and relies on the slow variable $\boldsymbol{x}$ and the time. The transformed macro-displacement field is supposed to be the mean value of the transformed micro-displacement field over the unit cell $\mathcal{Q}$  
\begin{flalign}
\label{upsca}
&\hat{U}_{h}^{M}(\boldsymbol{x},s)= \Big \langle \hat{u}_{h}\Big(\boldsymbol{x},\frac{\boldsymbol{x}}{\varepsilon}+\boldsymbol{\zeta},s\Big) \Big \rangle.&
\end{flalign}
Eq. $\eqref{upsca}$ is said to be the up-scaling relation and it links the transformed macro-displacement field with the transformed micro-displacement field. In Eq. $\eqref{upsca}$ the variable $\boldsymbol{\zeta} \in \mathcal{Q}$ identifies a family of translations of the heterogeneous domain respect to the $\mathcal{L}-$periodic body forces $\boldsymbol{b}(\boldsymbol{x},t)$, see  \cite{Smyshlyaev2000}, \cite{Bacigalupo2014}. Therefore, the transformed body forces in the Laplace space $\hat{\boldsymbol{b}}(\boldsymbol{x},s)$ are $\mathcal{L}-$periodic.\\ 
Replacing the down-scaling relation $~\eqref{eqn:dsr}$ into the micro-field Eq. $\eqref{eqn:fe}$ and assembling the terms with equal powers of $\varepsilon$, the average field equations of infinite order read
% $\varepsilon^{2}$
%\begin{flalign*}
%\label{secondorderinf}
%&n^{(2,0)}_{ipq_{1}q_{2}}\frac{\partial^{2}\hat{U}^{M}_{p}}{\partial x_{q_{1}} %\partial x_{q_{2}}}-n_{ip}^{(2,2)}s^{2}\hat{U}^{M}_{p}+\varepsilon\Big %(n^{(3,0)}_{ipq_{1}q_{2}q_{3}}\frac{\partial^{3}\hat{U}^{M}_{p}}{\partial %x_{q_{1}}\partial x_{q_{2}}\partial %x_{q_{3}}}+n^{(3,2)}_{ipq_{1}}s^{2}\frac{\partial\hat{U}^{M}_{p}}{\partial %x_{q_{1}}}\Big)+&
%\end{flalign*} 
%\begin{flalign}
%&+\varepsilon^{2}\Big %(n^{(4,0)}_{ipq_{1}q_{2}q_{3}q_{4}}\frac{\partial^{4}\hat{U}^{M}_{p}}{\partial %x_{q_{1}}\partial x_{q_{2}}\partial x_{q_{3}}\partial %x_{q_{4}}}+n^{(4,2)}_{ipq_{1}q_{2}}s^{2}\frac{\partial^{2}\hat{U}^{M}_{p}}{\par%tial x_{q_{1}}\partial %x_{q_{2}}}-n^{(4,4)}_{ip}s^{4}\hat{U}^{M}_{p}\Big)...+\boldsymbol{b}(\boldsymbo%l{x}),&
%\end{flalign}
%\begin{flalign*}
%n_{ipq_{1}q_{2}}^{(4,2)}&=\frac{1}{2}\langle %\rho^{m}N^{(2,0)}_{ipq_{1}q_{2}}-\hat{C}^{m}_{iq_{2}hq_{1}}N^{(2,2)}_{hp}-\hat{%C}^{m}_{iq_{2}hk}N^{(3,2)}_{hpq_{1},k}+\rho^{m}N^{(2,0)}_{ipq_{2}q_{1}}+&
%\end{flalign*}
%\begin{flalign}
%&-\hat{C}^{m}_{iq_{1}hq_{2}}N^{(2,2)}_{hp}-\hat{C}^{m}_{iq_{1}hk}N^{(3,2)}_{hpq%_{2},k}\rangle,&
%\end{flalign}
%where $w\in \mathbb{Z}$ and $w \geq 1$.
\begin{flalign*}
&n^{(2,0)}_{ipq_{1}q_{2}}\frac{\partial^{2}\hat{U}^{M}_{p}}{\partial x_{q_{1}} \partial x_{q_{2}}}-n_{ip}^{(2,2)}s^{2}\hat{U}^{M}_{p}+\sum_{n=0}^{+\infty}\varepsilon^{n+1}\sum_{|q|=n+3}n_{ipq}^{(n+3,0)} \frac{\partial^{n+3}\hat{U}^{M}_{p}}{\partial x_{q}}+\sum_{n=0}^{+\infty}\varepsilon^{n+1}\sum_{|q|=n+1}n_{ipq}^{(n+3,2)}s^{2} \frac{\partial^{n+1}\hat{U}^{M}_{p}}{\partial x_{q}}+&
\end{flalign*} 
\begin{flalign}
\label{eqn:infeq}
&-\sum_{\tilde{n}=0}^{+\infty}\varepsilon^{2\tilde{n}+2}n_{ip}^{(2\tilde{n}+4,2\tilde{n}+4)}s^{2\tilde{n}+4}\hat{U}^{M}_{p}-\sum_{\tilde{n,n}=0}^{+\infty}\varepsilon^{2\tilde{n}+n+3}\sum_{|q|=n+1}n_{ipq}^{(2\tilde{n}+n+5,2\tilde{n}+4)}s^{2\tilde{n}+4}\frac{\partial^{n+1}\hat{U}^{M}_{p}}{\partial x_{q}}+\hat{b}_{i}(\boldsymbol{x},s)=0,&
\end{flalign} 
where the coefficients of the gradients of the transformed macro-displacement are the known terms of the corresponding cell problems. Therefore it results 
\begin{flalign}
n_{ipq_{1}q_{2}}^{(2,0)}&=\frac{1}{2}\langle \hat{C}^{m}_{iq_{2}hq_{1}}+\hat{C}^{m}_{iq_{2}hk}N^{(1,0)}_{hpq_{1},k}+\hat{C}^{m}_{iq_{1}hq_{2}}+\hat{C}^{m}_{iq_{1}hk}N^{(1,0)}_{hpq_{2},k}\rangle,&
\end{flalign}
\begin{flalign}
n_{ip}^{(2,2)}&=\delta_{ip}\langle \rho^{m}\rangle,&
\end{flalign}
\begin{flalign}
n_{ipq_{1}...q_{w+2}}^{(w+2,0)}&=\frac{1}{w+2}\sum_{P^{*}(q)}\langle \hat{C}^{m}_{iq_{w+2}hj}N^{(w+1,0)}_{hpq_{1}...q_{w+1},j}+\hat{C}^{m}_{iq_{w+2}hq_{w+1}}N^{(w,0)}_{hpq_{1}...q_{w}} \rangle,&
\end{flalign}
\begin{flalign}
n_{ipq_{1}}^{(3,2)}&=\langle \rho^{m}N^{(1,0)}_{ipq_{1}}-\hat{C}^{m}_{iq_{1}hj}N^{(2,2)}_{hp,j}\rangle,&
\end{flalign}
\begin{flalign}
n_{ip}^{(2\tilde{n}+4,2\tilde{n}+4)}&= \langle \rho^{m}N^{(2\tilde{n}+2,2\tilde{n}+2)}_{ip}\rangle,&
\end{flalign}
\begin{flalign}
n_{ipq_{1}}^{(2\tilde{n}+5,2\tilde{n}+4)}& = \langle \rho^{m}N^{(2\tilde{n}+3,2\tilde{n}+2)}_{ipq_{1}}-\hat{C}^{m}_{iq_{1}hk}N^{(2\tilde{n}+4,2\tilde{n}+4)}_{hp,k}\rangle,&
\end{flalign}
\begin{flalign}
n_{ipq_{1}q_{2}}^{(2\tilde{n}+6,2\tilde{n}+4)}&=\frac{1}{2}\langle \rho^{m}N^{(2\tilde{n}+4,2\tilde{n}+4)}_{ipq_{1}q_{2}}-\hat{C}^{m}_{iq_{2}hq_{1}}N^{(2\tilde{n}+4,2\tilde{n}+4)}_{hp}-\hat{C}^{m}_{iq_{2}hj}N^{(2\tilde{n}+5,2\tilde{n}+4)}_{hpq_{1},j}+&
\end{flalign}
\begin{flalign*}
\quad \quad \quad \quad \quad \quad 
&+\rho^{m}N^{(2\tilde{n}+4,2\tilde{n}+2)}_{ipq_{2}q_{1}}-\hat{C}^{m}_{iq_{1}hq_{2}}N^{(2\tilde{n}+4,2\tilde{n}+2)}_{hp}-\hat{C}^{m}_{iq_{1}hj}N^{(2\tilde{n}+5,2\tilde{n}+4)}_{hpq_{2},j} \rangle,&
\end{flalign*}
\begin{flalign}
n_{ipq_{1}...q_{w+2}}^{(2\tilde{n}+w+6,2\tilde{n}+w+4)}&=\frac{1}{w+2}\sum_{P^{*}(q)}\langle \rho^{m}N^{(2\tilde{n}+w+4,2\tilde{n}+2)}_{ipq_{1}...q_{w+2}} -\hat{C}^{m}_{iq_{w+2}hq_{w+1}}N^{(2\tilde{n}+w+4,2\tilde{n}+4)}_{hpq_{1}...q_{w}}+&
\end{flalign}
\begin{flalign*}
\quad \quad \quad \quad \quad \quad \quad \quad &-\hat{C}^{m}_{iq_{w+2}hj}N^{(2\tilde{n}+w+5,2\tilde{n}+4)}_{hpq_{1}...q_{w+1},j} \rangle,&
\end{flalign*}
with $w\in \mathbb{Z}$, $w \geq 1$, $\tilde{n}\in \mathbb{Z}$ and $\tilde{n} \geq 0$.
%It is important to observe that $n, \tilde{n} \in \mathbb{Z}$ with $n \geq 2$ and  $\tilde{n} \geq %0$. 
The average field equations of infinite order $~\eqref{eqn:infeq}$ are formally solved by performing an asymptotic expansion of the transformed macro-displacement $\hat{U}^{M}_{p}(\boldsymbol{x})$ in power of $\varepsilon$, namely
\begin{flalign}
\label{eqn:esam}
\hat{U}^{M}_{p}(\boldsymbol{x})=\sum_{j=0}^{+\infty}\varepsilon^{j}U^{j}_{p}(\boldsymbol{x}).
\end{flalign}
The substitution of Eq. $~\eqref{eqn:esam}$ into Eq. $~\eqref{eqn:infeq}$ leads to
\begin{flalign}
\label{eqn:rse}
&n_{ipq_{1}q_{2}}^{(2,0)}\Big (\varepsilon^{0}\frac{\partial^{2}\hat{U}^{(0)}_{p}}{\partial x_{q_{1}} \partial x_{q_{2}}}+\varepsilon \frac{\partial^{2}\hat{U}^{(1)}_{p}}{\partial x_{q_{1}} \partial x_{q_{2}}}+...\Big )-n_{ip}^{(2,2)}s^{2}(\hat{U}^{(0)}_{p}+\varepsilon \hat{U}^{(1)}_{p}+...)+&
\end{flalign} 
\begin{flalign*}
&+\varepsilon n_{ipq_{1}...q_{3}}^{(3,0)}\Big (\frac{\partial^{3}\hat{U}^{(0)}_{p}}{\partial x_{q_{1}}...\partial x_{q_{3}}}+\varepsilon \frac{\partial^{3}\hat{U}^{(1)}_{p}}{\partial x_{q_{1}}...\partial x_{q_{3}}}+...\Big )+  \varepsilon^{2} n_{ipq_{1}...q_{4}}^{(4,0)}\Big (\frac{\partial^{4}\hat{U}^{(0)}_{p}}{\partial x_{q_{1}}...\partial x_{q_{4}}}+\varepsilon \frac{\partial^{4}\hat{U}^{(1)}_{p}}{\partial x_{q_{1}}...\partial x_{q_{4}}}+...\Big )+&
\end{flalign*} 
\begin{flalign*}
&...-\varepsilon n_{ipq_{1}}^{(3,2)}s^{2}\Big (\frac{\partial\hat{U}^{(0)}_{p}}{\partial x_{q_{1}}}+\varepsilon \frac{\partial \hat{U}^{(1)}_{p}}{\partial x_{q_{1}}}+...\Big )-\varepsilon^{2} n_{ipq_{1}q_{2}}^{(4,2)}s^{2}\Big (\frac{\partial^{2}\hat{U}^{(0)}_{p}}{\partial x_{q_{1}}\partial x_{q_{2}}}+\varepsilon \frac{\partial^{2} \hat{U}^{(1)}_{p}}{\partial x_{q_{1}}\partial x_{q_{2}}}+...\Big )+&
\end{flalign*}
\begin{flalign*}
&...-\varepsilon^{2} n_{ip}^{(4,4)}s^{4}\Big (\hat{U}^{(0)}_{p}+\varepsilon  \hat{U}^{(1)}+...\Big )-\varepsilon^{4}n_{ip}^{(6,6)}s^{6}\Big (\hat{U}^{(0)}_{p}+\varepsilon  \hat{U}^{(1)}+...\Big )+...+&
\end{flalign*}
\begin{flalign*}
&\varepsilon^{3} n_{ipq_{1}}^{(5,4)}s^{4}\Big (\frac{\partial\hat{U}^{(0)}_{p}}{\partial x_{q_{1}}}+\varepsilon \frac{\partial \hat{U}^{(1)}_{p}}{\partial x_{q_{1}}}+...\Big )-  \varepsilon^{4} n_{ipq_{1}q_{2}}^{(6,4)}s^{4}\Big (\frac{\partial^{2}\hat{U}^{(0)}_{p}}{\partial x_{q_{1}}\partial x_{q_{2}}}+\varepsilon \frac{\partial^{2}\hat{U}^{(1)}_{p}}{\partial x_{q_{1}}\partial x_{q_{2}}}+...\Big )+...+&
\end{flalign*}
\begin{flalign*}
&\varepsilon^{5} n_{ipq_{1}}^{(7,6)}s^{6}\Big (\frac{\partial^{8}\hat{U}^{(0)}_{p}}{\partial x_{q_{1}}}+\varepsilon \frac{\partial \hat{U}^{(1)}_{p}}{\partial x_{q_{1}}}+...\Big )-  \varepsilon^{6} n_{ipq_{1}q_{2}}^{(8,6)}s^{6}\Big (\frac{\partial^{2}\hat{U}^{(0)}_{p}}{\partial x_{q_{1}}\partial x_{q_{2}}}+\varepsilon \frac{\partial^{2}\hat{U}^{(1)}_{p}}{\partial x_{q_{1}}\partial x_{q_{2}}}+...\Big )+...+\hat{b}_{i}(\boldsymbol{x},s)=0,&
\end{flalign*}
which provides the following macroscopic recursive problems for the different orders of $\varepsilon$. Namely, at the order $\varepsilon^{0}$ it results 
\begin{flalign}
&n_{ipq_{1}q_{2}}^{(2,0)} \frac{\partial^{2}\hat{U}^{(0)}_{p}}{\partial x_{q_{1}} \partial x_{q_{2}}}-n_{ip}^{(2,2)}s^{2}\hat{U}^{(0)}_{p}+\hat{b}_{i}(\boldsymbol{x},s)=0,&
\end{flalign}
at the order $\varepsilon$ the problem is
\begin{flalign}
&n_{ipq_{1}q_{2}}^{(2,0)} \frac{\partial^{2}\hat{U}^{(0)}_{p}}{\partial x_{q_{1}} \partial x_{q_{2}}}-n_{ip}^{(2,2)}s^{2}\hat{U}^{(0)}_{p}+n_{ipq_{1}...q_{3}}^{(3,0)} \frac{\partial^{3}\hat{U}^{(0)}_{p}}{\partial x_{q_{1}}...\partial x_{q_{3}}}-n_{ipq_{1}}^{(3,2)}s^{2}\frac{\partial \hat{U}^{(0)}_{p}}{\partial x_{q_{1}}}=0,&
\end{flalign}
instead at the order $\varepsilon^{2}$ it reads
\begin{flalign}
&n_{ipq_{1}q_{2}}^{(2,0)} \frac{\partial^{2}\hat{U}^{(0)}_{p}}{\partial x_{q_{1}} \partial x_{q_{2}}}-n_{ip}^{(2,2)}s^{2}\hat{U}^{(2)}_{p}+n_{ipq_{1}...q_{3}}^{(3,0)} \frac{\partial^{3}\hat{U}^{(1)}_{p}}{\partial x_{q_{1}}...\partial x_{q_{3}}}+n_{ipq_{1}...q_{4}}^{(4,0)} \frac{\partial^{4}\hat{U}^{(0)}_{p}}{\partial x_{q_{1}}...\partial x_{q_{4}}}+&
\end{flalign}
\begin{flalign*}
&-n_{ipq_{1}}^{(3,2)}s^{2}\frac{\partial \hat{U}^{(0)}_{p}}{\partial x_{q_{1}}}-n_{ipq_{1}q_{2}}^{(4,2)}s^{2} \frac{\partial^{2}\hat{U}^{(0)}_{p}}{\partial x_{q_{1}}\partial x_{q_{2}}}-n_{ip}^{(4,4)}s^{4}\hat{U}^{0}_{p}=0.&
\end{flalign*}
A generic recursive problem, at odd order $\varepsilon^{2\tilde{w}-1}$, with  $\tilde{w}\in \mathbb{Z}$ and $\tilde{w} \geq 2$ is
\begin{flalign}
&n_{ipq_{1}q_{2}}^{(2\tilde{w}-1,0)} \frac{\partial^{2}\hat{U}^{(2\tilde{w}-1)}_{p}}{\partial x_{q_{1}} \partial x_{q_{2}}}-n_{ip}^{(2\tilde{w}-1,2\tilde{w}-1)}s^{2}\hat{U}^{(2\tilde{w}-1)}_{p}+\sum_{r=3}^{2\tilde{w}+1}\sum_{|q|=r}^{}n^{(r,0)}_{ipq}\frac{\partial^{r}\hat{U}^{(2\tilde{w}+1-r)}_{p}}{\partial x_{q}}+&
\end{flalign}
\begin{flalign*}
&-s^{2}\sum_{r=3}^{2\tilde{w}+1}\sum_{|q|=r-2}^{}n^{(r,2)}_{ipq}\frac{\partial^{r-2}\hat{U}^{(2\tilde{w}+1-r)}_{p}}{\partial x_{q}}-(1-\delta_{2\tilde{w}})\sum_{n=0}^{\tilde{w}-3+\delta_{2\tilde{w}}}s^{2n+4}\Big(n_{ip}^{(2n+4,2n+4)}\hat{U}_{p}^{(2\tilde{w}-3-2n)}\Big )+&
\end{flalign*}
\begin{flalign*}
&+\sum_{r=3}^{2\tilde{w}-1-2n}\sum_{|q|=r-2} n_{ipq}^{(r+2+2n,2n+4)}\frac{\partial^{r-2}\hat{U}_{p}^{(2\tilde{w}-1-r-2n)}}{\partial x_{q}}-n_{ip}^{(2\tilde{w},2\tilde{w})}s^{2\tilde{w}}\hat{U}^{(1)}_{p}-n_{ipq_{1}}^{(2\tilde{w},2\tilde{w})}s^{2\tilde{w}}\frac{\hat{U}^{(0)}_{p}}{\partial x_{q_{1}}}=0,&
\end{flalign*}
whereas a generic recursive problem at even order $\varepsilon^{2\tilde{w}}$ is 
\begin{flalign}
&n_{ipq_{1}q_{2}}^{(2\tilde{w},0)} \frac{\partial^{2}\hat{U}^{(2\tilde{w})}_{p}}{\partial x_{q_{1}} \partial x_{q_{2}}}-n_{ip}^{(2\tilde{w},2\tilde{w})}s^{2}\hat{U}^{(2\tilde{w})}_{p}+\sum_{r=3}^{2\tilde{w}+2}\sum_{|q|=r}^{}n^{(r,0)}_{ipg}\frac{\partial^{r}\hat{U}^{(2\tilde{w}+1-r)}_{p}}{\partial x_{q}}+&
\end{flalign}
\begin{flalign*}
&-s^{2}\sum_{r=3}^{2\tilde{w}+2}\sum_{|q|=r-2}^{}n^{(r,2)}_{ipq}\frac{\partial^{r-2}\hat{U}^{(2\tilde{w}+2-r)}_{p}}{\partial x_{q}}-\sum_{n=0}^{\tilde{w}-2}s^{2n+4}\Big(n_{ip}^{(2n+4,2n+4)}\hat{U}_{p}^{(2\tilde{w}-2-2n)}+&
\end{flalign*}
\begin{flalign*}
&+\sum_{r=3}^{2\tilde{w}-2n}\sum_{|q|=r-2} n_{ipq}^{(r+2+2n,2n+4)}\frac{\partial^{r-2}\hat{U}_{p}^{(2\tilde{w}-r-2n)}}{\partial x_{q}}-n_{ip}^{(2\tilde{w}+2,2\tilde{w}+2)}s^{2\tilde{w}+2}\hat{U}^{(0)}_{p}=0,&
\end{flalign*}
where $n \in \mathbb{Z}$ and $n \geq 2$. 
There is no point in managing the averaged equation of infinite order \eqref{eqn:infeq}. In addition, the ellipticity of the differential problem could be not guaranteed if Eq. \eqref{eqn:infeq} is truncated at a certain order. To overcome such a disadvantage, an asymptotic-variational approach is pursued.

\section{Asymptotic expansion of the energy and second order homogenization}
\label{sec:six}
In this Section a finite order governing equation %,whose ellepticity is guaranteed,
is provided by exploiting a variational-asymptotic procedure, see \cite{Smyshlyaev2000}, \cite{BacigalupoGambarotta2014}. Let ${\Lambda}$ be the energy-like functional written in terms of the energy-like density $\lambda_{m}$ at the microscale and referred to the periodic domain $L$, \cite{fabrizio1992mathematical},
\begin{flalign}
&{\Lambda}=\int_{L} \lambda_{m}\Big (\boldsymbol{x},\frac{\boldsymbol{x}}{\varepsilon}  \Big) d\boldsymbol{x}=\int_{L}\Big ( \frac{1}{2} \rho^{m} \dot{\boldsymbol{u}}\ast \dot{\boldsymbol{u}} +\frac{1}{2}
\nabla \boldsymbol{u} \ast (\mathbb{G}^{m}\ast \nabla \dot{\boldsymbol{u}})-\boldsymbol{u}\ast \boldsymbol{b} \Big )   d\boldsymbol{x}.&
\end{flalign} 
Let $\mathcal{L}({\Lambda})$ be the energy-like functional in the Laplace domain, which is expressed in terms of the energy-like density $\hat{\lambda}_{m}$ in the Laplace domain     
\begin{flalign}
\label{enefun}
&\hat{\Lambda}=\mathcal{L}({\Lambda})=\int_{L} \hat{\lambda}_{m}\Big (\boldsymbol{x},\frac{\boldsymbol{x}}{\varepsilon}  \Big) d\boldsymbol{x}=\int_{L} \Big ( \frac{1}{2} \rho^{m} s^{2}\hat{\boldsymbol{u}}\cdot \hat{\boldsymbol{u}} +\frac{1}{2} \nabla \hat{\boldsymbol{u}} : \hat{\mathbb{G}}^{m} \nabla \hat{\boldsymbol{u}} - \hat{\boldsymbol{u}} \cdot \hat{\boldsymbol{b}} \Big ) d\boldsymbol{x},&
\end{flalign}
where the symbol $:$ denotes the second order inner product. Specifically, 
the tranformed energy-like functional $\hat{\Lambda}$ and its corresponding energy-like density $\hat{\lambda}_{m}$ are influenced by the translation variable $\boldsymbol{\zeta} \in \mathcal{Q}$. Such a variable is introduced because the actual "phase" of the microstructure is undetectable and a family of translated microstructures is taken into account.\\ Therefore, the transformed micro-relaxation tensor $\mathbb{\hat{G}}^{m}$ depends on the translation variable $\boldsymbol{\zeta}$ and it may be written as $\mathbb{\hat{G}}^{m,\zeta}\Big (\boldsymbol{x},\frac{\boldsymbol{x}}{\varepsilon}\Big ) = \mathbb{\hat{G}}^{m}\Big (\boldsymbol{x},\frac{\boldsymbol{x}}{\varepsilon}+\boldsymbol{\zeta}\Big )$ and the perturbation functions $N^{(1,0)}_{hpq_{1}}, N^{(2,0)}_{hpq_{1}q_{2}},...,$, which are solutions of the cell problems determined in Section \ref{sec:four}, reckon on variable $\boldsymbol{\zeta}$. In addition, the energy-like density $\hat{\lambda}_{m}$ in the Laplace domain  complies with the property $\hat{\lambda}^{\zeta}_{m}\Big (\boldsymbol{x},\frac{\boldsymbol{x}}{\varepsilon}  \Big)=\hat{\lambda}_{m}\Big (\boldsymbol{x},\frac{\boldsymbol{x}}{\varepsilon}+\boldsymbol{\zeta}  \Big) $ and so the Laplace transform of the energy-like functional $\Lambda$, depending on the parameter $\zeta$, is    
\begin{flalign}
\label{beli}
&\hat{\Lambda}^{\zeta}= \hat{\Lambda}(\boldsymbol{\zeta})= \int_{L} \hat{\lambda}^{\zeta}_{m}\Big (\boldsymbol{x},\frac{\boldsymbol{x}}{\varepsilon}  \Big) d\boldsymbol{x} = \int_{L} \hat{\lambda}_{m}\Big (\boldsymbol{x},\frac{\boldsymbol{x}}{\varepsilon}+\boldsymbol{\zeta}  \Big) d\boldsymbol{x}.&
\end{flalign}
Let $\hat{\Lambda}_{m}$ be the average transformed energy-like functional at the microscale
\begin{flalign}
\label{bela}
&\hat{\Lambda}_{m} \dot{=} \langle \hat{\Lambda}^{\zeta} \rangle = \frac{1}{|\mathcal{Q}|} \int_{\mathcal{Q}} \hat{\Lambda}^{\zeta} d\boldsymbol{\zeta} = \frac{1}{|\mathcal{Q}|} \int_{\mathcal{Q}} \hat{\Lambda}(\boldsymbol{\zeta}) d\boldsymbol{\zeta} = \int_{L} \Big \langle \hat{\lambda}_{m}\Big( \boldsymbol{x},\frac{\boldsymbol {x}}{\varepsilon}+\boldsymbol{\zeta} \Big )\Big  \rangle d\boldsymbol{x},&  
\end{flalign} 
where the Fubini theorem is applied. The average transformed energy-like functional $\langle \hat{\Lambda}^{\zeta} \rangle$ at the microscale does not rely on the translation variable $\zeta$ because the energy-like functional $\hat{\Lambda}^{\zeta}$ is averaged with regard to the translated realizations of the microstructure and so the transformed energy-like density at the micoscale satisfies   
\begin{flalign}
\label{sou}
&\Big \langle \hat{\lambda}_{m}\Big ( \boldsymbol{x},\frac{\boldsymbol{x}}{\varepsilon}+\boldsymbol{\zeta}\Big ) \Big \rangle = \frac{1}{|\mathcal{Q}|}\int_{\mathcal{Q}} \hat{\lambda}_{m}\Big ( \boldsymbol{x},\frac{\boldsymbol{x}}{\varepsilon}+\boldsymbol{\zeta}\Big ) d\boldsymbol{\zeta}=\frac{1}{|\mathcal{Q}|}\int_{\mathcal{Q}} \hat{\lambda}_{m}\Big ( \boldsymbol{x},\boldsymbol{\xi}\Big ) d\boldsymbol{\xi}=\langle \hat{\lambda}_{m}(\boldsymbol{x},\boldsymbol{\xi})\rangle.&
\end{flalign}
Two methods are herein proposed to determine the governing field equation at the macroscale and the overall constitutive and inertial tensors.  
\subsection*{Approximation of the energy-like functional through truncation of its asymptotic expansion}
\label{sub:met1}
Let us consider the down-scaling relation related to the transformed micro-displacement $\boldsymbol{\hat{u}}(\boldsymbol{x},\boldsymbol{\xi},s)$, i.e.
\begin{flalign}
\label{secorddis}
&\hat{u}_{h}(\boldsymbol{x},\boldsymbol{\xi},s)=\hat{U}_{h}(\boldsymbol{x},s)+\varepsilon N^{(1,0)}_{hpq_{1}}(\boldsymbol{\xi})\frac{\partial \hat{U}^{M}_{p}}{\partial x_{q_{1}}}+\varepsilon^{2}\Big(N^{(2,0)}_{hpq_{1}q_{2}}(\boldsymbol{\xi})\frac{\partial^{2}\hat{U}^{M}_{p}}{\partial x_{q_{1}}\partial x_{q_{2}}}+N^{(2,2)}_{hp}(\boldsymbol{\xi})s^{2}\hat{U}^{M}_{p}\Big)+&
\end{flalign}
\begin{flalign*}
&+\varepsilon^{3}\Big ( N^{(3,0)}_{hpq_{1}q_{2}q_{3}}(\boldsymbol{\xi})\frac{\partial^{3}\hat{U}^{M}_{p}}{\partial x_{q1}\partial x_{q2}\partial x_{q3}}+N^{(3,2)}_{hpq_{1}}(\boldsymbol{\xi})s^{2}\frac{\partial\hat{U}^{M}_{p}}{\partial x_{q1}}\Big)+O(\varepsilon^{4}).&
\end{flalign*}
Let us replace the down-scaling relation \eqref{secorddis} into the tranformed energy-like functional \eqref{beli} and let us suppose that $\hat{\Lambda}_{m}$ is truncated at the second order. After applying the divergence theorem, the transformed energy-like functional at the second order is
\begin{flalign}
\label{Lmmmmm}
&\hat{\Lambda}_{m}^{II} = \int_{L} \langle \hat{\lambda}_{m}^{II}(\boldsymbol{x},\boldsymbol{\xi})\rangle d\boldsymbol{x}= \frac{1}{2} s^{2}\langle \rho^{m} \rangle \int_{L}\hat{U}^{M}_{h}\hat{U}^{M}_{h} d\boldsymbol{x}+\varepsilon s^{2}\langle \rho^{m}N^{(1,0)}_{rpq_{1}}\rangle \int_{L}\frac{\partial \hat{U}^{M}_{p}}{\partial x_{q_{1}}}\hat{U}^{M}_{r} d\boldsymbol{x} +&
\end{flalign} 
\begin{flalign*}
&+\varepsilon s^{3}\langle \hat{G}^{m}_{hkij}B^{(1,0)}_{hkpq_{1}}B^{(2,2)}_{ijr}\rangle \int_{L} \frac{\partial \hat{U}^{M}_{p}}{\partial x_{q_{1}}}\hat{U}^{M}_{r} d\boldsymbol{x} +&
\end{flalign*} 
\begin{flalign*}
&+\varepsilon s \langle \hat{G}^{m}_{hkij}B^{(1,0)}_{hkpq_{1}}B^{(2,0)}_{ijrw_{1}w_{2}}\rangle \int_{L}\frac{\partial \hat{U}^{M}_{p}}{\partial x_{q_{1}}}\frac{\partial^{2}\hat{U}^{M}_{r}}{\partial x_{w_{1}} \partial x_{w_{2}}} d\boldsymbol{x}+\varepsilon^{2}s^{2}\Big \langle \frac{1}{2} \rho^{m}N^{(1,0)}_{hpq_{1}}N^{(1,0)}_{hrq_{2}}-\rho^{m}N^{(2,0)}_{rpq_{1}q_{2}}\Big \rangle \int_{L}\frac{\partial \hat{U}^{M}_{p}}{\partial x_{q_{1}}} \frac{\partial \hat{U}^{M}_{r}}{\partial x_{q_{2}}} d\boldsymbol{x}+&
\end{flalign*} 
\begin{flalign*}
&+\varepsilon^{2}s^{3}\langle \hat{G}^{m}_{hkij}B^{(1,0)}_{hkpq_{1}}B^{(3,2)}_{ijrq_{2}}-\hat{G}^{m}_{hkij}B^{(2,0)}_{hkpq_{1}q_{2}}B^{(2,2)}_{ijr}\rangle \int_{L}\frac{\partial \hat{U}^{M}_{p}}{\partial x_{q_{1}}} \frac{\partial \hat{U}^{M}_{r}}{\partial x_{q_{2}}} d\boldsymbol{x}+\varepsilon^{2}s^{4}\langle \rho^{m}N^{(2,2)}_{rp}\rangle \int_{L}\hat{U}^{M}_{p}\hat{U}^{M}_{r}d\boldsymbol{x}+&
\end{flalign*} 
\begin{flalign*}
&+\varepsilon^{2}s^{5}\frac{1}{2}\langle \hat{G}^{m}_{hkij}B^{(2,2)}_{hkp}B^{(2,2)}_{ijr}\rangle \int_{L}\hat{U}^{M}_{p}\hat{U}^{M}_{r}d\boldsymbol{x}+\frac{1}{2}s\langle \hat{G}^{m}_{hkij}B^{(1,0)}_{hkpq_{1}}B^{(1,0)}_{ijrw_{1}}\rangle \int_{L}\frac{\partial \hat{U}^{M}_{p}}{\partial x_{q_{1}}}\frac{\partial \hat{U}^{M}_{r}}{\partial x_{w_{1}}} d\boldsymbol{x}+&
\end{flalign*} 
\begin{flalign*}
&+\varepsilon^{2}s \Big \langle \frac{1}{2} \hat{G}^{m}_{hkij}B^{(2,0)}_{hkpq_{1}q_{2}}B^{(2,0)}_{ijrw_{1}w_{2}}-\hat{G}^{m}_{hkij}B^{(1,0)}_{ijrw_{1}}B^{(3,0)}_{hkpq_{1}q_{2}w_{2}}\Big \rangle \int_{L}\frac{\partial^{2}\hat{U}^{M}_{p}}{\partial x_{q_{1}} \partial x_{q_{2}}} \frac{\partial^{2}\hat{U}^{M}_{r}}{\partial x_{w_{1}} \partial x_{w_{2}}} d\boldsymbol{x}- \int_{L}\hat{U}^{M}_{h}\hat{b}_{h}d\boldsymbol{x}.&
\end{flalign*} 
It is important to note that second-order gradients of the transformed macro-displacement are takent into account as well as the first-order gradient of the displacement (i.e., strain), by generalizing the standard continuum mechanics.

The localization tensors, appearing in Eq. \eqref{Lmmmmm}, assume the form
\begin{flalign}
\label{locten}
&B^{(1,0)}_{hkpq_{1}}=\delta_{hp}\delta_{kq_{1}}+N^{(1,0)}_{hpq_{1},k},&
\end{flalign}
\begin{flalign}
\label{eq:fl:80}
&B^{(2,0)}_{hkpq_{1}q_{2}}=\frac{1}{2}\Big ( \delta_{kq_{2}}N^{(1,0)}_{hpq_{1}}+ \delta_{kq_{1}}N^{(1,0)}_{hpq_{2}}\Big )+N^{(2,0)}_{hpq_{1}q_{2},k},&
\end{flalign}
\begin{flalign}
&B^{(2,2)}_{hkp} = N^{(2,2)}_{hp,k},&
\end{flalign}
\begin{flalign}
&B^{(3,0)}_{hkpq_{1}q_{2}q_{3}}= \frac{1}{3} \Big (\delta_{kq_{3}}N^{(2,0)}_{hpq_{1}q_{2}}+\delta_{kq_{1}}N^{(2,0)}_{hpq_{2}q_{3}}+\delta_{kq_{2}}N^{(2,0)}_{hpq_{3}q_{1}}\Big )+N^{(3,0)}_{hpq_{1}q_{2}q_{3},k},&
\end{flalign}
\begin{flalign}
\label{eq:fl:83}
&B^{(3,2)}_{hkpq_{1}}=\delta_{kq_{1}}N^{(2,2)}_{hp}+N^{(3,2)}_{hpq_{1},k}.&
\end{flalign}
The localization tensors are periodic functions with regard to the fast coordinate $\boldsymbol{\xi}$ since the perturbation functions and their gradients are $\mathcal{Q}$- periodic functions.
Both $B^{(2,0)}_{hkpq_{1}q_{2}}$ and $B^{(3,0)}_{hkpq_{1}q_{2}q_{3}}$ are symmetrized with respect to the indices $q_{1}, q_{2}$ and $q_{1}, q_{2}, q_{3}$, respectively (see Appendix B). The governing equation of a non-local homogeneous continuum is delivered by determining  the stability condition of the transformed energy-like functional $ \hat{\Lambda}_{m}^{II}$, which is found to be the first variation of the average transformed energy-like functional $\delta \hat{\Lambda}_{m}^{II}$,
\begin{flalign}
\label{varprob}
&\delta \hat{\Lambda}_{m}^{II}(\hat{U}^{M}_{t},\delta \hat{U}^{M}_{t})=s^{2}\langle \rho^{m} \rangle \int_{L} \hat{U}^{M}_{t} \delta \hat{U}^{M}_{t} d\boldsymbol{x}+\varepsilon s^{2}\langle \rho^{m}(N^{(1,0)}_{rpq_{1}}-N^{(1,0)}_{prq_{1}})\rangle \int_{L}\frac{\partial \hat{U}^{M}_{p}}{\partial x_{q_{1}}}\delta \hat{U}^{M}_{r} d\boldsymbol{x}+&
\end{flalign}
\begin{flalign*}
&+\varepsilon s^{3}\langle \hat{G}^{m}_{hkij}B^{(1,0)}_{hkpq_{1}}B^{(2,2)}_{ijr}-\hat{G}^{m}_{hkij}B^{(1,0)}_{hkrq_{1}}B^{(2,2)}_{ijp}\rangle \int_{L}\frac{\partial \hat{U}^{M}_{p}}{\partial x_{q_{1}}}\delta \hat{U}^{M}_{r} d\boldsymbol{x}+&
\end{flalign*}
\begin{flalign*}
&+\varepsilon s \langle \hat{G}^{m}_{hkij}B^{(1,0)}_{hkpq_{1}}B^{(2,0)}_{ijrw_{1}w_{2}}-\hat{G}^{m}_{hkij}B^{(1,0)}_{hkrw_{1}}B^{(2,0)}_{ijpq_{1}w_{2}}\rangle \int_{L}\frac{\partial^{3}\hat{U}^{M}_{p}}{\partial x_{w_{1}} \partial x_{w_{2}} \partial x_{q_{1}}} \delta \hat{U}^{M}_{r} d\boldsymbol{x}+&
\end{flalign*}
\begin{flalign*}
&-\varepsilon^{2}s^{2}\langle \rho^{m}N^{(1,0)}_{hpq_{1}}N^{(1,0)}_{hrq_{2}}-\rho^{m}(N^{(2,0)}_{rpq_{1}q_{2}}+N^{(2,0)}_{prq_{2}q_{1}})\rangle \int_{L} \frac{\partial^{2} \hat{U}^{M}_{p}}{\partial x_{q_{1}}\partial x_{q_{2}}} \delta \hat{U}^{M}_{r} d\boldsymbol{x}-\varepsilon^{2}s^{3}\langle \hat{G}^{m}_{hkij}B^{(1,0)}_{hkpq_{1}}B^{(3,2)}_{ijrq_{2}}-\hat{G}^{m}_{hkij}B^{(2,0)}_{hkpq_{1}q_{2}}B^{(2,2)}_{ijp}+&
\end{flalign*}
\begin{flalign*}
&+\hat{G}^{m}_{hkij}B^{(1,0)}_{hkrq_{2}}B^{(3,2)}_{ijpq_{1}}-\hat{G}^{m}_{hkij}B^{(2,0)}_{hkrq_{2}q_{1}}B^{(2,2)}_{ijp} \rangle  \int_{L} \frac{\partial^{2} \hat{U}^{M}_{p}}{\partial x_{q_{1}}\partial x_{q_{2}}} \delta \hat{U}^{M}_{r} d\boldsymbol{x}+\varepsilon^{2}s^{4}\langle \rho^{m}(N^{(2,2)}_{rp}+N^{(2,2)}_{pr})\rangle \int_{L}\hat{U}^{M}_{p} \delta \hat{U}^{M}_{r} d\boldsymbol{x}+&
\end{flalign*}
\begin{flalign*}
&+\varepsilon^{2}s^{5}\langle \hat{G}^{m}_{hkij}B^{(2,2)}_{hkp}B^{(2,2)}_{ijr}\rangle \int_{L} \hat{U}^{M}_{p} \delta \hat{U}^{M}_{r} d\boldsymbol{x}-s \langle \hat{G}^{m}_{hkij}B^{(1,0)}_{hkpq_{1}}B^{(1,0)}_{ijrw_{1}}\rangle \int_{L} \frac{\partial^{2} \hat{U}^{M}_{p}}{\partial x_{w_{1}}\partial x_{q_{1}}} \delta \hat{U}^{M}_{r} d\boldsymbol{x}+&
\end{flalign*}
\begin{flalign*}
&+\varepsilon^{2}s \langle  \hat{G}^{m}_{hkij}B^{(2,0)}_{hkpq_{1}q_{2}}B^{(2,0)}_{ijrw_{1}w_{2}}-\hat{G}^{m}_{hkij}B^{(1,0)}_{hkrq_{1}}B^{(3,0)}_{ijpq_{1}q_{2}w_{2}}-\hat{G}^{m}_{hkij}B^{(1,0)}_{hkpq_{1}}B^{(3,0)}_{ijrw_{1}w_{2}q_{2}}\rangle \int_{L} \frac{\partial ^{4}\hat{U}^{M}_{p}}{\partial x_{w_{2}} \partial x_{w_{1}} \partial x_{q_{1}} \partial x_{q_{2}}} \delta \hat{U}^{M}_{r} d\boldsymbol{x}+&
\end{flalign*}
\begin{flalign*}
&-\int_{L} \delta \hat{U}^{M}_{t}\hat{b}_{t}d\boldsymbol{x},&
\end{flalign*}
where the $\mathcal{Q}-$periodicity of the functions $\hat{G}^{m}_{hkij}$,  $N^{(1,0)}_{ipq_{1}}$, $N^{(2,0)}_{ipq_{1}q_{2}}$ and the localization tensors $B^{(1,0)}_{hkpq_{1}}$, $B^{(2,0)}_{hkpq_{1}q_{2}}$,..., $B^{(3,2)}_{hkpq_{1}}$ is taken into account.  
The first variation $\delta \hat{\Lambda}_{m}^{II}$ must vanish for all admissible $\delta \hat{U}^{M}_{t}$ and so the Euler-Lagrangian differential equation associated with the variational problem $\eqref{varprob}$ in the Laplace domain is 
\begin{flalign*}
\label{macrolapdomI}
&s^{2}\rho \hat{U}^{M}_{t}+s^{2}\rho (\hat{I}_{tpq_{1}}- \hat{\tilde{I}}_{tq_{1}p}) \frac{\partial \hat{U}^{M}_{p}}{\partial x_{q_{1}}}-s^{2}\rho \hat{I}_{tq_{2}pq_{1}} \frac{\partial^{2} \hat{U}^{M}_{p}}{\partial x_{q_{1}} \partial x_{q_{2}}}+s^{4}\rho \hat{I}_{tp}^{\sharp}\hat{U}^{M}_{p}=&
\end{flalign*}
\begin{flalign*}
&-s^{3}(\hat{J}_{tpq_{1}}-\hat{\tilde{J}}_{tq_{1}p})\frac{\partial \hat{U}^{M}_{p}}{\partial x_{q_{1}}}-s^{3}\hat{J}_{tq_{2}pq_{1}}^{1}\frac{\partial^{2}\hat{U}^{M}_{p}}{\partial x_{q_{1}}\partial x_{q_{2}}}+&
\end{flalign*}
\begin{flalign*}
&-s^{5}\hat{J}_{tp}^{\sharp}\hat{U}^{M}_{p}+s\hat{G}_{tr_{1}pq_{1}}\frac{\partial^{2}\hat{U}^{M}_{p}}{\partial x_{q_{1}}\partial x_{r_{1}}}+s(\hat{Y}_{tr_{1}pq_{1}r_{2}}-\hat{\tilde{Y}}_{tr_{1}r_{2}pq_{1}})\frac{\partial^{3}\hat{U}^{M}_{p}}{\partial x_{r_{1}} \partial x_{r_{2}} \partial x_{q_{1}}}+&
\end{flalign*}
\begin{flalign}
&-s\hat{S}_{tr_{1}r_{2}pq_{1}q_{2}}^{1}\frac{\partial^{4}\hat{U}^{M}_{p}}{\partial x_{q_{1}} \partial x_{q_{2}} \partial x_{r_{1}} \partial x_{r_{2}}}+ \hat{b}_{t},&
\end{flalign}
which is formulated in terms of the transformed macro-displacement and its gradients up to the fourth order. The components of the constitutive tensors in the Laplace domain related to the homogenized continuum are defined as 
\begin{flalign}
\label{visco}
&\hat{G}_{tr_{1}pq_{1}}=\langle \hat{G}^{m}_{hkij}B^{(1,0)}_{hkpq_{1}}B^{(1,0)}_{ijtr_{1}}\rangle,&
\end{flalign}
\begin{flalign}
\label{visco1}
&\hat{Y}_{tr_{1}pq_{1}r_{2}}=\varepsilon \langle \hat{G}^{m}_{hkij}B^{(1,0)}_{hktr_{1}}B^{(2,0)}_{ijpq_{1}r_{1}}\rangle,&
\end{flalign}
\begin{flalign}
&\hat{\tilde{Y}}_{tr_{1}r_{2}pq_{1}}=\hat{Y}_{pq_{1}tr_{1}r_{2}}=\varepsilon \langle \hat{G}^{m}_{hkij}B^{(1,0)}_{hkpq_{1}}B^{(2,0)}_{ijtr_{1}r_{2}}\rangle,&
\end{flalign}
\begin{flalign}
&\hat{S}_{tr_{1}r_{2}pq_{1}q_{2}}^{1} = \varepsilon^{2} \langle  \hat{G}^{m}_{hkij}B^{(2,0)}_{hkpq_{1}q_{2}}B^{(2,0)}_{hktr_{1}r_{2}}-\hat{G}^{m}_{hkij}B^{(1,0)}_{ijtr_{1}}B^{(3,0)}_{hkpq_{1}q_{2}r_{2}}-\hat{G}^{m}_{hkij}B^{(1,0)}_{ijpq_{1}}B^{(3,0)}_{hktr_{1}r_{2}q_{2}}\rangle,&
\end{flalign} 
\begin{flalign}
\label{visco2}
&\hat{J}_{tpq_{1}}=\varepsilon \langle \hat{G}^{m}_{hkij}B^{(1,0)}_{hkpq_{1}}B^{(2,2)}_{ijt}\rangle,&
\end{flalign}
\begin{flalign}
&\hat{\tilde{J}}_{tq_{1}p}=\hat{J}_{ptq_{1}}=\varepsilon \langle \hat{G}^{m}_{hkij}B^{(1,0)}_{hktq_{1}}B^{(2,2)}_{ijp}\rangle,&
\end{flalign}
\begin{flalign}
\label{visco3}
&\hat{J}_{tp}^{\sharp}=\hat{J}_{tp}+\hat{\tilde{J}}_{pt}=\frac{\varepsilon^{2}}{2}\langle \hat{G}^{m}_{hkij}B^{(2,2)}_{hkp}B^{(2,2)}_{ijt}\rangle+\frac{\varepsilon^{2}}{2}\langle \hat{G}^{m}_{hkij}B^{(2,2)}_{hkt}B^{(2,2)}_{ijp}\rangle=\varepsilon^{2}\langle \hat{G}^{m}_{hkij}B^{(2,2)}_{hkp}B^{(2,2)}_{ijt}\rangle,&
\end{flalign}
\begin{flalign}
&\hat{J}_{tq_{2}pq_{1}}^{1}=\hat{J}_{pq_{1}tq_{2}}^{1}=- \varepsilon^{2} \langle -\hat{G}^{m}_{hkij}B^{(1,0)}_{hkpq_{1}}B^{(3,2)}_{ijtq_{2}}+\hat{G}^{m}_{hkij}B^{(2,0)}_{hkpq_{1}q_{2}}B^{(2,2)}_{ijp}-\hat{G}^{m}_{hkij}B^{(1,0)}_{hktq_{2}}B^{(3,2)}_{ijpq_{1}}+\hat{G}^{m}_{hkij}B^{(2,0)}_{hktq_{2}q_{1}}B^{(2,2)}_{ijt}\rangle,&
\end{flalign}
where the components $\hat{G}^{m}_{tr_{1}pq_{1}}$, $\hat{Y}_{tr_{1}pq_{1}r_{2}}$ and  $\hat{S}_{tr_{1}r_{2}pq_{1}q_{2}}^{1}$ of the constitutive tensors in the Laplace domain are computed due to the micro-fluctuation functions $N^{(1,0)}_{ikl}$, $N^{(2,0)}_{iklp}$ and $N^{(3,0)}_{iklpq}$. Such tensors are in accordance with the ones determined in \cite{Bacigalupo2014}.\\    
The transformed inertial tensor components are given as 
\begin{flalign}
\label{in}
&\rho=\langle \rho^{m}\rangle,&
\end{flalign}
\begin{flalign}
\label{in1}
&\hat{I}_{tpq_{1}}=\varepsilon \langle \rho^{m}N^{(1,0)}_{tpq_{1}}\rangle \frac{1}{\rho},&
\end{flalign}
\begin{flalign}
&\hat{\tilde{I}}_{tq_{1}p}=\hat{I}_{ptq_{1}}=\varepsilon \langle \rho^{m}N^{(1,0)}_{ptq_{1}}\rangle \frac{1}{\rho},&
\end{flalign}
\begin{flalign}
\label{in2}
&\hat{I}^{\sharp}_{pt}=\hat{I}_{tp}+\hat{\tilde{I}}_{pt}= \varepsilon^{2} \langle \rho^{m}(N^{(2,2)}_{tp}+N^{(2,2)}_{pt})\rangle \frac{1}{\rho},&
\end{flalign}
\begin{flalign}
\label{2.95}
&\hat{I}_{tp}=\frac{\varepsilon^{2}}{\rho}\langle \rho^{m}N^{(2,2)}_{tp}\rangle,&
\end{flalign}
\begin{flalign}
&\hat{\tilde{I}}_{pt}=\frac{\varepsilon^{2}}{\rho}\langle \rho^{m}N^{(2,2)}_{pt}\rangle,&
\end{flalign}
\begin{flalign}
\label{in3}
&\hat{I}_{tq_{2}pq_{1}}=\varepsilon^{2}\langle \rho^{m}N^{(1,0)}_{hpq_{1}} N^{(1,0)}_{htq_{2}}-\rho^{m}(N^{(2,0)}_{tpq_{1}q_{2}}+N^{(2,0)}_{tpq_{2}q_{1}})\rangle \frac{1}{\rho}.&
\end{flalign} 
 \subsection*{Approximation of the energy-like functional through truncation of the down-scaling relation}
\label{sub:met2}
An alternative approach is here presented to evaluate the overall constitutive and inertial tensors. To this purpose, the gradient referred to the down-scaling relation $\eqref{secorddis}$ is approximated at the first order as
\begin{flalign}
\label{devprio}
&\Big (\frac{D \hat{u}_{h}}{D x_{k}}\Big )^{I}=\frac{\partial \hat{U}_{h}}{\partial x_{k}}+N^{(1,0)}_{hpq_{1},k}\frac{\partial \hat{U}^{M}_{p}}{\partial x_{q_{1}}}+ \varepsilon \Big(N^{(1,0)}_{hpq_{1}}\frac{\partial^{2} \hat{U}^{M}_{p}}{\partial x_{q_{1}} \partial x_{q_{2}}}++N^{(2,0)}_{hpq_{1}q_{2},k}(\boldsymbol{\xi})\frac{\partial^{2}\hat{U}^{M}_{p}}{\partial x_{q_{1}}\partial x_{q_{2}}}+N^{(2,2)}_{hp,k}(\boldsymbol{\xi})s^{2}\hat{U}^{M}_{p}\Big).&
\end{flalign} 
In addition, the transformed micro-displacement at the second order is formulated as
\begin{flalign}
\label{secordu}
&\hat{u}^{II}_{h}(\boldsymbol{x},\boldsymbol{\xi},s)=\hat{U}_{h}(\boldsymbol{x},s)+\varepsilon N^{(1,0)}_{hpq_{1}}(\boldsymbol{\xi})\frac{\partial \hat{U}^{M}_{p}}{\partial x_{q_{1}}}+\varepsilon^{2}\Big(N^{(2,0)}_{hpq_{1}q_{2}}(\boldsymbol{\xi})\frac{\partial^{2}\hat{U}^{M}_{p}}{\partial x_{q_{1}}\partial x_{q_{2}}}+N^{(2,2)}_{hp}(\boldsymbol{\xi})s^{2}\hat{U}^{M}_{p}\Big),&
\end{flalign}
where the perturbation functions are helpful to determine a consistent approximation of the gradient at the first order. Then the gradient approximation in Eq. \eqref{devprio} and the displacement approximation in Eq. \eqref{secordu} are replaced into the transformed energy-like functional \eqref{bela}, which is approximated at the second order as
\begin{flalign}
\label{Lm}
&\hat{\Lambda}_{m}^{II} = \int_{L} \langle \hat{\lambda}_{m}^{II}(\boldsymbol{x},\boldsymbol{\xi})\rangle d\boldsymbol{x}= \frac{1}{2} s^{2}\langle \rho^{m} \rangle \int_{L}\hat{U}^{M}_{h}\hat{U}^{M}_{h} d\boldsymbol{x}+\varepsilon s^{2}\langle \rho^{m}N^{(1,0)}_{rpq_{1}}\rangle \int_{L}\frac{\partial \hat{U}^{M}_{p}}{\partial x_{q_{1}}}\hat{U}^{M}_{r} d\boldsymbol{x} +&
\end{flalign} 
\begin{flalign*}
&+\varepsilon s^{3}\langle \hat{G}^{m}_{hkij}B^{(1,0)}_{hkpq_{1}}B^{(2,2)}_{ijr}\rangle \int_{L} \frac{\partial \hat{U}^{M}_{p}}{\partial x_{q_{1}}}\hat{U}^{M}_{r} d\boldsymbol{x} +&
\end{flalign*} 
\begin{flalign*}
&+\varepsilon s \langle \hat{G}^{m}_{hkij}B^{(1,0)}_{hkpq_{1}}B^{(2,0)}_{ijrw_{1}w_{2}}\rangle \int_{L}\frac{\partial \hat{U}^{M}_{p}}{\partial x_{q_{1}}}\frac{\partial^{2}\hat{U}^{M}_{r}}{\partial x_{w_{1}} \partial x_{w_{2}}} d\boldsymbol{x}+\varepsilon^{2}s^{2}\Big \langle \frac{1}{2} \rho^{m}N^{(1,0)}_{hpq_{1}}N^{(1,0)}_{hrq_{2}}-\rho^{m}N^{(2,0)}_{rpq_{1}q_{2}}\Big \rangle \int_{L}\frac{\partial \hat{U}^{M}_{p}}{\partial x_{q_{1}}} \frac{\partial \hat{U}^{M}_{r}}{\partial x_{q_{2}}} d\boldsymbol{x}+&
\end{flalign*} 
\begin{flalign*}
&+\varepsilon^{2}s^{3}\langle -\hat{G}^{m}_{hkij}B^{(2,0)}_{hkpq_{1}q_{2}}B^{(2,2)}_{ijr}\rangle \int_{L}\frac{\partial \hat{U}^{M}_{p}}{\partial x_{q_{1}}} \frac{\partial \hat{U}^{M}_{r}}{\partial x_{q_{2}}} d\boldsymbol{x}+\varepsilon^{2}s^{4}\langle \rho^{m}N^{(2,2)}_{rp}\rangle \int_{L}\hat{U}^{M}_{p}\hat{U}^{M}_{r}d\boldsymbol{x}+&
\end{flalign*} 
\begin{flalign*}
&+\varepsilon^{2}s^{5}\frac{1}{2}\langle \hat{G}^{m}_{hkij}B^{(2,2)}_{hkp}B^{(2,2)}_{ijr}\rangle \int_{L}\hat{U}^{M}_{p}\hat{U}^{M}_{r}d\boldsymbol{x}+\frac{1}{2}s\langle \hat{G}^{m}_{hkij}B^{(1,0)}_{hkpq_{1}}B^{(1,0)}_{ijrw_{1}}\rangle \int_{L}\frac{\partial \hat{U}^{M}_{p}}{\partial x_{q_{1}}}\frac{\partial \hat{U}^{M}_{r}}{\partial x_{w_{1}}} d\boldsymbol{x}+&
\end{flalign*} 
\begin{flalign*}
&+\varepsilon^{2}s \Big \langle \frac{1}{2} \hat{G}^{m}_{hkij}B^{(2,0)}_{hkpq_{1}q_{2}}B^{(2,0)}_{ijrw_{1}w_{2}}\Big \rangle \int_{L}\frac{\partial^{2}\hat{U}^{M}_{p}}{\partial x_{q_{1}} \partial x_{q_{2}}} \frac{\partial^{2}\hat{U}^{M}_{r}}{\partial x_{w_{1}} \partial x_{w_{2}}} d\boldsymbol{x}- \int_{L}\hat{U}^{M}_{h}\hat{b}_{h}d\boldsymbol{x}.&
\end{flalign*} 
In accordance with the procedure proposed earlier, the Euler-Lagrangian equation deriving from first variation of the transformed energy-like functional \eqref{Lm} in the Laplace domain is
\begin{flalign*}
\label{macrolapdom}
&s^{2}\rho \hat{U}^{M}_{t}+s^{2}\rho (\hat{I}_{tpq_{1}}- \hat{\tilde{I}}_{tq_{1}p}) \frac{\partial \hat{U}^{M}_{p}}{\partial x_{q_{1}}}-s^{2}\rho \hat{I}_{tq_{2}pq_{1}} \frac{\partial^{2} \hat{U}^{M}_{p}}{\partial x_{q_{1}} \partial x_{q_{2}}}+s^{4}\rho \hat{I}_{tp}^{\sharp}\hat{U}^{M}_{p}=&
\end{flalign*}
\begin{flalign*}
&=-s^{3}(\hat{J}_{tpq_{1}}-\hat{\tilde{J}}_{tq_{1}p})\frac{\partial \hat{U}^{M}_{p}}{\partial x_{q_{1}}}-s^{3}\hat{J}_{tq_{2}pq_{1}}^{2}\frac{\partial^{2}\hat{U}^{M}_{p}}{\partial x_{q_{1}}\partial x_{q_{2}}}+&
\end{flalign*}
\begin{flalign*}
&-s^{5}\hat{J}_{tp}^{\sharp}\hat{U}^{M}_{p}+s\hat{G}_{tr_{1}pq_{1}}\frac{\partial^{2}\hat{U}^{M}_{p}}{\partial x_{q_{1}}\partial x_{r_{1}}}+s(\hat{Y}_{tr_{1}pq_{1}r_{2}}-\hat{\tilde{Y}}_{tr_{1}r_{2}pq_{1}})\frac{\partial^{3}\hat{U}^{M}_{p}}{\partial x_{r_{1}} \partial x_{r_{2}} \partial x_{q_{1}}}+&
\end{flalign*}
\begin{flalign}
&-s\hat{S}_{tr_{1}r_{2}pq_{1}q_{2}}^{2}\frac{\partial^{4}\hat{U}^{M}_{p}}{\partial x_{q_{1}} \partial x_{q_{2}} \partial x_{r_{1}} \partial x_{r_{2}}}+ \hat{b}_{t},&
\end{flalign}
where the overall inertial tensors are \eqref{in},\eqref{in1}, \eqref{in2} and \eqref{in3}, whereas the overall constitutive tensors are \eqref{visco},  \eqref{visco1}, \eqref{visco2}, \eqref{visco3}, respectively, and   
\begin{flalign}
&\hat{S}_{tr_{1}r_{2}pq_{1}q_{2}}^{2} = \varepsilon^{2} \langle  \hat{G}^{m}_{hkij}B^{(2,0)}_{hkpq_{1}q_{2}}B^{(2,0)}_{hktr_{1}r_{2}}\rangle,&
\end{flalign} 
\begin{flalign}
\label{J2}
&\hat{J}_{tq_{2}pq_{1}}^{2}=\hat{J}_{pq_{1}tq_{2}}^{2}=- \varepsilon^{2} \langle \hat{G}^{m}_{hkij}B^{(2,0)}_{hkpq_{1}q_{2}}B^{(2,2)}_{ijp}+\hat{G}^{m}_{hkij}B^{(2,0)}_{hktq_{2}q_{1}}B^{(2,2)}_{ijt}\rangle.&
\end{flalign} 
By applying the inverse Laplace transform $\mathcal{L}^{-1}$ to Eq. $\eqref{macrolapdomI}$ and Eq. $\eqref{macrolapdom}$, the field equation at the macro-scale corresponding to Eq. \eqref{eqn:din} is recast in the time domain as
\begin{flalign*}
\label{riteq}
&\rho \ddot{U}^{M}_{t}+\rho (I_{tpq_{1}}-\tilde{I}_{tq_{1}p}) \ast \frac{\partial \ddot{U}^{M}_{p}}{\partial x_{q_{1}}}-\rho I_{tq_{2}pq_{1}} \ast \frac{\partial^{2} \ddot{U}^{M}_{p}}{\partial x_{q_{1}} \partial x_{q_{2}}}+\rho I^{\sharp}_{tp} \ast \ddddot{U}^{M}_{p}=&
\end{flalign*}
\begin{flalign*}
&=-(\dot{J}_{tpq_{1}}-\dot{\tilde{J}}_{tq_{1}p}) \ast\frac{\partial \ddot{U}^{M}_{p}}{\partial x_{q_{1}}}-\dot{J}_{tq_{2}pq_{1}}^{i} \ast \frac{\partial^{2}\ddot{U}^{M}_{p}}{\partial x_{q_{1}}\partial x_{q_{2}}}+&
\end{flalign*}
\begin{flalign*}
&-\dot{J}_{tp}^{\sharp} \ast \ddddot{U}^{M}_{p}+G_{tr_{1}pq_{1}} \ast \frac{\partial^{2}\dot{U}^{M}_{p}}{\partial x_{q_{1}}\partial x_{r_{1}}}+(Y_{tr_{1}pq_{1}r_{2}}-\tilde{Y}_{tr_{1}r_{2}pq_{1}}) \ast \frac{\partial^{3}\dot{U}^{M}_{p}}{\partial x_{r_{1}} \partial x_{r_{2}} \partial x_{q_{1}}}+&
\end{flalign*}
\begin{flalign}
&-S_{tr_{1}r_{2}pq_{1}q_{2}}^{i}\ast \frac{\partial^{4}\dot{U}^{M}_{p}}{\partial x_{q_{1}} \partial x_{q_{2}} \partial x_{r_{1}} \partial x_{r_{2}}}+ b_{t},&
\end{flalign}
where the superscript $i=1,2$ and the symbol $\ast$ stands for the convolution and the time derivative can be moved from the constitutive tensor to the variable.\\
The constitutive tensor components in the time domain are 
\begin{flalign}
\label{tensor_ce}
&G^{m}_{tr_{1}pq_{1}}=\mathcal{L}^{-1}(\hat{G}^{m}_{tr_{1}pq_{1}}),&  
\end{flalign}
\begin{flalign}
&(Y_{tr_{1}pq_{1}r_{2}}-\tilde{Y}_{tr_{1}r_{2}pq_{1}}) = \mathcal{L}^{-1}(\hat{Y}_{tr_{1}pq_{1}r_{2}}-\hat{\tilde{Y}}_{tr_{1}r_{2}pq_{1}}),&
\end{flalign}
\begin{flalign}
&S_{tr_{1}r_{2}pq_{1}q_{2}}^{i}=\mathcal{L}^{-1} (\hat{S}_{tr_{1}r_{2}pq_{1}q_{2}}^{i}),&
\end{flalign}
\begin{flalign}
&\dot{J}_{tpq_{1}}-\dot{\tilde{J}}_{tq_{1}p}=\mathcal{L}^{-1}(s(\hat{J}_{tpq_{1}}-\hat{\tilde{J}}_{tq_{1}p})),&   
\end{flalign}
\begin{flalign}
&\dot{J}_{tq_{2}pq_{1}}^{i}=\mathcal{L}^{-1}(s\hat{J}_{tq_{2}pq_{1}}^{i}),&
\end{flalign}
\begin{flalign}
&\dot{J}_{tp}^{\sharp}=\mathcal{L}^{-1}(s\hat{J}_{tp}^{\sharp}),&
\end{flalign}
with $i=\{1,2\}$, whereas in the time domain the inertial tensor components and the acceleration result to be 
\begin{flalign}
\label{tensor_cd}
&(I_{tpq_{1}}-\tilde{I}_{tq_{1}p})=\mathcal{L}^{-1}(\hat{I}_{tpq_{1}}-\hat{\tilde{I}}_{tq_{1}p}),&
\end{flalign}
\begin{flalign}
&I_{tq_{2}pq_{1}}=\mathcal{L}^{-1}(\hat{I}_{tq_{2}pq_{1}}),&
\end{flalign}
\begin{flalign}
&I_{tp}^{\sharp}=\mathcal{L}^{-1}(\hat{I}_{tp}^{\sharp}),&
\end{flalign}
\begin{flalign}
& \ddot{U}^{M}_{t}=\mathcal{L}^{-1}(s^{2}\hat{U}^{M}_{t}).&
\end{flalign} 
In case of a locally homogeneous material, i.e. if the miscrostructure disappears, the perturbation functions $N^{(1,0)}_{ikl}$, $N^{(2,0)}_{iklp}$,...., $N^{(3,0)}_{iklpq}$ are zero and the components of the localization tensors defined in \eqref{eq:fl:80}-\eqref{eq:fl:83} vanish except for $B^{(1,0)}_{hkpq_{1}}$, which becomes $B^{(1,0)}_{hkpq_{1}}= \frac{1}{2}\Big (\delta_{hp}\delta_{kq_{1}}+\delta_{hq_{1}}\delta_{kp}\Big )$ and so the equation of motion of a classical homogeneous continuum is retrieved. 
\subsection{Dispersive wave propagation}
In this Subsection, the Laplace and the Fourier transforms are applied to Eq. \eqref{riteq}  with respect to time $t$ and to the slow variable $\boldsymbol{x}$ to obtain the field equation at the macroscale within the frequency and the wave vector domain. In particular the two-sided Fourier transform of an arbitrary function $f$ is defined as, \cite{paley1934fourier}: 
\begin{equation}
\label{eqn:fou}
\mathcal{F}(f(\boldsymbol{x}))= \check{f}(k) = \int_{-\infty}^{+\infty} f(\boldsymbol{x}) e^{\iota \boldsymbol{k} \cdot \boldsymbol{x}} d\boldsymbol{x}= \int_{-\infty}^{+\infty} f(\boldsymbol{x}) e^{\iota k_{s}x_{s}} d\boldsymbol{x},\quad \boldsymbol{k}\in \mathbb{R}^{2},
\end{equation} 
where $\boldsymbol{k}$ is a bidimensional vector and so the field equation at the macroscale in the transformed space is rephrased as
\begin{flalign*}
\label{reteqlf1}
&s^{2}\rho \check{\hat{U}}^{M}_{t}+s^{2}\rho \iota (\hat{I}_{tpq_{1}}+\hat{\tilde{I}}_{tq_{1}p}) \check{\hat{U}}^{M}_{p}k_{q_{1}}+s^{2}\rho \hat{I}_{tq_{2}pq_{1}} \check{\hat{U}}^{M}_{p}k_{q_{1}}k_{q_{2}}+s^{4}\rho \hat{I}^{\sharp}_{tp} \check{\hat{U}}^{M}_{p}=&
\end{flalign*}
\begin{flalign*}
&-s^{3} \iota (\hat{J}_{tpq_{1}}-\hat{\tilde{J}}_{tq_{1}p}) \check{\hat{U}}^{M}_{p}k_{q_{1}}-s^{3}\hat{J}_{tq_{2}pq_{1}}^{}\check{\hat{U}}^{M}_{p}k_{q_{1}}k_{q_{2}}+&
\end{flalign*}
\begin{flalign*}
&-s^{5}\hat{J}_{tp}^{\sharp}  \check{\hat{U}}^{M}_{p}-s\hat{G}_{tr_{1}pq_{1}} \check{\hat{U}}^{M}_{p}k_{q_{1}}k_{r_{2}}-s\iota(\hat{Y}_{tr_{1}pq_{1}r_{2}}-\hat{\tilde{Y}}_{tr_{1}r_{2}pq_{1}}) \check{\hat{U}}^{M}_{p}k_{r_{1}}k_{r_{2}}k_{q_{1}}+&
\end{flalign*}
\begin{flalign}
&-s\hat{S}_{tr_{1}r_{2}pq_{1}q_{2}}^{i} \check{\hat{U}}^{M}_{p}k_{q_{1}}k_{q_{2}}k_{r_{1}}k_{r_{2}}+ \check{\hat{b}}_{t}, \quad i=\{1,2\}.&
\end{flalign}
The vector  $\boldsymbol{k}$ is written respect to $\boldsymbol{n}$ as $\boldsymbol{k}=k\boldsymbol{n}$ ($k=||\boldsymbol{k}||$ and $||\boldsymbol{n}||=1$), therefore the Eq. \eqref{reteqlf1} becomes
\begin{flalign*}
\label{reteqlf}
&s^{2}\rho \check{\hat{U}}^{M}_{t}+s^{2}\rho \iota (\hat{I}_{tpq_{1}}+\hat{\tilde{I}}_{tq_{1}p}) \check{\hat{U}}^{M}_{p}kn_{q_{1}}+s^{2}\rho \hat{I}_{tq_{2}pq_{1}} \check{\hat{U}}^{M}_{p}k^{2}n_{q_{1}}n_{q_{2}}+s^{4}\rho \hat{I}^{\sharp}_{tp} \check{\hat{U}}^{M}_{p}=&
\end{flalign*}
\begin{flalign*}
&=-s^{3} \iota(\hat{J}_{tpq_{1}}- \hat{\tilde{J}}_{tq_{1}p}) \check{\hat{U}}^{M}_{p}kn_{q_{1}}-s^{3}\hat{J}_{tq_{2}pq_{1}}^{i}\check{\hat{U}}^{M}_{p}k^{2}n_{q_{1}}n_{q_{2}}-s^{5}\hat{J}_{tp}^{\sharp}  \check{\hat{U}}^{M}_{p}+&
\end{flalign*}
\begin{flalign*}
&-s\hat{G}_{tr_{1}pq_{1}} \check{\hat{U}}^{M}_{p}k^{2}n_{q_{1}}n_{r_{2}}-s \iota (\hat{Y}_{tr_{1}pq_{1}r_{2}}-\hat{\tilde{Y}}_{tr_{1}r_{2}pq_{1}})\check{\hat{U}}^{M}_{p}k^{3}n_{r_{1}}n_{r_{2}}n_{q_{1}}+&
\end{flalign*}
\begin{flalign}
&-s\hat{S}_{tr_{1}r_{2}pq_{1}q_{2}}^{i} \check{\hat{U}}^{M}_{p}k^{4}n_{q_{1}}n_{q_{2}}n_{r_{1}}n_{r_{2}}+ \check{\hat{b}}_{t},\quad i=\{1,2\}. &
\end{flalign}
In case of an orthotropic material the motion equation in the Laplace domain $\eqref{macrolapdom}$ along the direction $\boldsymbol{e}_{\beta}$ ($\beta=1,2$) is rephased as 
\begin{flalign}
\label{tdeq}
&s^2\rho \hat{U}^{M}_{\alpha}-s^2 \rho \hat{I}_{\alpha \beta \alpha \beta} \frac{\partial^{2} \hat{U}_{\alpha}}{\partial x_{\beta}^2}+s^4\rho \hat{I}_{\alpha \alpha} \hat{U}^{M}_{\alpha}=&
\end{flalign}
\begin{flalign*}
&=s^3 \hat{J}_{\alpha \beta \alpha \beta}^{i} \frac{\partial^{2} \hat{U}^{M}_{\alpha}}{\partial x_{\beta}^2}- s^5 \hat{J}_{\alpha \alpha} \hat{U}^{M}_{\alpha}+s \hat{G}_{\alpha \beta \alpha \beta} \frac{\partial^{2} \hat{U}^{M}_{\alpha}}{\partial x_{\beta}^2}-s \hat{S}_{\alpha \beta \beta \alpha \beta \beta}^{i} \frac{\partial^{4} \hat{U}^{M}_{\alpha}}{\partial x_{\beta}^4},&
\end{flalign*}
with $i=\{1,2\}$ and $\alpha=1,2$.
The Fourier transform $\eqref{eqn:fou}$ is applied to Eq. $\eqref{tdeq}$ with respect to the slow variable $\boldsymbol{x}$ to retrieve the Christoffel equation depending on the complex angular frequency $s$ and the wave vector $k_{\beta}$,
\begin{flalign}
\label{latrasdi}
&(s^2\rho+s^2\rho \hat{I}_{\alpha \beta \alpha \beta} k_{\beta}^2+s^4 \rho \hat{I}_{\alpha \alpha}+s^3 \hat{J}_{\alpha \beta \alpha \beta}^{i}k_{\beta}^2+s^5 \hat{J}_{\alpha \alpha}+s\hat{G}_{\alpha \beta \alpha \beta}k_{\beta}^2+s\hat{S}_{\alpha \beta \beta \alpha \beta \beta}^{i}k_{\beta}^4)\check{\hat{U}}^{M}_{\alpha}(k_{\beta},s)=0, \quad i=\{1,2\},&
\end{flalign}
where $\check{\hat{U}}(k_{\beta},s)$ stands for the Fourier transform of the transformed macro-displacement $\hat{U}(x,s)$.
The Christoffel equation $\eqref{latrasdi}$ defines the wave propagation in the viscoelastic medium that is embedded in the Laplace-Fourier space. The dispersion function stemmed from Eq. $\eqref{latrasdi}$ is
\begin{flalign}
\label{latrasdidue}
&s^2\rho+s^2\rho \hat{I}_{\alpha \beta \alpha \beta} k_{\beta}^2+s^4 \rho \hat{I}_{\alpha \alpha}+s^3 \hat{J}_{\alpha \beta \alpha \beta}^{i}k_{\beta}^2+s^5 \hat{J}_{\alpha \alpha}+s\hat{G}_{\alpha \beta \alpha \beta}k_{\beta}^2+s\hat{S}_{\alpha \beta \beta \alpha \beta \beta}^{i}k_{\beta}^4=0,\quad i=\{1,2\}.&
\end{flalign}
The dispersion function describes the longitudinal and the transverse oscillatory motion of the viscoelastic homogeneous continuum.  
\section{Homogenization of a bi-phase layered material}
\begin{figure}[b!]
	\centering
	\includegraphics[height=0.35\textheight]{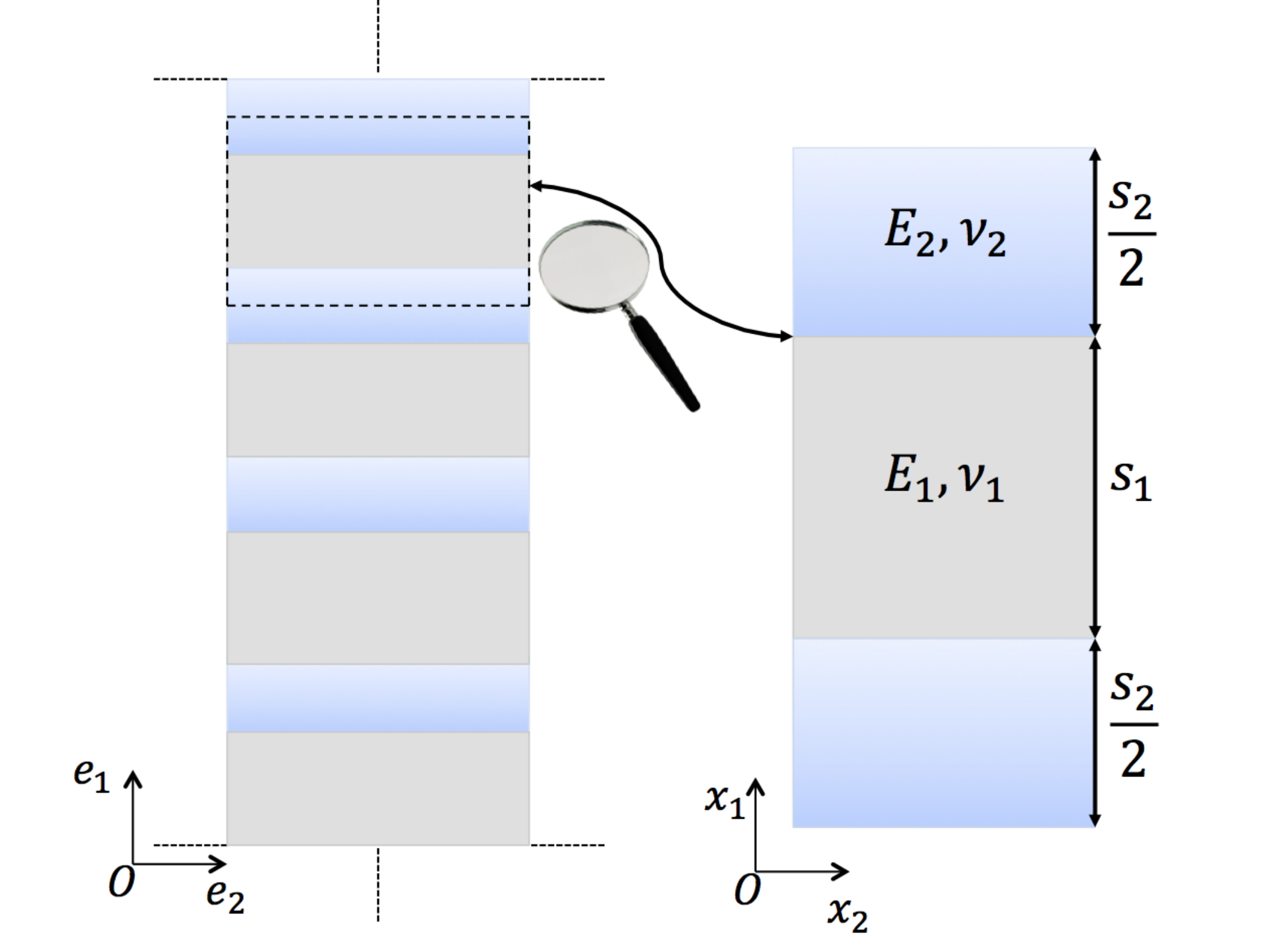}
	\caption{Heterogeneous bidimensional domain with its layered periodic cell.}
	\label{stratificato}
\end{figure}

The model proposed in Section \ref{sec:six} is herein applied to a domain made of two layered materials, which have thickness $s_{1}$ and $s_{2}$, and subject to $\mathcal{L}$-periodic body forces $\boldsymbol{b}(\boldsymbol{x})$. The domain displays orthotropic phases and the orthotropic axis is supposed to be parallel to the direction $\boldsymbol{e}_{1}$, see Fig. \ref{stratificato}. In case of isotropic phases, for the plane-stress state we have $\tilde{E}=E$ and $\tilde{\nu}=\nu$, whereas for the plane-strain state $\tilde{E}=\frac{E}{1-\nu^{2}}$ and $\tilde{\nu}=\frac{\nu}{1-\nu}$, where $E$ is the Young's modulus and $\nu$ is the Poisson's ratio.
\\For sake of simplicity but without loss of generality, the components of the viscoelastic tensor are
\begin{flalign*}
\label{compvisco}
&G_{1111}^{i}=G_{2222}^{i}=G_{1111}^{i,\infty}(e^{-\frac{t}{\tau_{r}}}+1), \quad G_{1122}^{i}=G_{1122}^{i, \infty}(e^{-\frac{t}{\tau_{r}}}+1),&
\end{flalign*}
\begin{flalign}
&G_{1212}^{i}=G_{1212}^{i,\infty}(e^{-\frac{t}{\tau_{r}}}+1),\quad i\in \{1,2\},&
\end{flalign}
where $G_{1111}^{i,\infty}=\frac{\tilde{E}}{1-\tilde{\nu}^{2}}$, $G_{1122}^{i, \infty}=\frac{\tilde{E}\tilde{\nu}}{1-\tilde{\nu}^{2}}$ and $G_{1212}^{i,\infty}=\frac{\tilde{E}}{2(1+\tilde{\nu})}$ are the equilibrium elastic modulus with $G_{1111}^{i, \infty}=G_{2222}^{i,\infty}$. It can be noticed that the viscoelastic tensor can be deemed as a term of the Prony series \citep{ferry1980viscoelastic} and so the proposed homogenization technique can be applied to any kind of sufficiently regular kernel, since the overall constitutive and inertial tensor components have a general structure.\\ 
In addition, $\tau_{r}$ stands for the relaxation time and the superscript $i$ represents either phase 1 or the phase 2. The Laplace transform \eqref{eqn:lap} applied to the components of the viscoelastic tensor \eqref{compvisco} leads to
\begin{flalign*}
&\hat{G}_{1111}^{i}= \hat{G}_{2222}^{i}=\frac{\tilde{E}}{1-\tilde{\nu}^{2}}\frac{2\tau_{r} s+1}{s(s\tau_{r}+1)}, \quad \hat{G}_{1122}^{i}=\frac{\tilde{E}\tilde{\nu}}{1-\tilde{\nu}^{2}}\frac{2\tau_{r} s+1}{s(s\tau_{r}+1)},&
\end{flalign*}
\begin{flalign}
&\hat{G}_{1212}^{i}=\frac{\tilde{E}}{2(1+\tilde{\nu})}\frac{2\tau_{r} s+1}{s(s\tau_{r}+1)}, \quad i\in \{1,2\}.&
\end{flalign}
Finally the relation $\mathbb{\hat{C}}^{m}\Big(\frac{\boldsymbol{x}}{\varepsilon},s \Big) =s\mathbb{\hat{G}}^{m}\Big(\frac{\boldsymbol{x}}{\varepsilon},s \Big)$ provides the viscoelastic tensor components as 
\begin{flalign*}
\label{comptrans}
&\hat{C}_{1111}^{i}= \hat{G}_{2222}^{i}=\frac{\tilde{E}}{1-\tilde{\nu}^{2}}\frac{2\tau_{r} s+1}{(s\tau_{r}+1)}, \quad \hat{C}_{1122}^{i}=\frac{\tilde{E}\tilde{\nu}}{1-\tilde{\nu}^{2}}\frac{2\tau_{r} s+1}{(s\tau_{r}+1)},& 
\end{flalign*}
\begin{flalign}
&\hat{C}_{1212}^{i}=\frac{\tilde{E}}{2(1+\tilde{\nu})}\frac{2\tau_{r} s+1}{(s\tau_{r}+1)}, \quad i\in \{1,2\}.&
\end{flalign}
The transformed components of the viscoelastic tensor \eqref{comptrans} are employed to determine the perturbation functions of first, second and third order, obtained by solving the cell problems \eqref{cps1}, \eqref{eqn:icsym}, \eqref{intconN32}, (see Appendix C for the structure of the perturbation functions).\\
The following dimensionless quantities are introduced
\begin{equation*}
r_{\rho}=\frac{\rho_{1}}{\rho_{2}},\quad r_{E}=\frac{\tilde{E}_{1}}{\tilde{E}_{2}}, \quad  \tau_{\varsigma} =\frac{\tau_{r}}{s_{2}}\sqrt{\frac{\tilde{E}_{2}}{\rho_{2}}}, \quad \eta=\frac{s_{1}}{s_{2}},
\end{equation*}
where $r_{\rho}$ stands for the ratio between the densities, $r_{E}$ is the ratio between the Young's moduli related to phase $1$ and phase $2$, $\tau_{\varsigma}$ is the relaxation time and $\eta$ is the ratio between the thicknesses of the layers $s_{1}$ and $s_{2}$.\\
\begin{figure}
	\centering
	\subfigure{\includegraphics[width=0.90\columnwidth]{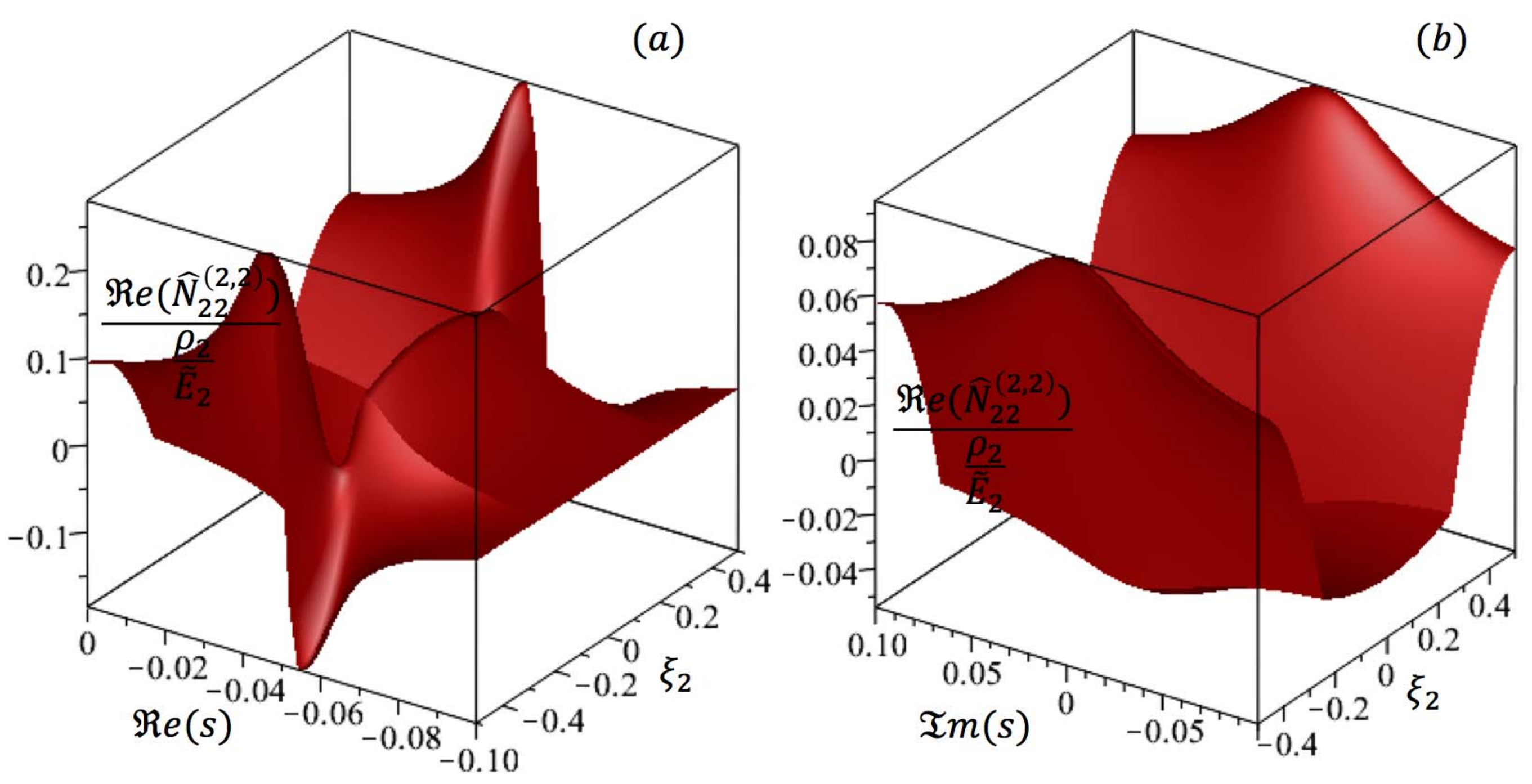}\label{3}}
	\subfigure{\includegraphics[width=0.90\columnwidth]{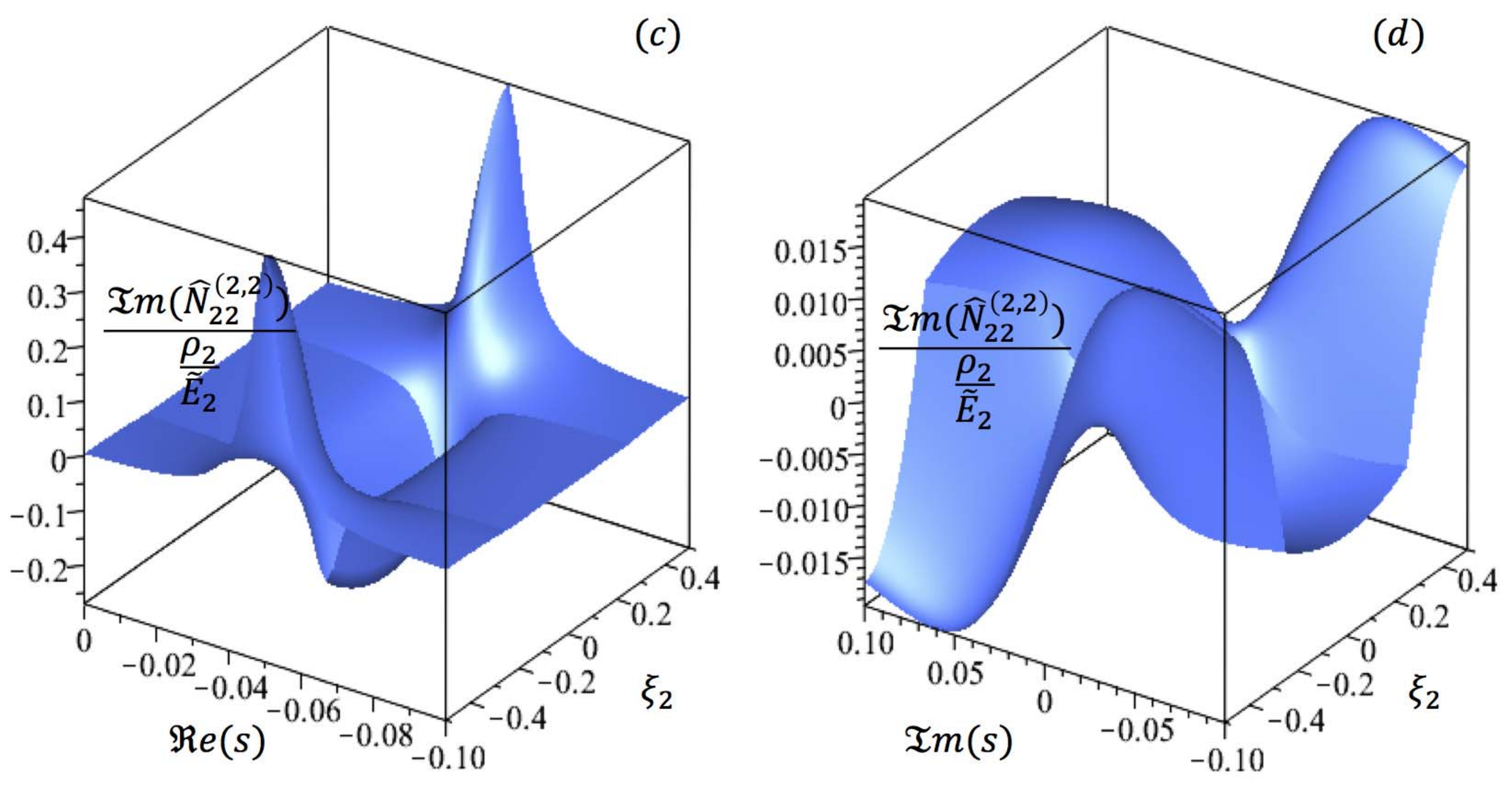}\label{4}}
	\caption{(a) Dimensionless real part $\mathfrak{R}e(\hat{N}^{(2,2)}_{22})\tilde{E}_{2}/\rho_{2}$ vs. the real part of $s$, $\mathfrak{R}e(s)$, and the coordinate $\xi_{2}$. (b) Dimensionless real part $\mathfrak{R}e(\hat{N}^{(2,2)}_{22})\tilde{E}_{2}/\rho_{2}$ vs. the imaginary part of $s$, $\mathfrak{I}m(s)$, and $\xi_{2}$. (c) Dimensionless imaginary part $\mathfrak{I}m(\hat{N}^{(2,2)}_{22})\tilde{E}_{2}/\rho_{2}$ vs. the real part of $s$, $\mathfrak{R}e(s)$, and the coordinate $\xi_{2}$. (d) Dimensionless imaginary part $\mathfrak{I}m(\hat{N}^{(2,2)}_{22})\tilde{E}_{2}/\rho_{2}$ vs. the imaginary part of $s$, $\mathfrak{I}m(s)$, and $\xi_{2}$ obtained for $r_{\rho}=10$, $r_{E}=10$, $\tau_{\varsigma}=10$, $\tilde{\nu}_{1}=\tilde{\nu}_{2}=0.2$ and $\eta=1$.}
\end{figure}
The perturbation function $N^{(2,2)}_{22}$ is analytically computed by solving the cell problem \eqref{22N} supplied with the interface conditions \eqref{22icN} in Sec. \ref{sec:four}, with respect to the phase $1$ and the phase $2$, and the structure of $N^{(2,2)}_{22}$ is reported in terms of the geometric and mechanical properties of the periodic domain in Eq. \eqref{zia}. Such function depends on the fast variable $\xi_{2}$, which is perpendicular to the transversal direction $\boldsymbol{e}_{1}$, as well as on the complex parameter $s$.
The perturbation function is non-dimensionalized as $\frac{N^{(2,2)}_{22}\tilde{E}_{2}}{\rho_{2}}$ and its real part $\frac{\mathfrak{R}e(N^{(2,2)}_{22})\tilde{E}_{2}}{\rho_{2}}$ and its imaginary part $\frac{\mathfrak{I}m(N^{(2,2)}_{22})\tilde{E}_{2}}{\rho_{2}}$ are taken into account since $s$ is a complex number.\\
 Fig. 3-(a) shows the real part of $\frac{N^{(2,2)}_{22}\tilde{E}_{2}}{\rho_{2}}$ along the periodic cell vs. the real part of $s$, $\mathfrak{R}e(s)$, and the vertical coordinate $\xi_{2}$.\\
  Fig. 3-(b) depicts the dependence of $\frac{\mathfrak{R}e(N^{(2,2)}_{22})\tilde{E}_{2}}{\rho_{2}}$ with respect to the imaginary part of $s$,  $\mathfrak{I}m(s)$, and $\xi_{2}$. The imaginary part of $\frac{N^{(2,2)}_{22}\tilde{E}_{2}}{\rho_{2}}$ is shown in Fig. $3$-(c) vs. $\mathfrak{R}e(s)$ and in Fig. $3$-(d) vs. $\mathfrak{I}m(s)$, by varying the vertical coordinate $\xi_{2}$.\\
The cell problem \eqref{32N} equipped with the interface conditions \eqref{intconN32} provides the perturbation function $N^{(3,2)}_{222}$ and its formulation is explicitely expressed in Eq. \eqref{sta}, where there is point in noticing they take into account of the effect of the microstructural heterogeneities of the domain.
The real part and the imaginary part of the dimensionless perturbation function $\frac{N^{(3,2)}_{222}\tilde{E}_{2}}{\rho_{2}}$ are considered, which are $\frac{\mathfrak{R}e(N^{(3,2)}_{222})\tilde{E}_{2}}{\rho_{2}}$ and $\frac{\mathfrak{I}m(N^{(3,2)}_{222})\tilde{E}_{2}}{\rho_{2}}$.\\
 Fig. 4-(a) and Fig. 4-(b) show how the real part of $\frac{N^{(3,2)}_{222}\tilde{E}_{2}}{\rho_{2}}$ depends on the real and imaginary parts of $s$ by varying the vertical coordinate $\xi_{2}$.\\
  Fig. 4-(c) and Fig. 4-(d) show that the imaginary part of $\frac{N^{(3,2)}_{222}\tilde{E}_{2}}{\rho_{2}}$ depends on $\mathfrak{R}e(s)$ and $\mathfrak{I}m(s)$ and the vertical coordinate $\xi_{2}$.\\
   In particular, the poles of dimensionless perturbation functions are emphasized in Fig. $3$-a, Fig. $3$-c, Fig. $4$-a and Fig. $4$-c.    
In all the previous figures, it is straightforward that the dimensionless perturbation functions $\frac{\mathfrak{R}e(N^{(2,2)}_{22})\tilde{E}_{2}}{\rho_{2}}$, $\frac{\mathfrak{I}m(N^{(2,2)}_{22})\tilde{E}_{2}}{\rho_{2}}$,  $\frac{\mathfrak{R}e(N^{(3,2)}_{222})\tilde{E}_{2}}{\rho_{2}}$ and  $\frac{\mathfrak{I}m(N^{(3,2)}_{222})\tilde{E}_{2}}{\rho_{2}}$ are $\mathcal{Q}$-periodic, they have vanishing mean values over the unit cell $\mathcal{Q}$ and they are smooth along the boundaries of $\mathcal{Q}$, as expected.\\
 The Poisson ratios are assumed to be equal for both phases $\tilde{\nu}_{1}=\tilde{\nu}_{2}=0.2$ and the dimensionless relaxation time is $\tau_{\varsigma}=10$. In addition, the dimensionless Young's modulus is $r_{E}=10$, the density is $r_{\rho}=10$ and the ratio between the thicknesses is $\eta=1$.\\
 \begin{figure}
 	\centering
 	\includegraphics[width=0.90\columnwidth]{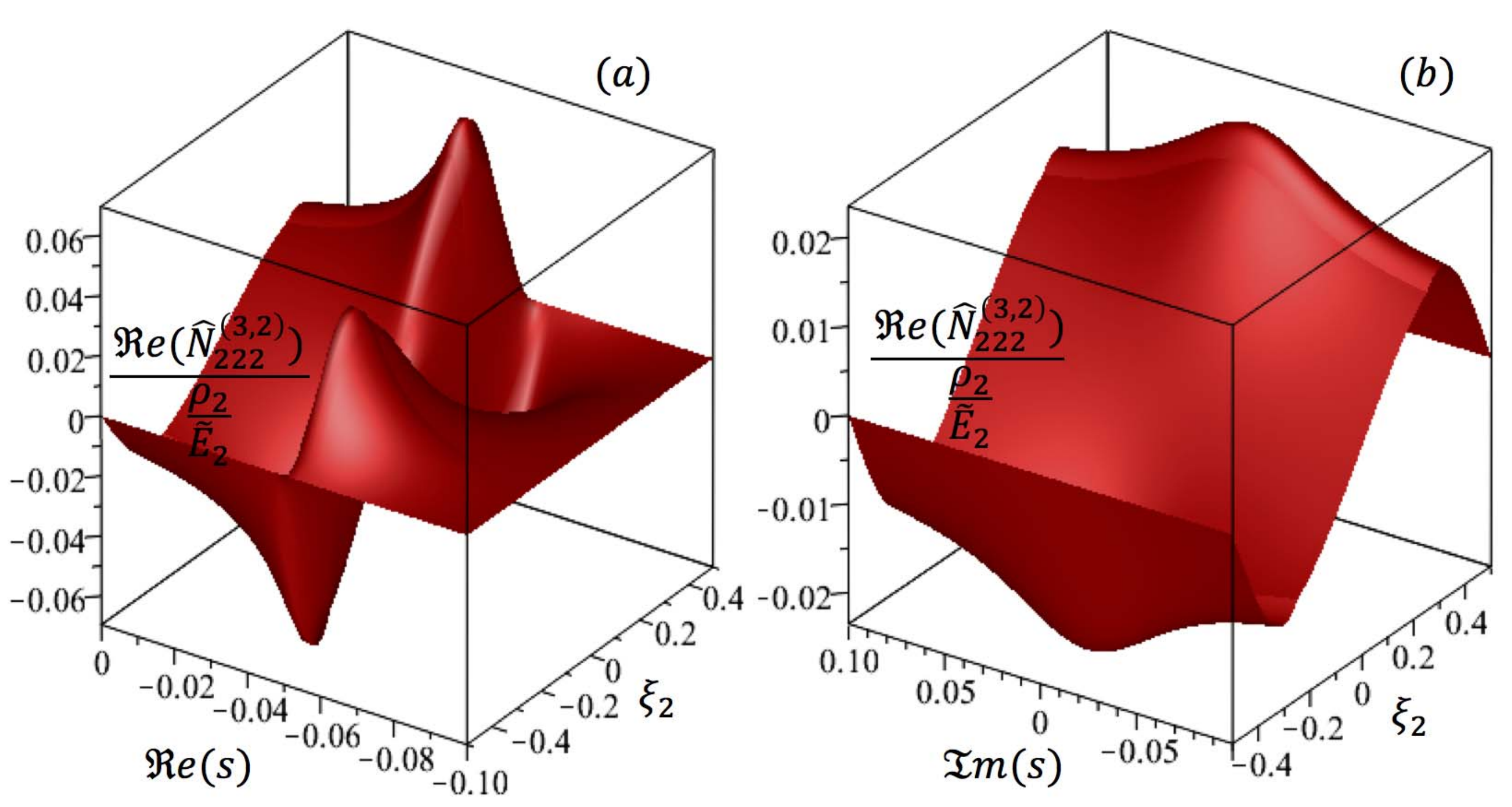}\label{3}	
 	\includegraphics[width=0.90\columnwidth]{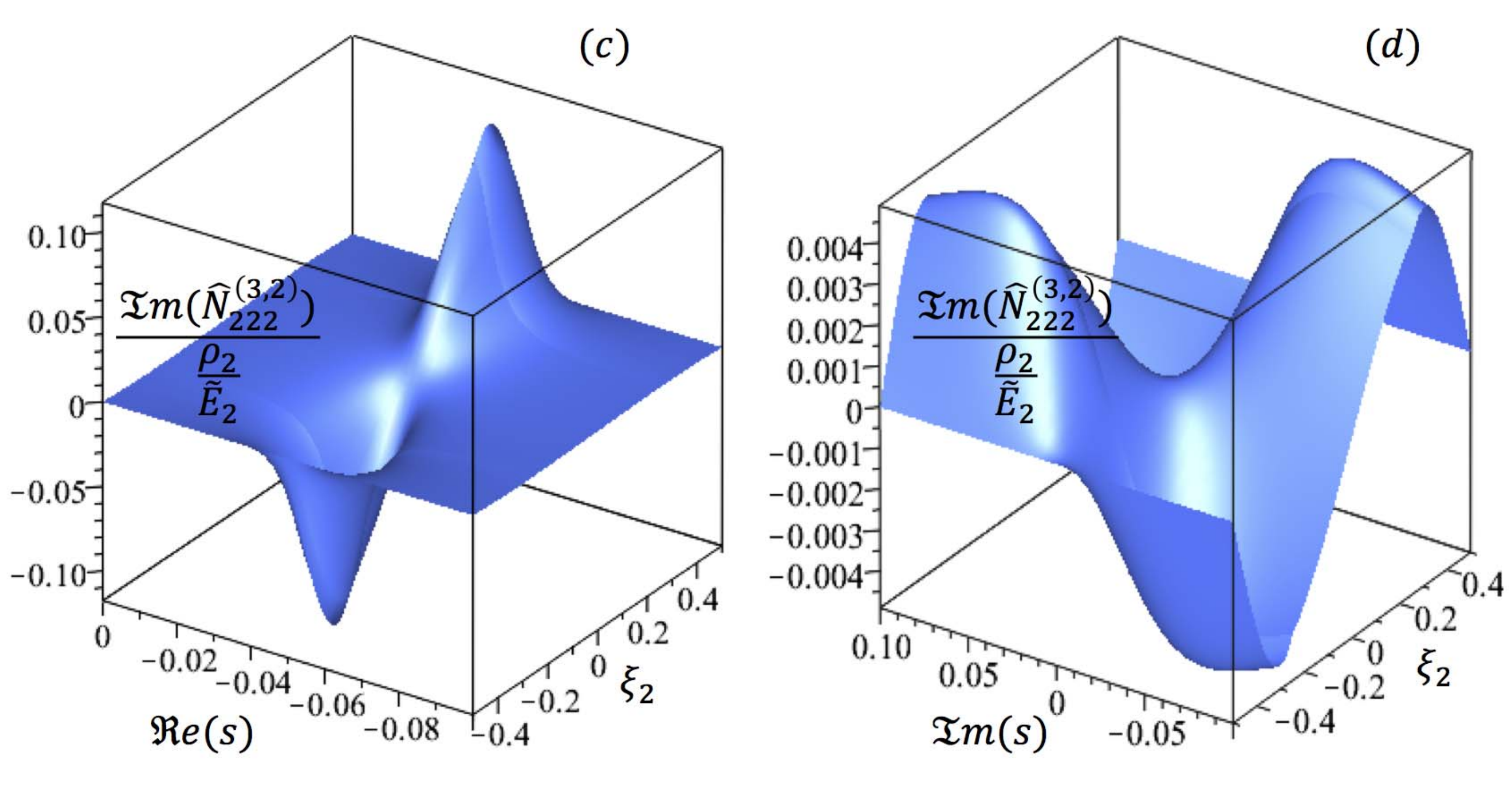}\label{4}
 	\caption{(a) Dimensionless real part of the perturbation function $\hat{N}^{(3,2)}_{222}$, $\mathfrak{R}e(\hat{N}^{(3,2)}_{222})\tilde{E}_{2}/\rho_{2}$, vs. the real part of $s$, $\mathfrak{R}e(s)$, and the coordinate $\xi_{2}$. (b) Dimensionless real part  $\mathfrak{R}e(\hat{N}^{(3,2)}_{222})\tilde{E}_{2}/\rho_{2}$ vs. the imaginary part of $s$, $\mathfrak{I}m(s)$, and $\xi_{2}$. (c) Dimensionless imaginary part  $\mathfrak{R}e(\hat{N}^{(3,2)}_{222})\tilde{E}_{2}/\rho_{2}$ vs. $\mathfrak{R}e(s)$, and $\xi_{2}$.(d) Dimensionless imaginary part  $\mathfrak{R}e(\hat{N}^{(3,2)}_{222})\tilde{E}_{2}/\rho_{2}$ vs. $\mathfrak{R}e(s)$, and $\xi_{2}$, obtained for $r_{\rho}=10$, $r_{E}=10$, $\tau_{\varsigma}=10$, $\tilde{\nu}_{1}=\tilde{\nu}_{2}=0.2$ and $\eta=1$.}
 \end{figure}
The transformed inertial tensor components $\hat{J}_{11}$, $\hat{J}_{1111}^{2}$ and $\hat{I}_{11}$, referred to the wave propagation along the transversal direction $\boldsymbol{e}_{1}$, are taken into account. They are affected by the complex angular frequency $s$ and their dimensionless components are written as 
$\frac{\hat{J}_{11}\sqrt\frac{\tilde{E}_{2}}{\rho_{2}}}{\varepsilon^{2}s_{2}\rho_{2}}$, $\frac{\hat{J}_{1111}^{2}\sqrt\frac{\tilde{E}_{2}}{\rho_{2}}}{\varepsilon^{2}s_{2}\rho_{2}}$ and $\frac{\hat{I}_{11}\tilde{E}_{2}}{\varepsilon^{2}\rho_{2}}$. The $\mathfrak{R}e(s)$-axis is assumed to be negative and therefore the poles of the dimensionless components are visible in Figs. $5$, $6$ and $7$.\\
 The second order transformed inertial tensor component $\frac{\hat{J}_{11}\sqrt\frac{\tilde{E}_{2}}{\rho_{2}}}{\varepsilon^{2}s_{2}\rho_{2}}$ is computed by means of Eq. \eqref{visco3} and it is decomposed into its real part $\frac{\mathfrak{R}e(\hat{J}_{11})\sqrt\frac{\tilde{E}_{2}}{\rho_{2}}}{\varepsilon^{2}s_{2}\rho_{2}}$ and its imaginary part $\frac{\mathfrak{I}m(\hat{J}_{11})\sqrt\frac{\tilde{E}_{2}}{\rho_{2}}}{\varepsilon^{2}s_{2}\rho_{2}}$, due to the presence of the complex frequency $s$.\\
  Fig. $5$-(a) shows the behaviour of $\frac{\mathfrak{R}e(\hat{J}_{11})\sqrt\frac{\tilde{E}_{2}}{\rho_{2}}}{\varepsilon^{2}s_{2}\rho_{2}}$ by varying the real part and the imaginary part of $s$. In particular it can be observed that the real part of $\frac{\hat{J}_{11}\sqrt\frac{\tilde{E}_{2}}{\rho_{2}}}{\varepsilon^{2}s_{2}\rho_{2}}$ is symmetric with respect to the $\mathfrak{R}e(s)$-axis.\\ 
In Fig. $5$-(b) the imaginary part $\frac{\mathfrak{I}m(\hat{J}_{11})\sqrt\frac{\tilde{E}_{2}}{\rho_{2}}}{\varepsilon^{2}s_{2}\rho_{2}}$ is plotted  with respect to the real part of $s$, $\mathfrak{R}e(s)$, and the imaginary part of $s$, $\mathfrak{I}m(s)$. The function assumes either positive or negative values and it is symmetric respect to zero.\\
Eq. \eqref{2.95} allows computing the transformed second order inertial tensor component $\frac{\hat{I}_{11}\tilde{E}_{2}}{\varepsilon^{2}\rho_{2}}$, which is decomposed into its real part $\frac{\mathfrak{R}e(\hat{I}_{11})\tilde{E}_{2}}{\varepsilon^{2}\rho_{2}}$ and its imaginary part $\frac{\mathfrak{I}m(\hat{I}_{11})\tilde{E}_{2}}{\varepsilon^{2}\rho_{2}}$.\\
The dependency of $\frac{\mathfrak{R}e(\hat{I}_{11})\tilde{E}_{2}}{\varepsilon^{2}\rho_{2}}$ and $\frac{\mathfrak{I}m(\hat{I}_{11})\tilde{E}_{2}}{\varepsilon^{2}\rho_{2}}$  on the real part and the imaginary part of $s$ is shown in Figs. $6$-(a) and $6$-(b). The symmetry of $\frac{\mathfrak{R}e(\hat{I}_{11})\tilde{E}_{2}}{\varepsilon^{2}\rho_{2}}$ with respect to $\mathfrak{R}e(s)$-axis and of $\frac{\mathfrak{I}m(\hat{I}_{11})\tilde{E}_{2}}{\varepsilon^{2}\rho_{2}}$ with respect to zero are shown.\\ 
Finally, in Figs. $7$-(a) and $7$-(b), the real part of the dimensionless fourth order inertial tensor component $\frac{\mathfrak{R}e(\hat{J}_{1111}^{2})\sqrt\frac{\tilde{E}_{2}}{\rho_{2}}}{\varepsilon^{2}s_{2}\rho_{2}}$ and its imaginary 
part $\frac{\mathfrak{I}m(\hat{J}_{1111}^{2})\sqrt\frac{\tilde{E}_{2}}{\rho_{2}}}{\varepsilon^{2}s_{2}\rho_{2}}$ are plotted vs. the real and the imaginary parts of the complex frequency $s$.
It is worth to noticing that the inertial tensor component $\hat{J}_{1111}^{1}$ stemming from the approach $1$ is vanishing.\\
	\begin{figure}
	\centering
	\includegraphics[height=0.31\textheight]{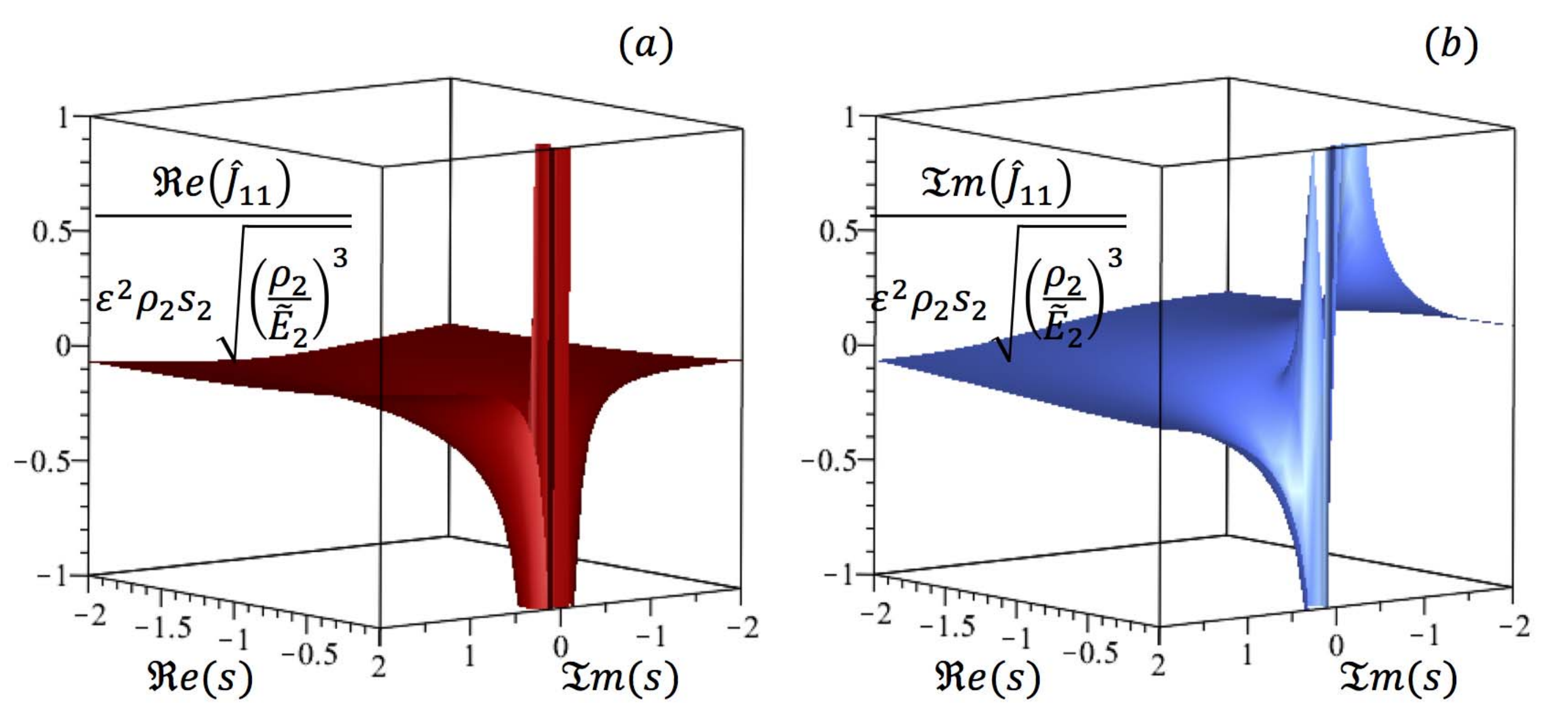}\label{77}
	\caption{(a) Dimensionless real part component $\mathfrak{R}e(\hat{J}_{11})\sqrt\frac{\tilde{E}_{2}}{\rho_{2}}/\varepsilon^{2}s_{2}\rho_{2}$ vs. the real part of $s$, $\mathfrak{R}e(s)$, and the imaginary part of $s$, $\mathfrak{I}m(s)$. (b) Dimensionless imaginary part component
		$\mathfrak{I}m(\hat{J}_{11}^{2})\sqrt\frac{\tilde{E}_{2}}{\rho_{2}}/\varepsilon^{2}s_{2}\rho_{2}$ vs. the real part of $s$, $\mathfrak{R}e(s)$, and the imaginary part of $s$, $\mathfrak{I}m(s)$, with $r_{\rho}=10$, $r_{E}=10$, $\tau_{\varsigma}=10$, $\tilde{\nu}_{1}=\tilde{\nu}_{2}=0.2$ and $\eta=1$.}
\end{figure}
\begin{figure}
	\centering
	\includegraphics[height=0.31\textheight]{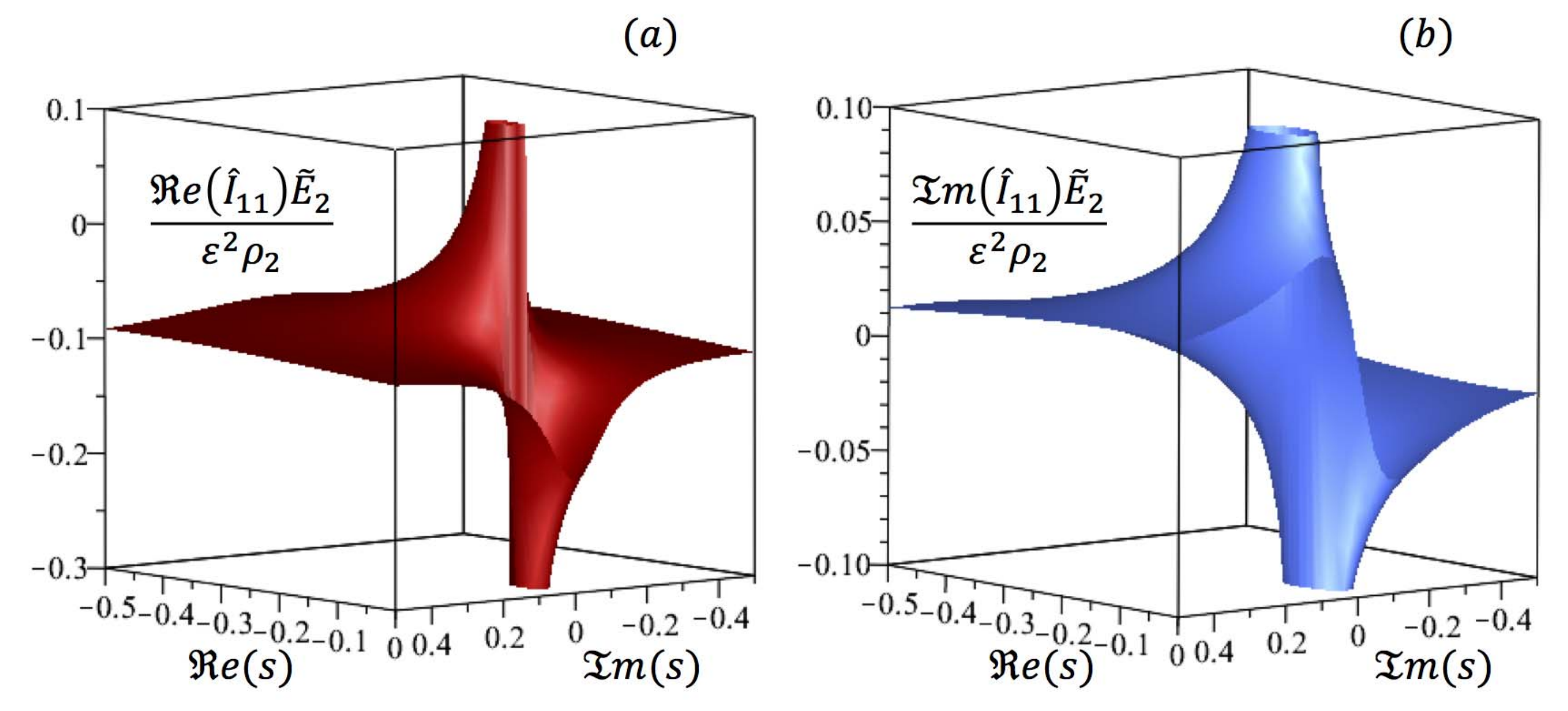}\label{6}
	\caption{(a) Dimensionless real part component $\mathfrak{R}e(\hat{I}_{11})\tilde{E}_{2}/\varepsilon^{2}s_{2}$ vs. the real part of $s$, $\mathfrak{R}e(s)$, and the imaginary part of $s$, $\mathfrak{I}m(s)$. (b) Dimensionless component 
		$\mathfrak{I}m(\hat{I}_{11})\tilde{E}_{2}/\varepsilon^{2}s_{2}$ vs. the real part of $s$, $\mathfrak{R}e(s)$, and the imaginary part of $s$, $\mathfrak{I}m(s)$, obtained for $r_{\rho}=10$, $r_{E}=10$, $\tau_{\varsigma}=10$, $\tilde{\nu}_{1}=\tilde{\nu}_{2}=0.2$ and $\eta=1$.}
\end{figure} 
	\begin{figure}[b!]
	\centering
	\includegraphics[height=0.31\textheight]{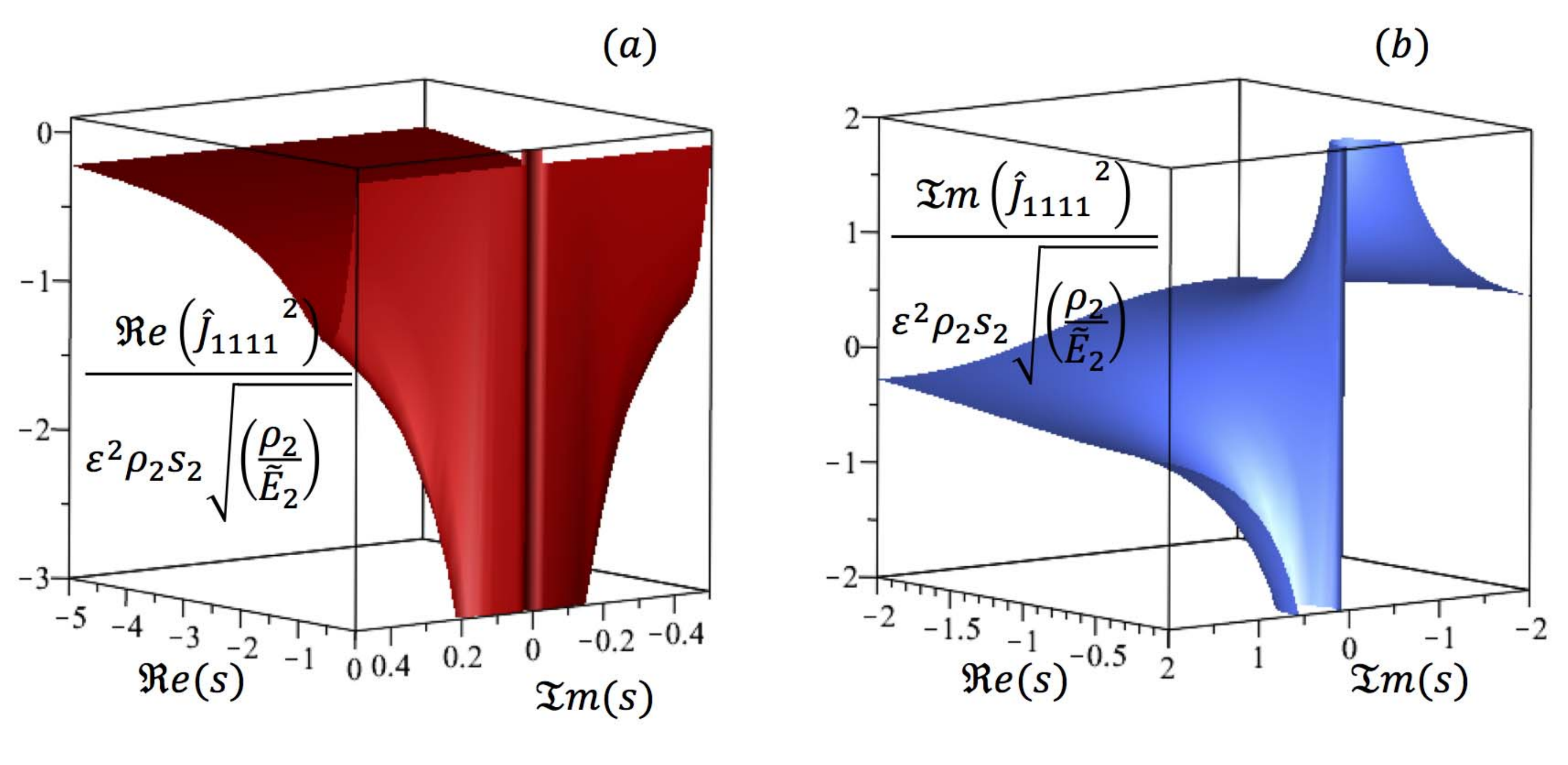}\label{7}
	\caption{(a) Dimensionless real part component $\mathfrak{R}e(\hat{J}_{1111}^{2})\sqrt\frac{\tilde{E}_{2}}{\rho_{2}}/\varepsilon^{2}s_{2}\rho_{2}$ vs. the real part of $s$, $\mathfrak{R}e(s)$, and the imaginary part of $s$, $\mathfrak{I}m(s)$. (a) Dimensionless imaginary part component
		$\mathfrak{I}m(\hat{J}_{1111}^{2})\sqrt\frac{\tilde{E}_{2}}{\rho_{2}}/\varepsilon^{2}s_{2}\rho_{2}$ vs. the real part of $s$, $\mathfrak{R}e(s)$, and the imaginary part of $s$, $\mathfrak{I}m(s)$. Both are retrieved for $r_{\rho}=10$, $r_{E}=10$, $\tau_{\varsigma}=10$, $\tilde{\nu}_{1}=\tilde{\nu}_{2}=0.2$ and $\eta=1$.}
\end{figure} 
Fig. $8$-(a) shows that the magnitudes of the constitutive tensor components $\hat{S}_{111111}^{1}$ and $\hat{S}_{111111}^{2}$, determined for the compressional waves along the direction $\boldsymbol{e}_{1}$, are compared in terms of the magnitude of $s$
and the non-dimensional Young's modulus $r_{E}$. $|\hat{S}_{111111}^{1}|$ decreases more steadily than $|\hat{S}_{111111}^{2}|$, as $\frac{1}{10}<r_{E}<1$,
 and both are equal to zero when the material is homogeneous, i.e. when $r_{E}=1$.
   With increasing the dimensionless Young's modulus $r_{E}$, a significant rise in $|\hat{S}_{111111}^{2}|$ is observed as compared to $|\hat{S}_{111111}^{1}|$. Finally as $|s|$ goes up, $|\hat{S}_{111111}^{1}|$ decreases more slowly than 
$|\hat{S}_{111111}^{2}|$.\\
 In Fig. $8$-(b) the magnitudes of $\hat{S}_{211211}^{1}$ and $\hat{S}_{211211}^{2}$ are plotted with respect to the magnitude of the complex angular frequency $|s|$ and the non-dimensional Young's modulus $r_{E}$. They are computed for the shear waves travelling along the layering direction $\boldsymbol{e}_{1}$. It can be observed that if $\frac{1}{10}<r_{E}<1$, the magnitudes of $\hat{S}_{211211}^{1}$ and $\hat{S}_{211211}^{2}$ decrease down to zero when the material is homogeneous. A rise in $r_{E}$
(with $r_{E}>1$) makes $|\hat{S}_{211211}^{1}|$ and $|\hat{S}_{211211}^{2}|$ increasing rapidly. Moreover, a growth in the magnitude of $s$ leads $|\hat{S}_{211211}^{1}|$ and $|\hat{S}_{211211}^{2}|$ to decrease.\\ 
Fig. $8$-(c) depicts the behaviour of the magnitudes of the transformed viscoelastic component $\hat{G}_{1212}$ related to the shear wave travelling along $\boldsymbol{e}_{2}$. It is observed that 
the trend of $|\hat{G}_{1212}|$ steadely increases by a rise in the Young's modulus and by varying the magnitude of the complex frequency $s$.\\
 Fig. $8$-(d) shows the magnitude of the transformend viscoelastic component $\hat{G}_{1111}$ concerning with the compressional wave along $\boldsymbol{e}_{1}$. $|\hat{G}_{1111}|$ significantly grows up in the interval $\frac{1}{10}<r_{E}<4$ by varying $|s|$ and then its trend becomes a plateau.
 	\begin{figure}
 	\centering
 	\subfigure{\includegraphics[width=0.86\columnwidth]{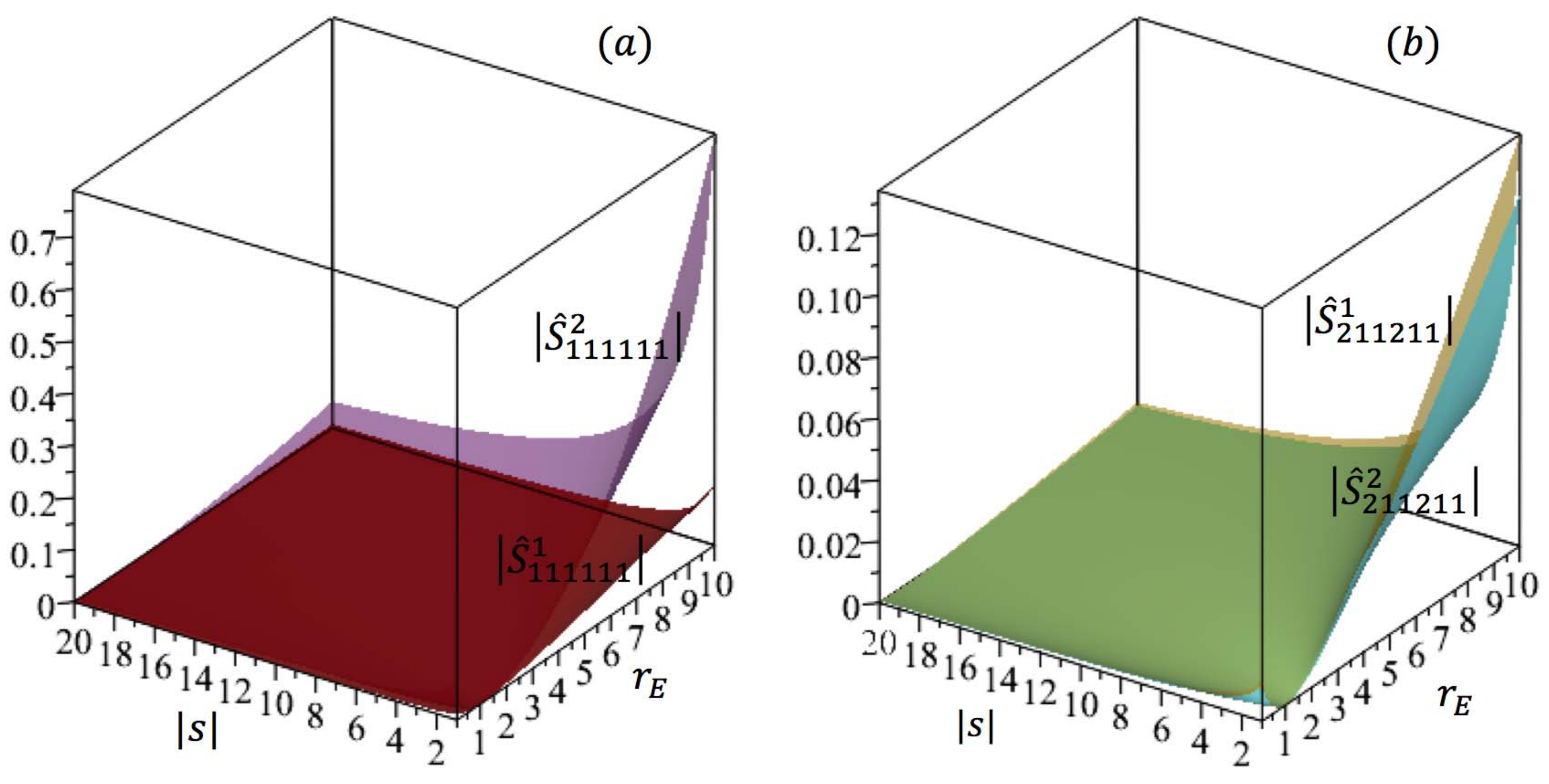}\label{3}}
 	\subfigure{\includegraphics[width=0.86\columnwidth]{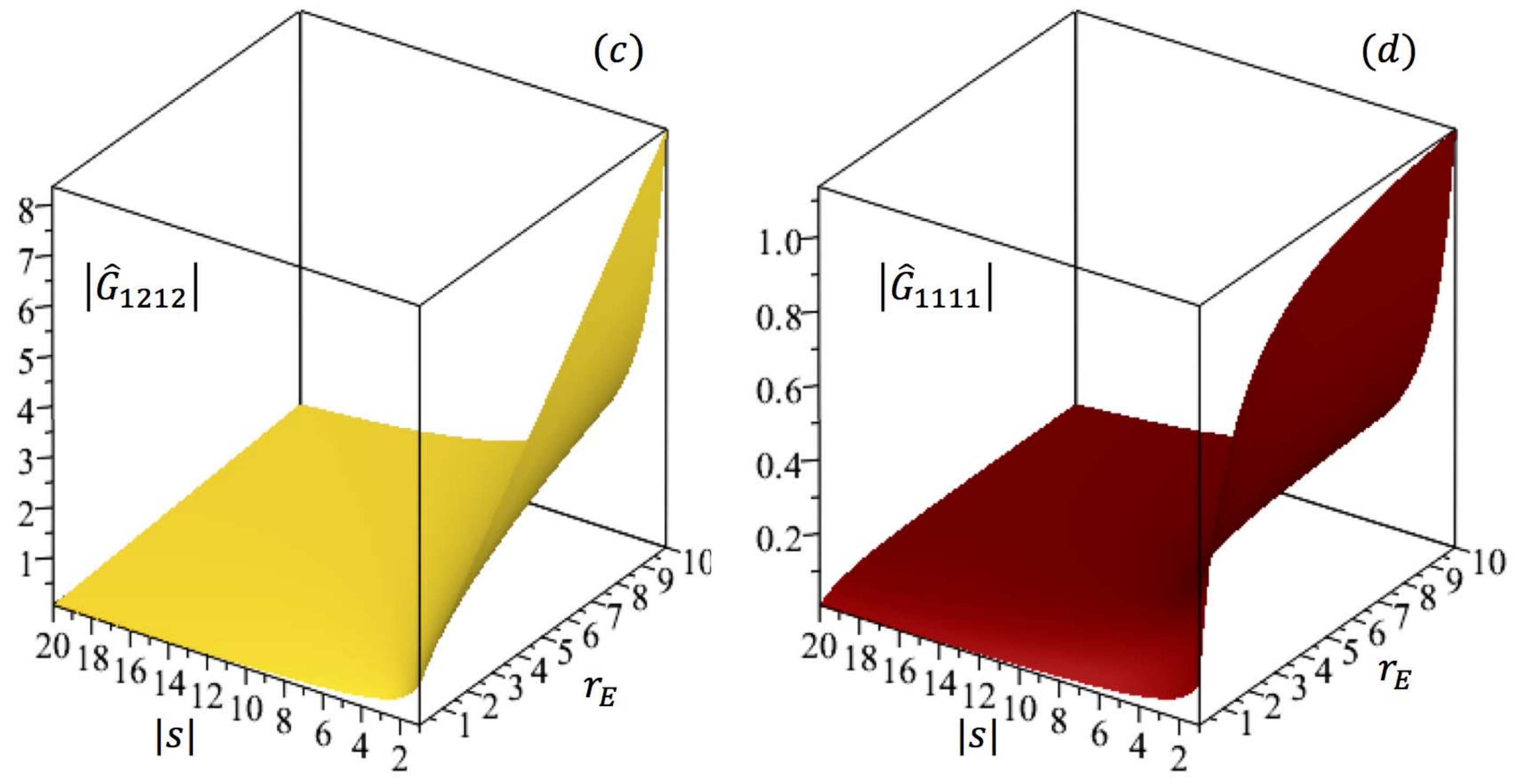}\label{4}}
 	\caption{(a) The magnitude of the constitutive tensor $\hat{S}_{111111}^{1}$ (red) is compared with $|\hat{S}_{111111}^{2}|$ (violet) by varying the dimensionless Young's modulus $r_{E}$ and the magnitude of $s$. (b) $|\hat{S}_{211211}^{1}|$ (gold) and $|\hat{S}_{211211}^{2}|$ (green) are compared with respect to $r_{E}$ and $|s|$.(c) $|\hat{G}_{1212}|$ is depicted by varying the Young's modulus $r_{E}$ and the magnitude of $s$. (d) $|\hat{G}_{1111}|$ is represented by changing $r_{E}$ and the magnitude of $s$. The graphs are obtained for $r_{\rho}=10$, $\tau_{\varsigma}=10$, $\tilde{\nu}_{1}=\tilde{\nu}_{2}=0.2$, $\eta=1$ and the argument of the s is $\theta=0$.}
 \end{figure}
 
\subsection{Benchmark test: homogenized problem vs. heterogeneous material}

A horizontal (or vertical) sample of length $L=L_{1}(=L_{2})$ is taken into account since the body
forces and the heterogeneous domain are periodic. The solution of the heterogeneous model is provided by a numerical procedure, accounting for the actual heterogeneous composition of the layered composite, and it is compared with the solution deriving from the homogenized model. The field equation \eqref{tdeq} is written as
\begin{flalign}
\label{soluzana}
&s \hat{G}_{\alpha \beta \alpha \beta} \frac{\partial^{2} \hat{U}^{M}_{\alpha}}{\partial x_{\beta}^2}-s \hat{S}_{\alpha \beta \beta \alpha \beta \beta}^{i} \frac{\partial^{4} \hat{U}^{M}_{\alpha}}{\partial x_{\beta}^4}=-\hat{b}_{\alpha},&
\end{flalign}
where $i=\{1,2\}$ and the indices $\alpha$ and $\beta$ are not summed. The shear problem takes place for $\alpha \ne\beta$ whereas the compressional problem for $\alpha = \beta$. The transformed macro-displacement is determined from Eq. \eqref{soluzana}
	\begin{flalign}
	\label{solsol}
	&\hat{U}^{M}_{\alpha}(x_{\beta})=\Big(\frac{L_{\beta}}{2\pi}\Big)^{2}\frac{\hat{b}_{\alpha}}{s \hat{G}_{\alpha \beta \alpha \beta}\Big [1+\Big (\frac{2\pi k}{L_{\beta}} \Big)^{2}\frac{\hat{S}_{\alpha \beta \beta \alpha \beta \beta}^{i}}{\hat{G}_{\alpha \beta \alpha \beta}}\Big]},&
	\end{flalign}
	with $i=1,2$ and $k=1,2,..$ The dimensionless transformed macro-displacement is written as
	\begin{flalign}
	\label{soluzioni}
	&\tilde{\hat{U}}^{M}_{\alpha}=\frac{\hat{U}^{M}_{\alpha}\tilde{E}_{1}}{\Upsilon_{\alpha}\Big(\frac{L_{\beta}}{2\pi}\Big)^{2}},&
	\end{flalign}  
	where $\tilde{E}_{1}$ is the Young's modulus related to the phase $1$.
	A FE method provides the macro-displacements $\hat{U}^{M}_\alpha$, computed for each individual unit cell composing the sample of length $L/\varepsilon=11$, with amplitute $\Upsilon_{\alpha}=1$ N/mm$^3$, $\tilde{E}_{1}=10000$ MPa, $\tilde{E}_{2}=1000$ MPa, $\tilde{\nu}_{1}= \tilde{\nu}_{2}=0.2$, $\tau_{\varsigma}=5$, $\eta=1$ and $r_{\rho}=10$.

	The homogenized and the heterogeneous problems are 
	supposed to be in a plane stress state.  In Fig. $9$-(a) the magnitude of the dimensionless transformed macro-displacement
	\eqref{soluzioni} with $\alpha=1$, obtained by solving analytically the field equation \eqref{soluzana} along the direction $x_{1}$, is compared with the solution provided by a finite element analysis of the heterogeneous domain equipped with proper periodic boundary conditions on the displacement. The continuous curve stands for the analytical dimensionless transformed macro-displacement $\tilde{\hat{U}}^{M}_{1}(x_{1}),$ which is derived from the method $1$ providing the overall constitutive tensor $\hat{S}_{111111}^{1}$, Eq. \eqref{S111111_1}.

The dotted curve represents the analytical dimensionless transformed macro-
	displacement $\tilde{\hat{U}}^{M}_{1}(x_{1})$ and derives from the alternative method $2$, which provides the overall constitutive tensor $\hat{S}_{111111}^{2}$, Eq. \eqref{S211211_2}. Finally, the diamonds stand for the numerical results related to the heterogeneous model and obtained from the corresponding microscopic solution through the up-scaling relation \eqref{upsca} and considering the imaginary part of the body force $\hat{b}_{1}= \Upsilon_{1}\sin{\frac{2\pi k}{L_{1}}x_{1}},$ with $k=1.$\\
	 All the methods are compared in Fig. $9$-(a) and three values for the dimensionless parameter $s_{\varsigma} =s s_{2}/\sqrt{\frac{\tilde{E}_{2}}{\rho_{2}}}$ are considered. The red curves and diamonds are obtained for $s_{\varsigma}=-2$, the blue curves and diamonds are given for $s_{\varsigma}=-0.5$ and finally the green ones stand for $s_{\varsigma}=-0.3$. By increasing the values of $s_{\varsigma}$ a gradual reduction in the behaviour of the solution is emphasized. In Fig. $9$-(b) the magnitude of the analytical dimensionless transformed macro-displacement $\tilde{\hat{U}}^{M}_{2}(x_{1}),$ which is obtained from the method $1$ (continuous curve), and the transformed macro-displacement stemming from the method $2$ (dotted curve) are put in relation with the heterogeneous solution (diamonds) for three increasing values of $s_{\varsigma}$, which are $s_{\varsigma}=-2$ (red one), $s_{\varsigma}=-0.5$  (blue one) and $s_{\varsigma}=-0.3$ (green one) respectively. The dotted and the continuous curves are almost coincident.\\
	  Both for the compressional and shear problems $9$-(a) and $9$-(b), there is a very agreement deal between the solution of both homogenized models and the numerical solution of the heterogeneous approach in the static case. Along the orthotropic direction $x_{2}$, the 
	  
	overall constitutive tensors $\hat{S}_{122122}^{i}$
	and $\hat{S}_{222222}^{i}$, with $i=1,2$, are equal
	to zero and so the transformed macro-displacement \eqref{solsol} assumes the form $\hat{U}^{M}_{\alpha}(x_{2})=\Big(\frac{L_{2}}{2\pi}\Big)^{2}\frac{\Upsilon_{\alpha}\sin{\frac{2\pi }{L_{2}}x_{2}}}{s \hat{G}_{\alpha 2 \alpha 2}},$ with $\alpha=1,2.$ Therefore, the magnitude of the dimensionless macro-displacement field $\tilde{\hat{U}}^{M}_{1}(x_{2})$ (Fig. $9$-(c)) and the magnitude of the dimensionless macro-
	displacement $\hat{U}^{M}_{2}(x_{2})$ (Fig. $9$-(d)), represented by continuous curves, are compared with the numerical macro-displacements associated with the heterogeneous model (diamonds), by varying three values of the dimensionless parameter $s_{\varsigma}$, which are $s_{\varsigma}=-2$ (red one), $s_{\varsigma}=-0.5$  (blue one) and $s_{\varsigma}=-0.3$ (green one) respectively.
	\begin{figure}
	\centering
	\subfigure{\includegraphics[width=0.80\columnwidth]{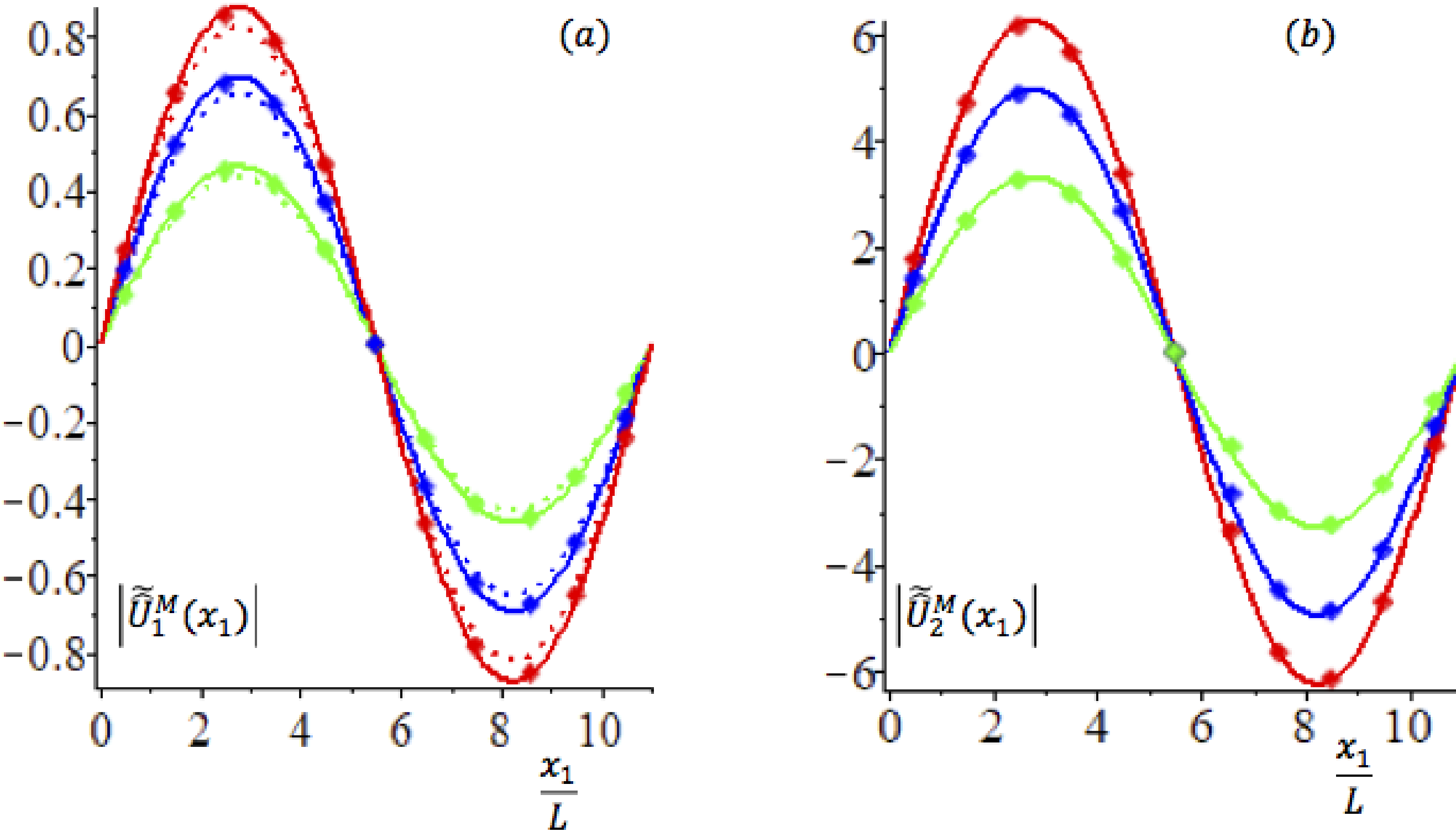}\label{5}}
	\subfigure{\includegraphics[width=0.80\columnwidth]{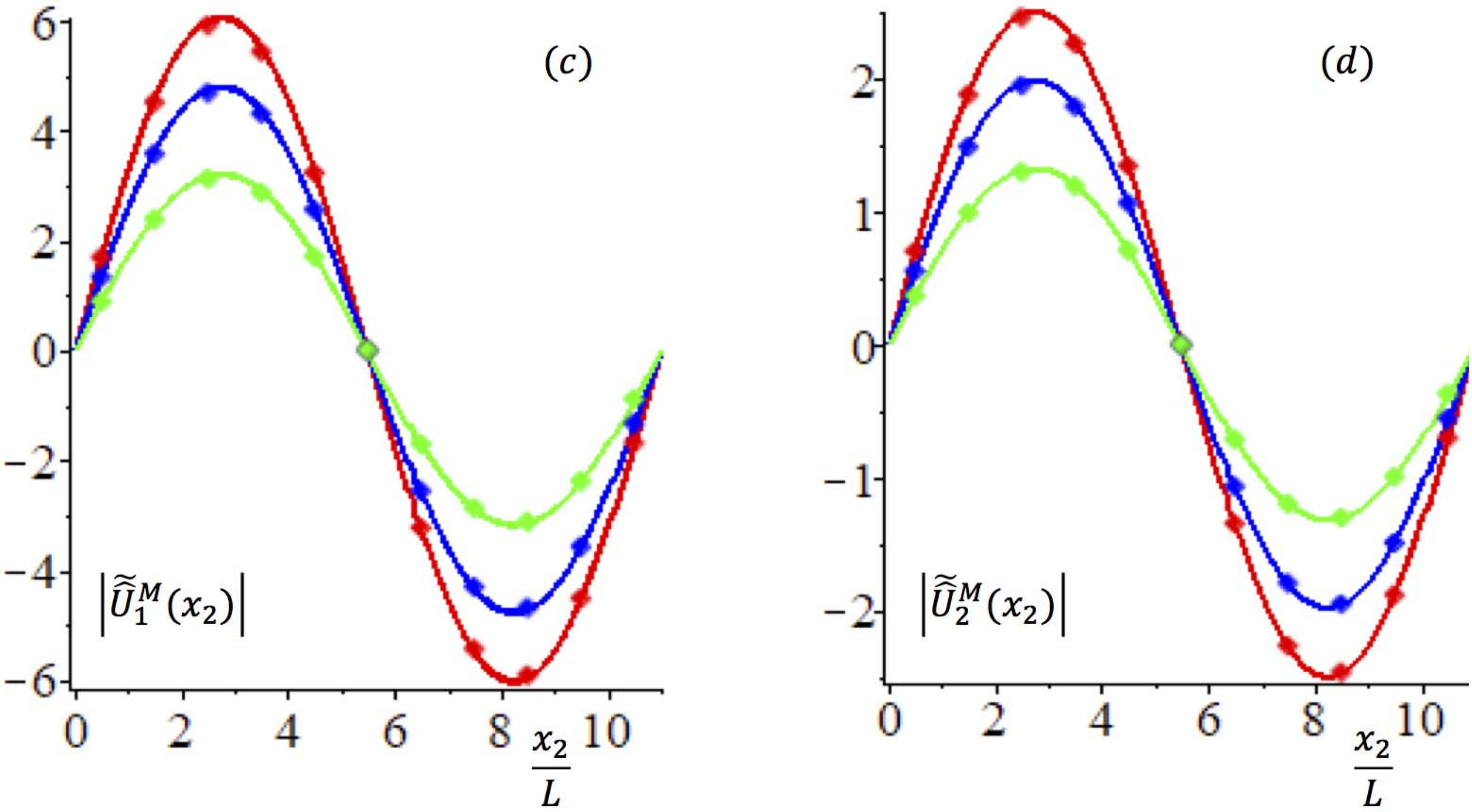}\label{6}}
	\caption{\small(a) Magnitude of the dimensionless macro-displacement component $\tilde{\hat{U}}^{M}_{1}(x_{1})$ induced by the harmonic body force $\hat{b}_{1}(x_{1})$ and (b) $|\tilde{\hat{U}}^{M}_{2}(x_{1})|$ caused by $\hat{b}_{2}(x_{1})$ along direction $x_{1}$.(c) Magnitude of the component $\tilde{\hat{U}}^{M}_{1}(x_{2})$ induced by $\hat{b}_{1}(x_{2})$ and (d) $|\tilde{\hat{U}}^{M}_{2}(x_{2})|$ produced by $\hat{b}_{2}(x_{2})$ along $x_{2}$. The solution given by the homogenized model $1$ (continuous curves) and the homogenized model $2$ (dotted curves) are compared with the solution given by the heterogeneous one (diamonds). The red curves are given for the dimensionless parameter $s_{\varsigma}=-2$, the blue curves for $s_{\varsigma}=-0.5$ and the green ones for $s_{\varsigma}=-0.3$.}
\end{figure}	
\subsection{Dispersion curves for layered materials}
	The dispersion relation \eqref{latrasdidue} is an implicit function depending on the real part of s,
	$\mathfrak{R}e(s)$, its imaginary part, $\mathfrak{I}m(s)$, 
	and the wave number $k_{\beta}$, $\beta=1,2$, which have been non-dimensionalized as $\frac{\mathfrak{R}e(s)s_{2}}{\sqrt{\frac{\tilde{E}_{2}}{\rho_{2}}}}$, $\frac{\mathfrak{I}m(s)s_{2}}{\sqrt{\frac{\tilde{E}_{2}}{\rho_{2}}}}$ and $k_{\beta}\varepsilon$. Eq. \eqref{latrasdidue} provides the dispersion curves related to the dynamic homogenization model and its simplified version presented in Sec. \eqref{sub:met2}. \\
	The dispersion curves obtained with two approaches are compared in Fig. $10$ and Fig. $11$. The curves given by the dynamic homogenization model are plotted with a thicker line than the ones obtained with the simplified version and it is observed 
	that the curves produced from both approaches are pretty coincident in the range of the considered wave-number. The dual nature of a viscoelastic material is reflected by the presence of the complex parameter $s$. Its real part is related to the viscosity of the material whereas its imaginary part deals with
	its elastic behaviour. Therefore, the dispersion curves lay both on the real axis of $s$, which describes the wave propagation, and its imaginary axis, which characterizes the damping.\\
	 Fig. $10$ and Fig. $11$ illustrate the variation of the dimensionless real part of $s$ in terms of the dimensionless imaginary part of $s$ and the dimensionless wave number. In case of dispersion curves that travel along the axis $\boldsymbol{e}_{\beta}$, the wave number $k_{\beta}\varepsilon$ belongs to the

	interval $[0,\pi]$, with
	\begin{figure}
		\centering
		\subfigure{\includegraphics[height=0.35\textheight]{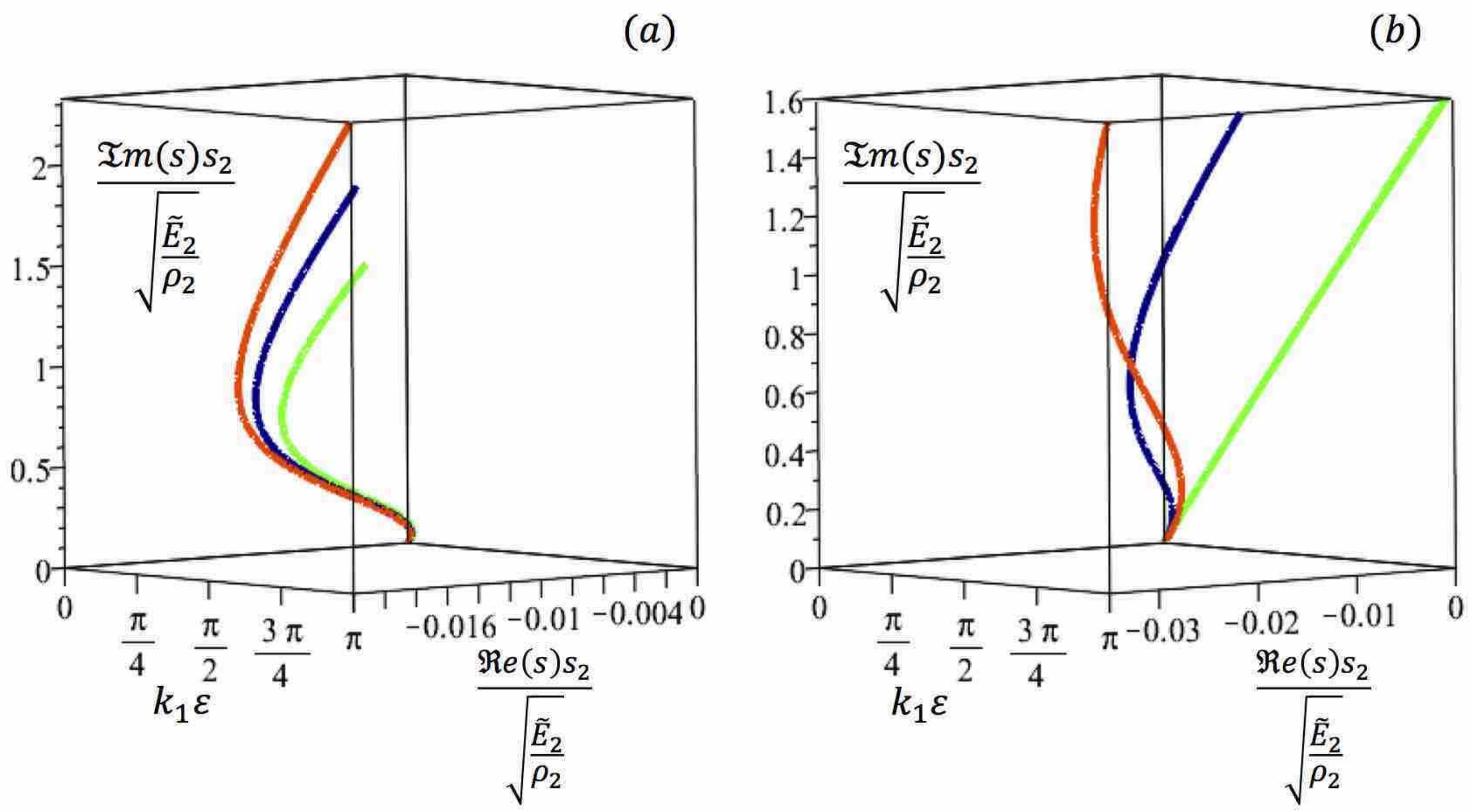}\label{10}}
		\subfigure{\includegraphics[height=0.35\textheight]{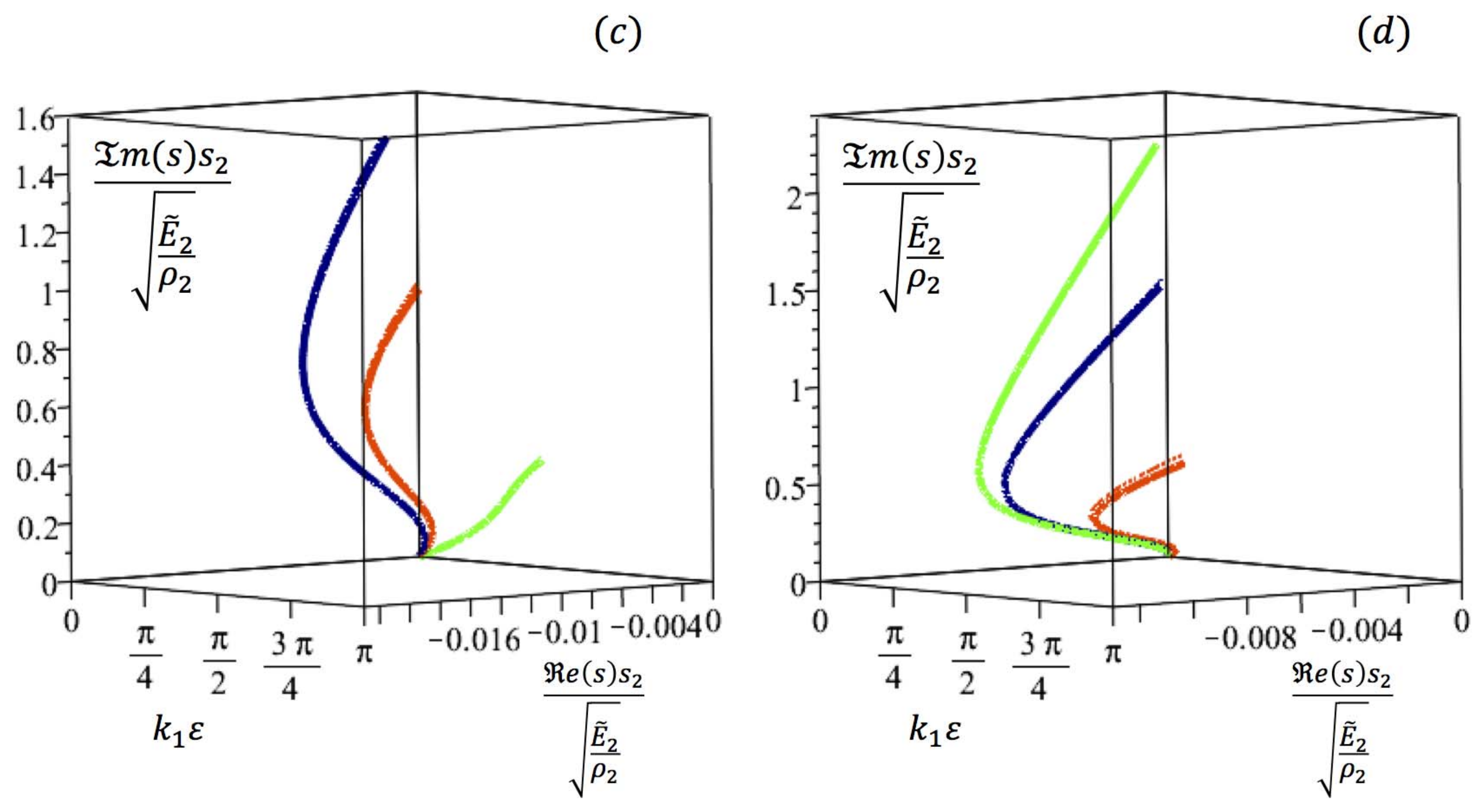}\label{11}}
		\caption{dispersion curves along the direction $\boldsymbol{e}_{1}$ derived with two approaches (a) by varying the dimensionless Young's modulus:  $r_{E}=20$ (orange), $r_{E}=15$ (blue) and $r_{E}=10$ (green), (b) by modifying the dimensionless relaxation time: $\tau_{\varsigma}=1$ (orange), $\tau_{\varsigma}=2$ (blue) and $\tau_{\varsigma}=\infty$ (green), (c) with different values of the dimensionless thickness ratio: $\eta=2$ (orange), $\eta=1$ (blue) and $\eta=6$ (green), (d) with different values of the dimensionless density: $r_{\rho}=50$ (orange), $r_{\rho}=10$ (blue) and $r_{\rho}=5$ (green).}
	\end{figure}
	$\beta=1,2$.\\
Fig. $10$-(a) shows the curves obtained for three values of the Young's modulus. The orange one corresponds to $r_{E}=20$, the blue one to $r_{E}=15$ and the green one to $r_{E}=10$. The values for $\tau_{\varsigma}=2$, $\eta=1$ and $r_{\rho}=10$ are supposed to be fixed. The curves are monotonically increasing and as $r_{E}$ goes down they gradually decline.\\
Fig. $10$-(b) shows curves obtained for three values of the relaxation time $\tau_{\varsigma}$. 
For low values of $\tau_{\varsigma}$, namely $\tau_{\varsigma}=1$, which is represented by the orange curve, and $\tau_{\varsigma}=2$, identified by the blue curve, the viscoelasticity strongly affects the dispersion curves, which are enterely embedded in the real and the imaginary plane. By increasing the relaxation time $\tau_{\varsigma}=\infty$, the elastic behaviour is retrieved (green curve), infact the curve is squeezed in the imaginary axes of the complex frequency.\\
Fig. $10$-(c) highlights the trends of three curves developped by setting three different values of the dimensionless ratio between the thicknesses of the material. The blue curve is obtained for $\eta=1$, the orange one for $\eta=2$ and the green one for $\eta=6$, with fixed values of the dimensionless parameters $r_{E}=10$, $\tau_{\varsigma}=2$ and $r_{\rho}=10$. A rise in the ratio implies that the corresponding curves become more flatten.\\ Finally, Fig. $10$-(d) represents how for three increasing values of the dimensionless density the related curves are influenced. The  green curve is given with $r_{\rho}=5$, the blue one with $r_{\rho}=10$ and the orange one corresponds to $r_{\rho}=50$, by fixing $r_{E}=10$, $\tau_{\varsigma}=2$ and $\eta=1$.\\
In Fig. $11$ dispersion curves that travel along the axis $\boldsymbol{e}_{2}$ are taken into account. In Fig. $11$ the curves achieved from approach $1$ and approach $2$ are compared.\\
	 	\begin{figure}
		\centering
		\subfigure{\includegraphics[height=0.35\textheight]{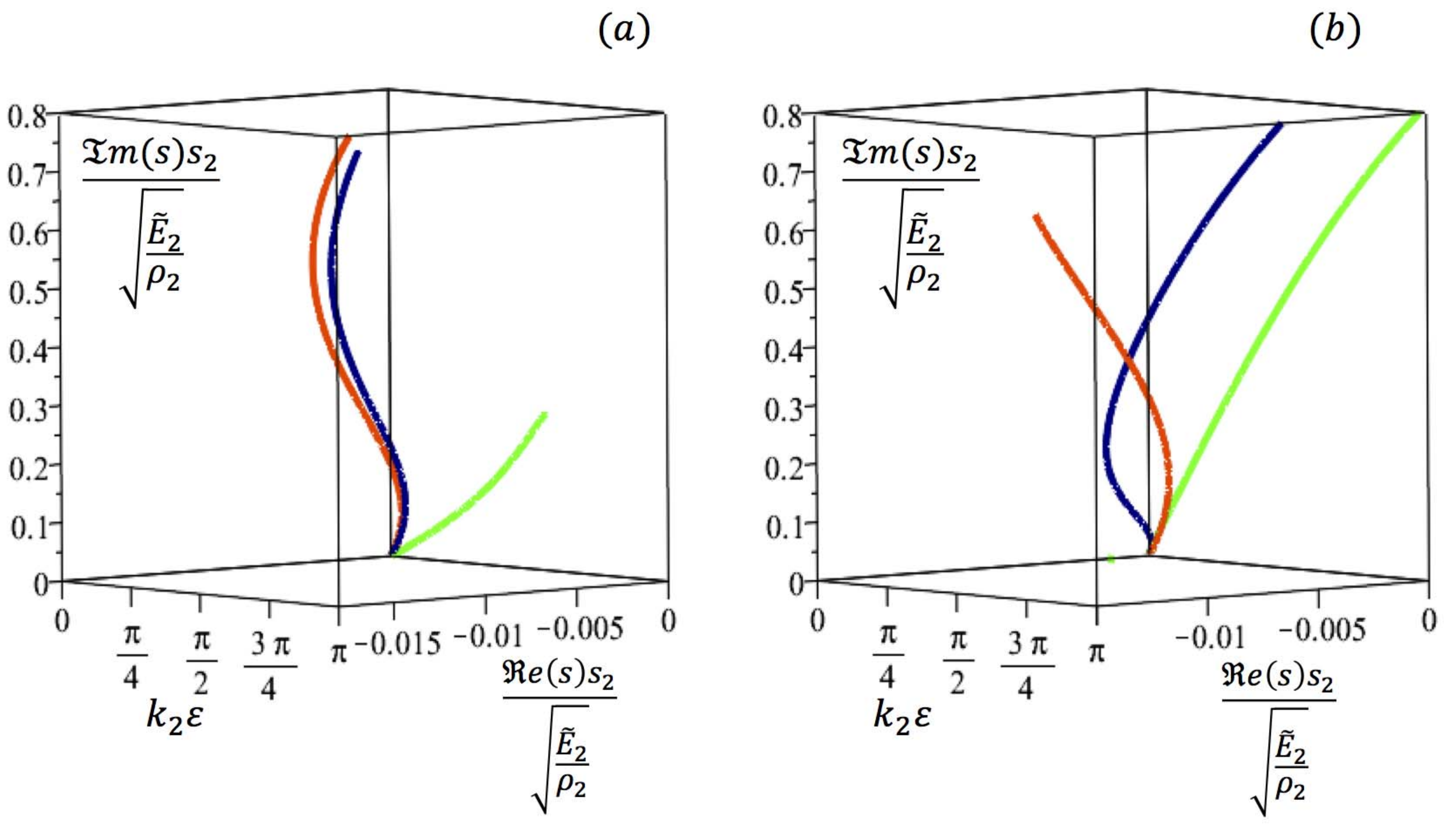}\label{12}}
		\subfigure{\includegraphics[height=0.36\textheight]{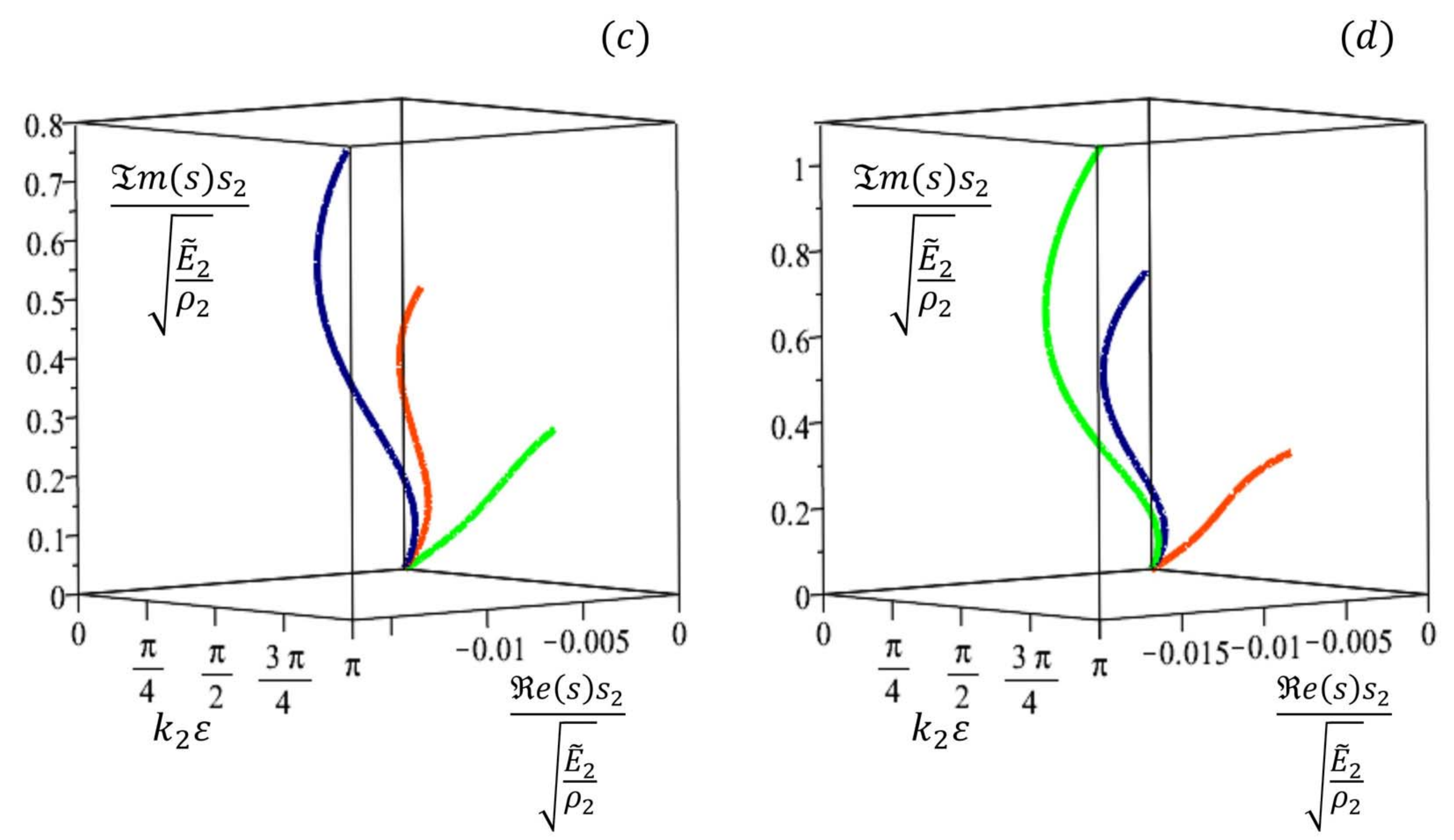}\label{13}}
		\caption{dispersion curves along  $\boldsymbol{e}_{2}$ derived with two approaches, (a) by varying the Young's modulus: $r_{E}=100$ (orange), $r_{E}=3$ (blue) and $r_{E}=0.1$ (green), (b) by modifying the nondimensional relaxation time: $\tau_{\varsigma}=1$ (orange), $\tau_{\varsigma}=7$ (blue) and $\tau_{\varsigma}=\infty$ (green), (c) with three values of the thickness ratio: $\eta=2$ (orange), $\eta=1$ (blue) and $\eta=6$ (green), (d) with different values of the dimensionless density: $r_{\rho}=50$ (orange), $r_{\rho}=10$ (blue) and $r_{\rho}=5$ (green).}
	\end{figure}
In Fig. $11$-(a), the values of the dimensionless parameters are set as $\tau_{\varsigma}=2$, $\eta=1$ and $r_{\rho}=10$. The orange curve obtained for $r_{E}=100$ and the blue one for $r_{E}=3$ have a similar trend. By considering a lower dimensionless Young's modulus $r_{E}=0.1$, the green curve is recovered, which is strongly overwhelmed with respect to the previous ones. Fig. $11$-(b) shows that by increasing the relaxation time $\tau_{\varsigma}$ up to high values ($\tau_{\varsigma}=\infty$) the green curve is achieved and so the viscoelastic effect is negligible. As the relaxation time assumes low values ($\tau_{\varsigma}=1$ and $\tau_{\varsigma}=7$) the corresponding curves (the orange and the blue one) are strongly influenced by the viscoelastic response. The dimensionless parameters are set as $r_{E}=10$, $\tau_{\varsigma}=2$ and $r_{\rho}=10$.\\
In Fig. $11$-(c) by setting three values for the dimensionless ratio between the thicknesses of the material, the following curves are accomplished: the blue curve is stemmed for $\eta=1$, the orange one for $\eta=2$ and the green one for $\eta=6$, with fixed values of the non-dimensional parameters $r_{E}=10$, $\tau_{\varsigma}=2$ and $r_{\rho}=10$. As the ratio increases the curves become more and more flatten. Fig. $11$-(d) shows the curves obtained for three values of the dimensionless density. The green one relates to $r_{\rho}=5$, the blue curve corresponds with the value $r_{\rho}=10$ and the orange one $r_{\rho}=50$, with the constant parameters $r_{E}=10$, $\tau_{\varsigma}=2$ and $\eta=1$. The diagram emphasises that the higher the density is, the more sqeezed the curve is.

\newpage
\section{Conclusions}
The paper proposed a variational-asymptotic homogenization model for viscoelastic materials having a periodic microstructure. Specifically, the field equations at the micro-scale have been derived and transposed into the Laplace domain to treat viscoelasticity, relying on the complex frequency and the micro-relaxation tensor. Next, the down-scaling relation and the up-scaling relation have been detailed. In particular, the down-scaling relation relates the transformed micro-displacement field to the transformed macro-displacement field and its gradients by means of the perturbation functions, which are the solutions of the cell problems defined over the unit cell $\mathcal{Q}$. Perturbation functions are $\mathcal{Q}$-periodic and have vanishing mean values over the unit cell. On the other hand, the up-scaling relation defines the transformed macro-displacement field as the mean value of the transformed micro-displacement field over the unit cell $\mathcal{Q}$.\\
By introducing the down-scaling relation into the field equations at the micro-scale of the viscoelastic material, the average field equations of infinite order have been determined. Their truncation at an arbitrary order could not ensure the ellipticity of the differential problem.\\
To avoid such an issue, two methods, based on a variational-asymptotic approach, have been invoked. According to the first method, the down-scaling relation is replaced into the transformed energy-like functional, which is truncated at the second order. By imposing the first variation of the truncated energy-like functional equal to zero, the field equation at the macro-scale and the overall inertial and constitutive tensors have been determined. In the second method, the gradient related to the down-scaling relation has been approximated at the first order and the transformed micro-displacement has been truncated at second order. Both are introduced into the transformed energy-like functional and, from its first variation, the field equation at the macro-scale and the corresponding overall inertial and constitutive tensors have been derived. In both methods, the overall constitutive tensors depend on the localization functions, whereas the overall inertial tensors are expressed through the perturbation functions. In the limit case of an homogeneous material without heterogeneities, the perturbation functions consistently become identically equal to zero, the components of the localization tensors vanish and the equation of motion of the classical continuum has been retrieved.\\
The proposed model has been applied to a bi-phase layered material with isotropic phases, subject to $\mathcal{L}$-periodic body forces. In such a case, some of the perturbation functions and some of the components of the overall inertial tensors have been computed and displayed to remark their dependence on the real part and on the imaginary part of the complex frequency.\\
Moreover, the analytical solutions for the transformed macro-displacement field derived from the two homogenized models and subjected to harmonic volume forces have been compared with the reference numerical solution obtained from a finite element analysis of the heterogeneous model, for three different values of the complex frequency. A good agreement between the three models has been achieved, proving the validity of the proposed homogenized methods.\\
Finally, a detailed parametric study of the dispersion function, derived from the generalized Christoffel equation, allowed determining the behaviour of the dispersion curves through the viscoelastic medium along two orthogonal axis, for both homogenized methods. An optimal matching between the dispersion curves derived from the two methods has been pointed out. The variational-asymptotic homogenization method herein described is helpful to detect the effective viscoelastic properties of many composite materials and it may be adopted for the manufacture and the design of more efficient and sophisticated devices, for a large spectra of applications. 

{\small
	\bibliographystyle{abbrvnat}
	\bibliography{Bibliography}
}

\newpage
\section*{ \small Appendix A. Higher order recursive differential problems}
The recursive differential problems are established at order $\varepsilon$, $\varepsilon^{2}$,  $\varepsilon^{2\tilde{w}-1}$ and $\varepsilon^{2\tilde{w}}$, with $\tilde{w} \in \mathbb{Z}$ and $\tilde{w} \geq 2$ and they are helpful to formulate the cell problems in section $4$. \\
Taking account of the solutions $~\eqref{eqn:solp1}$ and $~\eqref{eqn:sol0}$ related to the differential problems at order $\varepsilon^{-1}$ and $\varepsilon^{0}$, respectively, the differential problem at order $\varepsilon$, stemmed from equation  $~\eqref{eqn:ce}$, is
\begin{equation}
\label{eqn:sole}
\Big (\hat{C}^{m}_{ijhk}\hat{u}^{(3)}_{h,k}\Big)_{,j}+\Big(\Big(\hat{C}^{m}_{ikhj}N^{(2,0)}_{hpq_{1}q_{2}}\Big)_{,k}+\hat{C}^{m}_{ikhj}N^{(2,0)}_{hpq_{1}q_{2},j}+\Big (\hat{C}^{m}_{ijhq_{2}}N^{(1,0)}_{hpq_{1}}\Big)\Big)\frac{\partial^{3}\hat{U}^{M}_{p}}{\partial x_{q_{1}}\partial x_{q_{2}}\partial x_{k}}+
\end{equation} 
\[
+\Big[ \Big ( \hat{C}^{m}_{ijhq_{1}}N^{(2,2)}\Big)_{,j}+\hat{C}^{m}_{iq_{1}hj}N^{(2,2)}_{hp,j} -\rho^{m}N^{(1,0)}_{ipq_{1}}       \Big ]s^{2}\frac{\partial \hat{U}^{M}_{p}}{\partial x_{q_{1}}} = f^{(3)}_{i}(\boldsymbol{x}),
\]
with interface conditions
\begin{equation}
\Big[\Big[\hat{u}^{(3)}_{h}\Big]\Big]\Big\vert_{\boldsymbol{\xi} \in \Sigma_{1}}=0, \quad \Big [\Big[ \Big ( \hat{C}^{m}_{ijhk}\Big (\hat{u}^{(3)}_{h,k}+N^{(2,0)}_{hpq_{1}q_{2}}\frac{\partial^{3}\hat{U}^{M}_{p}}{\partial x_{q1}\partial x_{q2}\partial x_{k}}+N^{(2,2)}s^{2}\frac{\partial \hat{U}^{M}_{p}}{\partial x_{q_{1}}} \Big) \Big )n_{j}\Big ]\Big]\Big \vert_{\boldsymbol{\xi} \in \Sigma_{1}}=0.
\end{equation}

The solvability condition of differential problem $~\eqref{eqn:sole}$ in the class of $\mathcal{Q}-$periodic functions and the divergence theorem yield to 

\begin{equation}
f^{(3)}_{i}(\boldsymbol{x}) =\Big \langle \hat{C}^{m}_{ikhj}N^{(2,0)}_{hpq_{1}q_{2},j}+\Big (\hat{C}^{m}_{ikhq_{2}}N^{(1,0)}_{hpq_{1}}\Big)\Big \rangle \frac{\partial^{3}\hat{U}^{M}_{p}}{\partial x_{q_{1}}\partial x_{q_{2}}\partial x_{k}}+\Big \langle \hat{C}^{m}_{iq_{1}hj}N^{(2,2)}_{hp,j} -\rho^{m}N^{(1,0)}_{ipq_{1}}       \Big \rangle s^{2}\frac{\partial \hat{U}^{M}_{p}}{\partial x_{q_{1}}},
\end{equation}    
and consequentely the solution of the differential problem at order $\varepsilon$ is
\begin{equation}
\label{eqn:eqsol}
\hat{u}^{(3)}_{h}(\boldsymbol{x},\boldsymbol{\xi},s) = N^{(3,0)}_{hpq_{1}q_{2}q_{3}}\frac{\partial^{3}\hat{U}^{M}_{p}}{\partial x_{q_{1}}\partial x_{q_{2}}\partial x_{q_{3}}}+N^{(3,2)}_{hpq_{1}} s^{2}\frac{\partial \hat{U}^{M}_{p}}{\partial x_{q_{1}}}.
\end{equation}  

Again, considering the solutions $~\eqref{eqn:sol0}$ and $~\eqref{eqn:eqsol}$  related to the differential problems at order $\varepsilon^{0}$ and $\varepsilon$, respectively, the differential problem at order $\varepsilon^{2}$, derived from equation  $~\eqref{eqn:ce}$, is
\begin{equation}
\label{eqn:nee2}
\Big (\hat{C}^{m}_{ijhk}\hat{u}^{(4)}_{h,k}\Big)_{,j}+\Big(\Big(\hat{C}^{m}_{ikhj}N^{(3,0)}_{hpq_{1}q_{2}q_{3}}\Big)_{,k}+\hat{C}^{m}_{iq_{3}hk}N^{(2,0)}_{hpq_{1}q_{2}}+\Big (\hat{C}^{m}_{ikhj}N^{(3,0)}_{hpq_{1}q_{2}q_{3},j}\Big)\Big)\frac{\partial^{4}\hat{U}^{M}_{p}}{\partial x_{q_{1}}\partial x_{q_{2}}\partial x_{q_{3}}\partial x_{k}}+
\end{equation} 
\[
+\Big[ \Big ( \hat{C}^{m}_{ijhq_{2}}N^{(3,2)}_{hpq_{1}}\Big)_{,j}+ \hat{C}^{m}_{iq_{2}hq_{1}}N^{(2,2)}_{hp}+\hat{C}^{m}_{iq_{2}hk}N^{(3,2)}_{hpq_{1},k} -\rho^{m}N^{(2,0)}_{ipq_{1}q_{2}}\Big ]s^{2}\frac{\partial \hat{U}^{M}_{p}}{\partial x_{q_{1}}} -\rho^{m}N^{(2,2)}_{ip}s^{4}\hat{U}^{M}_{p} = f^{(4)}_{i}(\boldsymbol{x}),
\]
with interface conditions
\begin{equation}
\Big[\Big[\hat{u}^{(4)}_{h}\Big]\Big]\Big\vert_{\boldsymbol{\xi} \in \Sigma_{1}}=0, 
\end{equation}
\[
\Big [\Big[ \Big ( \hat{C}^{m}_{ijhk}\Big (\hat{u}^{(4)}_{h,k}+N^{(3,0)}_{hpq_{1}q_{2}q_{3}}\frac{\partial^{4}\hat{U}^{M}_{p}}{\partial x_{q1}\partial x_{q2}\partial x_{q3}\partial x_{k}}+N^{(2,2)}s^{2}\frac{\partial \hat{U}^{M}_{p}}{\partial x_{q_{1}}}+N^{(3,2)}_{hpq_{1}}s^{2}\frac{\partial^{2} \hat{U}^{M}_{p}}{\partial x_{q_{1}}\partial x_{k}} \Big) \Big )n_{j}\Big ]\Big]\Big \vert_{\boldsymbol{\xi} \in \Sigma_{1}}=0.
\].

The solvability condition of differential problem $~\eqref{eqn:nee2}$ in the class of $\mathcal{Q}-$periodic functions and the divergence theorem lead to 
\begin{flalign*}
&f^{(4)}_{i}(\boldsymbol{x}) =\Big \langle \Big (\hat{C}^{m}_{iq_{3}hk}N^{(2,0)}_{hpq_{1}q_{2}}+ \hat{C}^{m}_{ikhj}N^{(3,0)}_{hpq_{1}q_{2}q_{3},j}\Big)\Big \rangle \frac{\partial^{4}\hat{U}^{M}_{p}}{\partial x_{q_{1}}\partial x_{q_{2}}\partial x_{q_{3}}\partial x_{k}}+&
\end{flalign*} 
\begin{flalign}
&+\Big \langle \hat{C}^{m}_{iq_{2}hq_{1}}N^{(2,2)}_{hp}+\hat{C}^{m}_{iq_{2}hk}N^{(3,2)}_{hpq_{1},k} -\rho^{m}N^{(2,0)}_{ipq_{1}q_{2}}\Big \rangle s^{2}\frac{\partial \hat{U}^{M}_{p}}{\partial x_{q_{1}}} -\langle \rho^{m}N^{(2,2)}_{ip}\rangle s^{4}\hat{U}^{M}_{p},&
\end{flalign} 
and consequentely the solution of the differential problem at order $\varepsilon^{2}$ is
\begin{equation}
\label{eqn:soleq}
\hat{u}^{(4)}_{h}(\boldsymbol{x},\boldsymbol{\xi},s) = N^{(4,0)}_{hpq_{1}q_{2}q_{3}q_{4}}\frac{\partial^{4}\hat{U}^{M}_{p}}{\partial x_{q_{1}}\partial x_{q_{2}}\partial x_{q_{3}}\partial x_{q_{4}}}+N^{(4,2)}_{hpq_{1}q_{2}} s^{2}\frac{\partial^{2} \hat{U}^{M}_{p}}{\partial x_{q_{1}}\partial x_{q_{2}}}+N^{(4,4)}_{hp}s^{4}\hat{U}^{M}_{p}.
\end{equation}  

The generic recursive differential problem of odd order $\varepsilon^{2\tilde{w}-1}$, with  $\tilde{w}\in \mathbb{Z}$ and $\tilde{w} \geq 2$,  is
\begin{flalign}
&(\hat{C}^{m}_{ijhk}u^{(2\tilde{w}+1)}_{h,k})_{j}+\frac{1}{2\tilde{w}+1}\sum_{P^{*}(q)}^{}\Big [\Big ( \hat{C}^{m}_{ijhq_{2\tilde{w}+1}}N^{(2\tilde{w},0)}_{hpq_{1}...q_{2\tilde{w}}}\Big )_{,j}+\hat{C}^{m}_{iq_{2\tilde{w}+1}hj}N^{(2\tilde{w},0)}_{hpq_{1}...q_{2\tilde{w}},j}+&
\end{flalign}
\begin{flalign*}
&+\hat{C}^{m}_{iq_{2\tilde{w}+1}hq_{2\tilde{w}}}N^{(2\tilde{w}-1,0)}_{hpq_{1}...q_{2\tilde{w}-1}}\Big ] \frac{\partial ^{2\tilde{w}+1}\hat{U}^{M}_{p}}{\partial x_{q_{1}}... \partial x_{q_{2\tilde{w}+1}}}+\frac{1}{2\tilde{w}-1}\sum_{P^{*}(q)}\Big [ \Big ( \hat{C}^{m}_{ijhq_{2\tilde{w}-1}}N^{(2\tilde{w},2)}_{hpq_{1}...q_{2\tilde{w}-2}}\Big )_{,j}+&
\end{flalign*}
\begin{flalign*}
&+\hat{C}^{m}_{iq_{2\tilde{w}-1}hq_{2\tilde{w}-2}}N^{(2\tilde{w}-1,2)}_{hpq_{1}...q_{2\tilde{w}-3}}+\hat{C}^{m}_{iq_{2\tilde{w}-1}hj}N^{(2\tilde{w},2)}_{hpq_{1}...q_{2\tilde{w}-2},j}-\rho^{m}N^{(2\tilde{w}-1,0)}_{ipq_{1}...q_{2\tilde{w}-1}}\Big ]  \frac{\partial ^{2\tilde{w}-1}\hat{U}^{M}_{p}}{\partial x_{q_{1}}... \partial x_{q_{2\tilde{w}-1}}}s^{2}+&
\end{flalign*}
\begin{flalign*}
&+\sum_{n=1}^{n=\tilde{w}-1}(1-\delta_{1n})\frac{1}{2\tilde{w}-2n+1}\sum_{P^{*}(q)}\Big [ \Big ( \hat{C}^{m}_{ijhq_{2\tilde{w}+1-2n}}N^{(2\tilde{w},2n)}_{hpq_{1}...q_{2\tilde{w}-2n}}\Big )_{,j}+\hat{C}^{m}_{iq_{2\tilde{w}+1-2n}hq_{2\tilde{w}-2n}}N^{(2\tilde{w}-1,2n)}_{hpq_{1}...q_{2\tilde{w}-1-2n}}+&
\end{flalign*}
\begin{flalign*}
&+\hat{C}^{m}_{iq_{2\tilde{w}+1-2n}hj}N^{(2\tilde{w},2n)}_{hpq_{1}...q_{2\tilde{w}-2n},j}-\rho^{m}N^{(2\tilde{w}-1,2n-2)}_{ipq_{1}...q_{2\tilde{w}+1-2n}}\Big ]  \frac{\partial ^{2\tilde{w}+1-2n}\hat{U}^{M}_{p}}{\partial x_{q_{1}}... \partial x_{q_{2\tilde{w}+1-2n}}}s^{2n}+&
\end{flalign*}
\begin{flalign*}
&+\Big [ \Big ( \hat{C}^{m}_{ijhq_{1}}N^{(2\tilde{w},2n)}_{hp}\Big)_{,j}+\hat{C}^{m}_{iq_{1}hk}N^{(2\tilde{w},2\tilde{w})}_{hp,k}-\rho^{m}N^{(2\tilde{w}-1,2\tilde{w}-2)}_{ipq_{1}}\Big ]\frac{\partial \hat{U}^{m}_{p}}{\partial x_{q_{1}}}s^{2\tilde{w}}=&
\end{flalign*}
\begin{flalign*}
&=\frac{1}{2\tilde{w}+1}\sum_{P^{*}(q)}\langle \hat{C}^{m}_{iq_{2\tilde{w}+1}hj}N^{(2\tilde{w})}_{hpq_{1}...q_{2\tilde{w}},j}+
\hat{C}^{m}_{iq_{2\tilde{w}+1}hq_{2\tilde{w}}}N^{(2\tilde{w}-1)}_{hpq_{1}...q_{2\tilde{w}-1}}\rangle \frac{\partial ^{2\tilde{w}+1}\hat{U}^{M}_{p}}{\partial x_{q_{1}}... \partial x_{q_{2\tilde{w}+1}}}+&
\end{flalign*}
\begin{flalign*}
&+\frac{1}{2\tilde{w}-2}\sum_{P^{*}(q)}\langle \hat{C}^{m}_{iq_{2\tilde{w}-1}hq_{2\tilde{w}-2}}N^{(2\tilde{w}-1,2)}_{hpq_{1}...q_{2\tilde{w}-3}}+\hat{C}^{m}_{iq_{2\tilde{w}-1}hj}N^{(2\tilde{w},2)}_{hpq_{1}...q_{2\tilde{w}-2},j}-\rho^{m}N^{(2\tilde{w}-1)}_{ipq_{1}...q_{2\tilde{w}-1}}\rangle  \frac{\partial ^{2\tilde{w}-1}\hat{U}^{M}_{p}}{\partial x_{q_{1}}... \partial x_{q_{2\tilde{w}-1}}}s^{2}+&
\end{flalign*}
\begin{flalign*}
&+\sum_{n=1}^{n=\tilde{w}-1}(1-\delta_{1n})\frac{1}{2\tilde{w}-2n+1}\sum_{P^{*}(q)}\langle \hat{C}^{m}_{iq_{2\tilde{w}+1-2n}hq_{2\tilde{w}-2n}}N^{(2\tilde{w}-1,2n)}_{hpq_{1}...q_{2\tilde{w}-1-2n}}+&
\end{flalign*}
\begin{flalign*}
&+\hat{C}^{m}_{iq_{2\tilde{w}+1-2n}hj}N^{(2\tilde{w},2n)}_{hpq_{1}...q_{2\tilde{w}-2n},j}-\rho^{m}N^{(2\tilde{w}-1,2n-2)}_{ipq_{1}...q_{2\tilde{w}+1-2n}} \rangle  \frac{\partial ^{2\tilde{w}+1-2n}\hat{U}^{M}_{p}}{\partial x_{q_{1}}... \partial x_{q_{2\tilde{w}+1-2n}}}s^{2n}+&
\end{flalign*}
\begin{flalign*}
&+\langle \hat{C}^{m}_{iq_{1}hk}N^{(2\tilde{w},2\tilde{w})}_{hp,k}-\rho^{m}N^{(2\tilde{w}-1,2\tilde{w}-2)}_{ipq_{1}}\rangle \frac{\partial \hat{U}^{m}_{p}}{\partial x_{q_{1}}}s^{2\tilde{w}},&
\end{flalign*}
and their interface conditions are
\begin{equation}
\Big[\Big[\hat{u}^{(2\tilde{w}+1)}_{h}\Big]\Big]\Big\vert_{\boldsymbol{\xi} \in \Sigma_{1}}=0
\end{equation}
and
\begin{flalign*}
&\Big [\Big[ \Big (\hat{C}^{m}_{ijhk}\hat{u}^{(2\tilde{w}+1)}_{h,k}+\frac{1}{2\tilde{w}+1}\sum_{P^{*}(q)}\Big (\hat{C}^{m}_{ijhq_{2\tilde{w}+1}}N^{(2\tilde{w},0)}_{hpq_{1}...q_{2\tilde{w}}}\Big ) \frac{\partial^{2\tilde{w}+1}\hat{U}^{M}_{p}}{\partial x_{q_{1}}...\partial x_{q_{2\tilde{w}+1}}}+&
\end{flalign*}
\begin{flalign*}
&+\frac{1}{2\tilde{w}-1}\sum_{P^{*}(q)}\Big (\hat{C}^{m}_{ijhq_{2\tilde{w}-1}}N^{(2\tilde{w},2)}_{hpq_{1}...q_{2\tilde{w}-2}}\Big ) \frac{\partial ^{2\tilde{w}-1}\hat{U}^{M}_{p}}{\partial x_{q_{1}}...\partial x_{q_{2\tilde{w}-1}}}s^{2}+&
\end{flalign*}
\begin{flalign*}
&+\sum_{n=1}^{n=\tilde{w}-1}(1-\delta_{1n})\frac{1}{2\tilde{w}-2n+1}\sum_{P^{*}(q)} \Big ( \hat{C}^{m}_{ijhq_{2\tilde{w}+1-2n}}N^{(2\tilde{w},2n)}_{hpq_{1}...q_{2\tilde{w}-2n}}\Big )\frac{\partial ^{2\tilde{w}+1-2n}\hat{U}^{M}_{p}}{\partial x_{q_{1}}... \partial x_{q_{2\tilde{w}+1-2n}}}s^{2n}+&
\end{flalign*}
\begin{flalign*}
&+\hat{C}^{m}_{ijhq_{1}}N^{(2\tilde{w},2\tilde{w})}_{hp}\frac{\partial \hat{U}^{M}_{p}}{\partial x_{q_{1}} }s^{2\tilde{w}}\Big ) n_{j}\Big ]\Big ]\Big \vert_{\boldsymbol{\xi} \in \Sigma_{1}}=0.&
\end{flalign*}

The generic recursive differential problem of even order $\varepsilon^{2\tilde{w}}$ is
\begin{flalign}
&(\hat{C}^{m}_{ijhk}u^{(2\tilde{w}+2)}_{h,k})_{j}+\frac{1}{2\tilde{w}+2}\sum_{P^{*}(q)}\Big [\Big ( \hat{C}^{m}_{ijhq_{2\tilde{w}+2}}N^{(2\tilde{w}+1)}_{hpq_{1}...q_{2\tilde{w}+1}}\Big )_{,j}+\hat{C}^{m}_{iq_{2\tilde{w}+2}hj}N^{(2\tilde{w}+1)}_{hpq_{1}...q_{2\tilde{w}+1},j}+&
\end{flalign}
\begin{flalign*}
&+\hat{C}^{m}_{iq_{2\tilde{w}+2}hq_{2\tilde{w}+1}}N^{(2\tilde{w})}_{hpq_{1}...q_{2\tilde{w}}}\Big ] \frac{\partial ^{2\tilde{w}+2}\hat{U}^{M}_{p}}{\partial x_{q_{1}}... \partial x_{q_{2\tilde{w}+2}}}+\frac{1}{2\tilde{w}}\sum_{P^{*}(q)}\Big [ \Big ( \hat{C}^{m}_{ijhq_{2\tilde{w}}}N^{(2\tilde{w}+1,2)}_{hpq_{1}...q_{2\tilde{w}-1}}\Big )_{,j}+&
\end{flalign*}
\begin{flalign*}
&+\hat{C}^{m}_{iq_{2\tilde{w}}hq_{2\tilde{w}-1}}N^{(2\tilde{w},2)}_{hpq_{1}...q_{2\tilde{w}-2}}+\hat{C}^{m}_{iq_{2\tilde{w}}hj}N^{(2\tilde{w}+1,2)}_{hpq_{1}...q_{2\tilde{w}-1},j}-\rho^{m}N^{(2\tilde{w})}_{ipq_{1}...q_{2\tilde{w}}}\Big ]  \frac{\partial ^{2\tilde{w}}\hat{U}^{M}_{p}}{\partial x_{q_{1}}... \partial x_{q_{2\tilde{w}}}}s^{2}+&
\end{flalign*}
\begin{flalign*}
&+\sum_{n=1}^{n=\tilde{w}-1}(1-\delta_{1n})\frac{1}{2^{2\tilde{w}-2n+2}}\sum_{P^{*}(q)}\Big [ \Big ( \hat{C}^{m}_{ijhq_{2\tilde{w}-2n+2}}N^{(2\tilde{w}+1,2n)}_{hpq_{1}...q_{2\tilde{w}-2n+1}}\Big )_{,j}+\hat{C}^{m}_{iq_{2\tilde{w}-2n+2}hq_{2\tilde{w}+1-2n}}N^{(2\tilde{w},2n)}_{hpq_{1}...q_{2\tilde{w}-2n}}+&
\end{flalign*}
\begin{flalign*}
&+\hat{C}^{m}_{iq_{2\tilde{w}+2-2n}hj}N^{(2\tilde{w}+1,2n)}_{hpq_{1}...q_{2\tilde{w}+1-2n},j}-\rho^{m}N^{(2\tilde{w},2n-2)}_{ipq_{1}...q_{2\tilde{w}+2-2n}}\Big ]  \frac{\partial ^{2\tilde{w}+2-2n}\hat{U}^{M}_{p}}{\partial x_{q_{1}}... \partial x_{q_{2\tilde{w}+2-2n}}}s^{2n}+&
\end{flalign*}
\begin{flalign*}
&+\frac{1}{2}\Big [ \Big ( \hat{C}^{m}_{ijhq_{2}}N^{(2\tilde{w}+1,2\hat{w})}_{hpq_{1}}\Big)_{,j}+\hat{C}^{m}_{iq_{2}hq_{1}}N^{(2\tilde{w},2\tilde{w})}_{hp}+\hat{C}^{m}_{iq_{2}hj}N^{(2\tilde{w}+1,2\tilde{w})}_{hpq_{1},j}-\rho^{m}N^{(2\tilde{w},2\tilde{w}-2)}_{ipq_{1}q_{2}}+&
\end{flalign*}
\begin{flalign*}
&+\hat{C}^{m}_{ijhq_{1}}N^{(2\tilde{w}+1,2\hat{w})}_{hpq_{2}}\Big)_{,j}+\hat{C}^{m}_{iq_{1}hq_{2}}N^{(2\tilde{w},2\tilde{w})}_{hp}+\hat{C}^{m}_{iq_{1}hj}N^{(2\tilde{w}+1,2\tilde{w})}_{hpq_{2},j}-\rho^{m}N^{(2\tilde{w},2\tilde{w}-2)}_{ipq_{2}q_{1}}\Big ] \frac{\partial^{2} \hat{U}^{m}_{p}}{\partial x_{q_{1}}\partial x_{q_{2}}}s^{2\tilde{w}}-\rho^{m}N^{(2\tilde{w},2\tilde{w})}_{ip}\hat{U}^{M}_{p}s^{2\tilde{w}}=&
\end{flalign*}
\begin{flalign*}
&=\frac{1}{2\tilde{w}+2}\sum_{P^{*}(q)}\langle \hat{C}^{m}_{iq_{2\tilde{m}+2}hj}N^{(2\tilde{w}+1)}_{hpq_{1}...q_{2\tilde{w}+1},j}+
\hat{C}^{m}_{iq_{2\tilde{w}+2}hq_{2\tilde{w}+1}}N^{(2\tilde{w})}_{hpq_{1}...q_{2\tilde{w}}}\rangle \frac{\partial ^{2\tilde{w}+2}\hat{U}^{M}_{p}}{\partial x_{q_{1}}... \partial x_{q_{2\tilde{w}+2}}}+&
\end{flalign*}
\begin{flalign*}
&+\frac{1}{2\tilde{w}}\sum_{P(q)}^{}\langle \hat{C}^{m}_{iq_{2\tilde{w}}hq_{2\tilde{w}-1}}N^{(2\tilde{w},2)}_{hpq_{1}...q_{2\tilde{w}-2}}+\hat{C}^{m}_{iq_{2\tilde{w}}hj}N^{(2\tilde{w}+1,2)}_{hpq_{1}...q_{2\tilde{w}-1},j}-\rho^{m}N^{(2\tilde{w})}_{ipq_{1}...q_{2\tilde{w}}}\rangle  \frac{\partial ^{2\tilde{w}}\hat{U}^{M}_{p}}{\partial x_{q_{1}}... \partial x_{q_{2\tilde{w}}}}s^{2}+&
\end{flalign*}
\begin{flalign*}
&+\sum_{n=1}^{n=\tilde{w}-1}(1-\delta_{1n})\frac{1}{2\tilde{w}-2n+2}\sum_{P^{*}(q)}^{}\langle \hat{C}^{m}_{iq_{2\tilde{w}-2n+2}hq_{2\tilde{w}+1-2n}}N^{(2\tilde{w},2n)}_{hpq_{1}...q_{2\tilde{w}-2n}}+&
\end{flalign*}
\begin{flalign*}
&+\hat{C}^{m}_{iq_{2\tilde{w}+2-2n}hj}N^{(2\tilde{w}+1,2n)}_{hpq_{1}...q_{2\tilde{w}+1-2n},j}-\rho^{m}N^{(2\tilde{w},2n-2)}_{ipq_{1}...q_{2\tilde{w}+2-2n}} \rangle  \frac{\partial ^{2\tilde{w}+2-2n}\hat{U}^{M}_{p}}{\partial x_{q_{1}}... \partial x_{q_{2\tilde{w}+2-2n}}}s^{2n}+&
\end{flalign*}
\begin{flalign*}
&+\frac{1}{2}\Big [ \Big ( \hat{C}^{m}_{iq_{2}hq_{1}}N^{(2\tilde{w},2\tilde{w})}_{hp}+\hat{C}^{m}_{iq_{2}hj}N^{(2\tilde{w}+1,2\tilde{w})}_{hpq_{1},j}-\rho^{m}N^{(2\tilde{w},2\tilde{w}-2)}_{ipq_{1}q_{2}}+&
\end{flalign*}
\begin{flalign*}
&+\hat{C}^{m}_{iq_{1}hq_{2}}N^{(2\tilde{w},2\tilde{w})}_{hp}+\hat{C}^{m}_{iq_{1}hj}N^{(2\tilde{w}+1,2\tilde{w})}_{hpq_{2},j}-\rho^{m}N^{(2\tilde{w},2\tilde{w}-2)}_{ipq_{2}q_{1}}\Big ] \frac{\partial^{2} \hat{U}^{m}_{p}}{\partial x_{q_{1}}\partial x_{q_{2}}}s^{2\tilde{w}}-\langle \rho^{m}N^{(2\tilde{w},2\tilde{w})}_{ip} \rangle \hat{U}^{M}_{p}s^{2\tilde{w}},&
\end{flalign*}
and their interface conditions are
\begin{equation}
\Big[\Big[\hat{u}^{(2\tilde{w}+2)}_{h}\Big]\Big]\Big\vert_{\boldsymbol{\xi} \in \Sigma_{1}}=0
\end{equation}
and
\begin{flalign}
&\Big [\Big[ \Big (\hat{C}^{m}_{ijhk}\hat{u}^{(2\tilde{w}+2)}_{h,k}+\frac{1}{2\tilde{w}+2}\sum_{P^{*}(q)}^{}\Big (\hat{C}^{m}_{ijhq_{2\tilde{w}+2}}N^{(2\tilde{w}+1,0)}_{hpq_{1}...q_{2\tilde{w}}+1}\Big ) \frac{\partial^{2\tilde{w}+2}\hat{U}^{M}_{p}}{\partial x_{q_{1}}...\partial x_{q_{2\tilde{w}+2}}}+&
\end{flalign}
\begin{flalign*}
&+\frac{1}{2\tilde{w}}\sum_{P^{*}(q)}\Big (\hat{C}^{m}_{ijhq_{2\tilde{w}}}N^{(2\tilde{w}+1,2)}_{hpq_{1}...q_{2\tilde{w}-1}}\Big ) \frac{\partial ^{2\tilde{w}}\hat{U}^{M}_{p}}{\partial x_{q_{1}}...\partial x_{q_{2\tilde{w}}}}s^{2}+&
\end{flalign*}
\begin{flalign*}
&+\sum_{n=1}^{n=\tilde{w}-1}(1-\delta_{1n})\frac{1}{2\tilde{w}-2n+2}\sum_{P^{*}(q)}^{} \Big ( \hat{C}^{m}_{ijhq_{2\tilde{w}-2n+2}}N^{(2\tilde{w}+1,2n)}_{hpq_{1}...q_{2\tilde{w}-2n+1}}\Big )\frac{\partial ^{2\tilde{w}+2-2n}\hat{U}^{M}_{p}}{\partial x_{q_{1}}... \partial x_{q_{2\tilde{w}+2-2n}}}s^{2n}+&
\end{flalign*}
\begin{flalign*}
&+\frac{1}{2}\Big [\Big( \hat{C}^{m}_{ijhq_{2}}N^{(2\tilde{w}+1,2\tilde{m})}_{hpq_{1}}+\hat{C}^{m}_{ijhq_{1}}N^{(2\tilde{w}+1,2\tilde{w})}_{hpq_{2}}\Big) \frac{\partial^{2}\hat{U}^{M}_{p}}{\partial x_{q_{1}}\partial x_{q_{2}}} s^{2\tilde{w}}\Big ]
\Big )n_{j}\Big ]\Big ]\Big \vert_{\boldsymbol{\xi} \in \Sigma_{1}}=0.&
\end{flalign*}
\section*{\small Appendix B. Symmetrization of the localization tensors}
In order to perform the symmetrization of a tensor $Z_{hkpq_{1}...q_{n}}$ with respect to the indices $q_{1}...q_{n}$, the set $P^{*}(q)$, which consists of all permutations with no fixed indices, is considered. For instance, if $|q|=n$, it results that $P^{*}(q)=\Big\{{}f_{1}= \Big ( q_{1}\rightarrow q_{1},q_{2}\rightarrow q_{2},...,q_{n}\rightarrow q_{n}\Big ),..., f_{n}= \Big ( q_{1}\rightarrow q_{n},q_{2}\rightarrow q_{1},...,q_{n}\rightarrow q_{2}\Big)\Big\}$ and the tensor $Z_{hkpq_{1}...q_{n}}$ is symmetrized with respect to $q_{1},...,q_{n}$ as
\begin{flalign*}
\frac{1}{n} \sum_{P^{*}(q)}^{} Z_{hkpq_{1}...q_{n}}=\frac{1}{n}\Big (Z_{hkpq_{1}...q_{n}}+...+Z_{hkpq_{n}q_{1}..q_{2}}\Big ). 
\end{flalign*}
In particular, if $|q|=2$ then the permutations set $P^{*}(q)$ with no fixed points is $P^{*}(q)=\Big\{{}f_{1}= \Big ( q_{1}\rightarrow q_{1},q_{2}\rightarrow q_{2}\Big ), f_{2}= \Big ( q_{1}\rightarrow q_{2},q_{2}\rightarrow q_{1}\Big)\Big\}$ and the symmetrization with respect to $q_{1}$ and $q_{2}$ of the localization tensor $\tilde{B}^{(2,0)}_{hkpq_{1}q_{2}}=\Big (\delta_{kq_{2}}N^{(1,0)}_{hpq_{1}}+N^{(2,0)}_{hpq_{1}q_{2},k}\Big )$ results
\begin{flalign*}
&B^{(2,0)}_{hkpq_{1}q_{2}}=\frac{1}{2} \Big (\delta_{kq_{2}}N^{(1,0)}_{hpq_{1}}+\delta_{kq_{1}}N^{(1,0)}_{hpq_{2}}+N^{(2,0)}_{hpq_{1}q_{2},k}+N^{(2,0)}_{hpq_{2}q_{1},k}\Big)=&
\end{flalign*} 
\begin{flalign}
&=\frac{1}{2} \Big (\delta_{kq_{2}}N^{(1,0)}_{hpq_{1}}+\delta_{kq_{1}}N^{(1,0)}_{hpq_{2}}\Big )+N^{(2,0)}_{hpq_{1}q_{2},k}.&
\end{flalign}   
In case of $|q|=3$, the permutations set $P^{*}(q)$ having no fixed points is $P^{*}(q)=\Big \{{}  f_{1}= \Big ( q_{1}\rightarrow q_{1},q_{2}\rightarrow q_{2},q_{3}\rightarrow q_{3}\Big ), f_{2}= \Big ( q_{1}\rightarrow q_{2},q_{2}\rightarrow q_{3},q_{3}\rightarrow q_{1}\Big),  f_{3}= \Big ( q_{1}\rightarrow q_{3},q_{2}\rightarrow q_{1},q_{3}\rightarrow q_{2}\Big)\Big\}$ and the localization tensor $\tilde{B}^{(3,0)}_{hkpq_{1}q_{2}q_{3}}=\Big (\delta_{kq_{3}}N^{(2,0)}_{hpq_{1}q_{2}}+N^{(3,0)}_{hpq_{1}q_{2}q_{3},k}\Big )$ is symmetrized with respect to $q_{1}$, $q_{2}$ and $q_{3}$ as
\begin{flalign*}
&B^{(3,0)}_{hkpq_{1}q_{2}q_{3}}=\frac{1}{3} \Big (\delta_{kq_{3}}N^{(2,0)}_{hpq_{1}q_{2}}+\delta_{kq_{1}}N^{(2,0)}_{hpq_{2}q_{3}}+\delta_{kq_{2}}N^{(2,0)}_{hpq_{3}q_{1}}+&
\end{flalign*} 
\begin{flalign*}
&+N^{(3,0)}_{hpq_{1}q_{2}q_{3},k}+N^{(3,0)}_{hpq_{2}q_{3}q_{1},k}+N^{(3,0)}_{hpq_{3}q_{1}q_{2},k}\Big)=&
\end{flalign*}
\begin{flalign}
&=\frac{1}{3} \Big (\delta_{kq_{3}}N^{(2,0)}_{hpq_{1}q_{2}}+\delta_{kq_{1}}N^{(2,0)}_{hpq_{2}q_{3}}+\delta_{kq_{2}}N^{(2,0)}_{hpq_{3}q_{1}}\Big )+N^{(3,0)}_{hpq_{1}q_{2}q_{3},k},&
\end{flalign} 
as it appears in Eq. \eqref{locten}.
\section*{ \small Appendix C. Perturbation functions of first, second and third order}
\normalsize \textit{Appendix C.1. Perturbation function of first order $N_{hpq}^{(1,0)}$ }\\ \\
Let $\Omega$ be a layered domain obtained as a $d_{2}-$ periodic arrangement of two different layers whose thickness is $s_{1}$ and $s_{2}$ (here $d_{2}=s_{1}+s_{2}$ and $\eta=s_{1}/s_{2}$ are defined).
The phases are supposed to be homogeneous and orthotropic, with an
orthotropic axis coincident with the layering direction $\boldsymbol{e}_{1}$. The micro-fluctuation functions $N_{hpq}^{(1,0)_{i}}$ are analytically obtained by solving the first cell problems \eqref{cps1}. The superscript $i=\{1,2\}$ stands for the phase 1 and the phase 2 and they are formulated as
\begin{flalign}
&N_{211}^{(1,0)_1}=-{\frac { \left( { {\hat{C}_{1122}^{1}}}-{ {\hat{C}_{1122}^{2}}} \right) { \xi_2}}{{
			{\hat{C}_{2222}^{2}}}\,\eta+{ {\hat{C}_{2222}^{1}}}}}
, \quad 	N_{211}^{(1,0)_2}={\frac {\eta \, \left( { {\hat{C}_{1122}^{1}}}-{ {\hat{C}_{1122}^{2}}}
		\right) { \xi_2}}{{ {\hat{C}_{2222}^{2}}}\, \eta +{ 
			{\hat{C}_{2222}^{1}}}}},&	 
\end{flalign}
\begin{flalign}
&N_{222}^{(1,0)_1}=-{\frac { \left( { {\hat{C}_{2222}^{1}}}-{ {\hat{C}_{2222}^{2}}} \right) { \xi_2}}{{
			{\hat{C}_{2222}^{2}}}\,\eta+{ {\hat{C}_{2222}^{1}}}}}
, \quad 	N_{222}^{(1,0)_2}={\frac {\eta \, \left( { {\hat{C}_{2222}^{1}}}-{ {\hat{C}_{2222}^{2}}}
		\right) { \xi_2}}{{ {\hat{C}_{2222}^{2}}}\, \eta +{ 
			{\hat{C}_{2222}^{1}}}}},&	 
\end{flalign}
\begin{flalign}
&N_{112}^{(1,0)_1}=N_{121}^{(1,0)_1}=-{\frac { \left( { {\hat{C}_{1212}^{1}}}-{ {\hat{C}_{1212}^{2}}} \right) { \xi_2}}{{
			{\hat{C}_{1212}^{2}}}\,\eta+{ {\hat{C}_{1212}^{1}}}}}
, \quad 	N_{112}^{(1,0)_2}=N_{121}^{(1,0)_2}={\frac {\eta \, \left( { {\hat{C}_{1212}^{1}}}-{ {\hat{C}_{1212}^{2}}}
		\right) { \xi_2}}{{ {\hat{C}_{1212}^{2}}}\, \eta +{ 
			{\hat{C}_{1212}^{1}}}}}.&	 
\end{flalign}
Such functions depend on the fast variable $\boldsymbol{\xi}$, since the microstructure enjoys the simmetry property. In the following it is assumed that the coordinate $\xi_{2}$ is centered in both layers.\\\\
\normalsize \emph{ Appendix C.2. Perturbation functions of second order $N_{hpqr}^{(2,0)}$} \\ \\
The perturbation functions $N_{hpqr}^{(2,0)_{i}}$, $i=\{1,2\}$, deriving from the cell problem \eqref{eqn:sym} are:	
\begin{flalign}
&N_{1111}^{(2,0)_1}=A_{1111}^{2} \xi_{2}^2+A_{1111}^{0}, \quad N_{1111}^{(2,0)_2}=B_{1111}^{2} \xi_{2}^2+B_{1111}^{0},&
\end{flalign}	
\begin{flalign}
&N_{2211}^{(2,0)_1}=A_{2211}^{2} \xi_{2}^2+A_{2211}^{0}, \quad N_{2211}^{(2,0)_2}=B_{2211}^{2} \xi_{2}^2+B_{2211}^{0},&
\end{flalign}
\begin{flalign}
&N_{2222}^{(2,0)_1}=A_{2222}^{2} \xi_{2}^2+A_{2222}^{0}, \quad N_{2222}^{(2,0)_2}=B_{2222}^{2} \xi_{2}^2+B_{2222}^{0},&
\end{flalign}
\begin{flalign}
&N_{1122}^{(2,0)_1}=A_{1122}^{2} \xi_{2}^2+A_{1122}^{0}, \quad N_{1122}^{(2,0)_2}=B_{1122}^{2} \xi_{2}^2+B_{1122}^{0},&
\end{flalign}
where the constants $A_{1111}^{2}$, $A_{1111}^{0}$, $B_{1111}^{2}$, $B_{1111}^{0}$, $A_{2211}^{2}$, $A_{2211}^{0}$, $B_{2211}^{2}$, $B_{2211}^{0}$, $A_{2222}^{2}$, $A_{2222}^{0}$, $B_{2222}^{2}$, $B_{2222}^{0}$, $A_{1122}^{2}$,\\ \\ $A_{1122}^{0}$, $B_{1122}^{2}$ and $B_{1122}^{0}$ are determined as follows
\begin{flalign}
&A_{1111}^{2}=-\frac{1}{2}\,{\frac {A_{1111}^{2,0}+\eta A_{1111}^{2,1} }{ \left( \eta+1 \right)  \left( { \hat{C}_{2222}^{{2}}}\,\eta+ \hat{C}_{2222}^{{1}} \right) { \hat{C}_{1212}^{{1}}}}},&
\end{flalign}	             
\begin{flalign}
&A_{1111}^{0}=\frac{1}{24}\,{\frac {\eta\, \left( { A_{1111}^{0,3}}\,{\eta}^{3}+{ A_{1111}^{0,2}}
		\,{\eta}^{2}+{ A_{1111}^{0,1}}\,\eta+{ A_{1111}^{0,0}} \right) }{ \left( 
		\eta+1 \right) ^{4} \left( { \hat{C}_{2222}^{{2}}}\,\eta+{ \hat{C}_{2222}^{{1}}}
		\right) { \hat{C}_{1212}^{{2}}}\,{ \hat{C}_{1212}^{{1}}}}},&
\end{flalign}

\begin{flalign}
&B_{1111}^{2}=\frac{1}{2}\,{\frac {\eta B_{1111}^{2,1}+\eta^2 B_{1111}^{2,2} }{ \left( \eta+1 \right)  \left( { \hat{C}_{2222}^{{2}}}\,\eta+ \hat{C}_{2222}^{{1}} \right) { \hat{C}_{1212}^{{2}}}}},&
\end{flalign}
\begin{flalign}
&B_{1111}^{0}=\frac{1}{24}\,{\frac {\eta\, \left( { -2 A_{1111}^{0,3}}\,{\eta}^{3}+{ B_{1111}^{0,2}}
		\,{\eta}^{2}+{ B_{1111}^{0,1}}\,\eta-{\frac{A_{1111}^{0,0}}{2}} \right) }{ \left( 
		\eta+1 \right) ^{4} \left( { \hat{C}_{2222}^{{2}}}\,\eta+{ \hat{C}_{2222}^{{1}}}
		\right) { \hat{C}_{1212}^{{1}}}\,{ \hat{C}_{1212}^{{2}}}}},&
\end{flalign}
\begin{flalign}
&A_{2211}^{2}=\frac{1}{2}\,{\frac {{ \hat{C}_{1122}^{{1}}}\, \left( { \hat{C}_{1212}^{{1}}}-{ \hat{C}_{1212}^{{2}}}
		\right) }{ \left( { \hat{C}_{1212}^{{2}}}\,\eta+{ \hat{C}_{1212}^{{1}}} \right) { 
			\hat{C}_{2222}^{{1}}}}},&
\end{flalign}
\begin{flalign}
&A_{2211}^{0}=-\frac{1}{24}\,{\frac { \left( { \hat{C}_{1212}^{{1}}}-{ \hat{C}_{1212}^{{2}}} \right) \eta\,
		\left( { C_{1122}^{{1}}}\,{\eta}^{2}{ \hat{C}_{2222}^{{2}}}+3\,{ C_{1122}^{{1}}}\,
		\eta\,{ \hat{C}_{2222}^{{2}}}+2\,{ C_{1122}^{{2}}}\,{ \hat{C}_{2222}^{{1}}} \right) }{
		\left( \eta+1 \right) ^{3} \left( { \hat{C}_{1212}^{{2}}}\,\eta+{ \hat{C}_{1212}^{{1}}}
		\right) { \hat{C}_{2222}^{{2}}}\,{ \hat{C}_{2222}^{{1}}}}},&
\end{flalign}	
\begin{flalign}
&B_{2211}^{2}=-\frac{\eta  C_{1122}^{2}  \hat{C}_{2222}^{{1}}}{ \hat{C}_{2222}^{{2}} C_{1122}^{1}}A_{2211}^{2},&
\end{flalign}	
\begin{flalign}
&B_{2211}^{0}=\frac{1}{24}\,{\frac { \left( 2\,{ C_{1122}^{{1}}}\,{\eta}^{2}{ \hat{C}_{2222}^{{2}}}+3\,
		\eta\,{ C_{1122}^{{2}}}\,{ \hat{C}_{2222}^{{1}}}+{ C_{1122}^{{2}}}\,{ \hat{C}_{2222}^{{1}}}
		\right)  \left( { \hat{C}_{1212}^{{1}}}-{ \hat{C}_{1212}^{{2}}} \right) \eta}{ \left( 
		\eta+1 \right) ^{3} \left( { \hat{C}_{1212}^{{2}}}\,\eta+{ \hat{C}_{1212}^{{1}}}
		\right) { \hat{C}_{2222}^{{2}}}\,{ \hat{C}_{2222}^{{1}}}}},&
\end{flalign}
\begin{flalign}
&A_{2222}^{2}={\frac {{ \hat{C}_{2222}^{{1}}}-{ \hat{C}_{2222}^{{2}}}}{2\,{ \hat{C}_{2222}^{{2}}}\,\eta+2\,{
			\hat{C}_{2222}^{{1}}}}}, \quad A_{2222}^{0}=-\frac{1}{24}\,{\frac { \left( { \hat{C}_{2222}^{{1}}}-{ \hat{C}_{2222}^{{2}}} \right) \eta\,
		\left( \eta+2 \right) }{ \left( { \hat{C}_{2222}^{{2}}}\,\eta+{ \hat{C}_{2222}^{{1}}}
		\right)  \left( \eta+1 \right) ^{2}}},&
\end{flalign}
\begin{flalign}
&B_{2222}^{2}=-\eta A_{2222}^{2}, \quad B_{2222}^{0}=-\frac{2\eta+1}{\eta+2}A_{2222}^{0},&
\end{flalign}
\begin{flalign}
&A_{1122}^{2}={\frac {{ \hat{C}_{1212}^{{1}}}-{ \hat{C}_{1212}^{{2}}}}{2\,{ \hat{C}_{1212}^{{2}}}\,\eta+2\,{
			\hat{C}_{1212}^{{1}}}}}, \quad A_{1122}^{0}=-\frac{1}{24}\,{\frac { \left( { \hat{C}_{1212}^{{1}}}-{ \hat{C}_{1212}^{{2}}} \right) \eta\,
		\left( \eta+2 \right) }{ \left( { \hat{C}_{1212}^{{2}}}\,\eta+{ \hat{C}_{1212}^{{1}}}
		\right)  \left( \eta+1 \right) ^{2}}},&
\end{flalign}
\begin{flalign}
&B_{1122}^{2}=-\eta A_{1122}^{2}, \quad B_{1122}^{0}=-\frac{2\eta+1}{\eta+2}A_{1122}^{0}.&
\end{flalign}
\normalsize \emph{Appendix C.3. Perturbation functions $N_{hp}^{(2,2)}$}\\\\
The perturbation functions $N_{hp}^{(2,2)_{i}}$, $i=\{1,2\}$, obtained by performing the cell problem \eqref{22N} are:
\begin{flalign}
&N_{11}^{(2,2)_1}=A_{11}^{2} \xi_{2}^2+A_{11}^{0}, \quad N_{11}^{(2,2)_2}=B_{11}^{2} \xi_{2}^2+B_{11}^{0},&
\end{flalign}		
\begin{flalign}
\label{zia}
&N_{22}^{(2,2)_1}=A_{22}^{2} \xi_{2}^2+A_{22}^{0}, \quad N_{22}^{(2,2)_2}=B_{22}^{2} \xi_{2}^2+B_{22}^{0},&
\end{flalign}
where the constants $A_{11}^{2}$, $A_{11}^{0}$, $B_{11}^{2}$, $B_{11}^{0}$, $A_{22}^{2}$, $A_{22}^{0}$, $B_{22}^{2}$ and $B_{22}^{0}$ are     
\begin{flalign}
&A_{11}^{2}=\frac{1}{2}\,{\frac { \left( { \rho_{1}}-{ \rho_{2}} \right) { }}{
		\left( \eta +1 \right) { \hat{C}_{1212}^{{1}}}}}, \quad A_{11}^{0}=-\frac{1}{24}\,{\frac { \left( { \rho_{1}}-{ \rho_{2}} \right) { }
		\, \eta \, \left( { \hat{C}_{1212}^{{2}}}\,{\eta }^{2}+3\,{
			\hat{C}_{1212}^{{2}}}\, \eta +2\,{ \hat{C}_{1212}^{{1}}} \right) }{{
			\hat{C}_{1212}^{{2}}}\, \left( \eta +1 \right) ^{4}{ \hat{C}_{1212}^{{1}}}}},&	
\end{flalign}
\begin{flalign}
&B_{11}^{2}=-\frac{1}{2}\,{\frac { \left( { \rho_{1}}-{ \rho_{2}} \right) \eta { }}{
		\left( \eta +1 \right) { \hat{C}_{1212}^{{2}}}}}, \quad B_{11}^{0}=\frac{1}{24}\,{\frac { \left( { \rho_{1}}-{ \rho_{2}} \right) { }
		\, \eta \, \left( { 2\hat{C}_{1212}^{{2}}}\,{\eta }^{2}+3\,{
			\hat{C}_{1212}^{{1}}}\, \eta +{ \hat{C}_{1212}^{{1}}} \right) }{{
			\hat{C}_{1212}^{{2}}}\, \left( \eta +1 \right) ^{4}{ \hat{C}_{1212}^{{1}}}}},&	
\end{flalign}
\begin{flalign}
&A_{22}^{2}=\frac{1}{2}\,{\frac { \left( { \rho_{1}}-{ \rho_{2}} \right) { }}{
		\left( \eta +1 \right) { \hat{C}_{2222}^{{1}}}}}, \quad A_{22}^{0}=-\frac{1}{24}\,{\frac { \left( { \rho_{1}}-{ \rho_{2}} \right) { }
		\, \eta \, \left( { \hat{C}_{2222}^{{2}}}\,{\eta }^{2}+3\,{
			\hat{C}_{2222}^{{2}}}\, \eta +2\,{ \hat{C}_{2222}^{{1}}} \right) }{{
			\hat{C}_{2222}^{{2}}}\, \left( \eta +1 \right) ^{4}{ \hat{C}_{2222}^{{1}}}}},&
\end{flalign}
\begin{flalign}
&B_{22}^{2}=-\frac{1}{2}\,{\frac { \left( { \rho_{1}}-{ \rho_{2}} \right) \eta { }}{
		\left( \eta +1 \right) { \hat{C}_{2222}^{{2}}}}}, \quad B_{22}^{0}=\frac{1}{24}\,{\frac { \left( { \rho_{1}}-{ \rho_{2}} \right) { }
		\, \eta \, \left( { 2\hat{C}_{2222}^{{2}}}\,{\eta }^{2}+3\,{
			\hat{C}_{2222}^{{1}}}\, \eta +{ \hat{C}_{2222}^{{1}}} \right) }{{
			\hat{C}_{2222}^{{2}}}\, \left( \eta +1 \right) ^{4}{ \hat{C}_{2222}^{{1}}}}}.&	
\end{flalign}

\normalsize \emph{Appendix C.4. Perturbation functions of third order $N_{hpqrs}^{(3,0)}$} \\ \\
The non-vanishing micro-fluctuation functions $N_{hpqrs}^{(3,0)_{i}}$, $i=\{1,2\}$, obtained by performing the cell problem \eqref{meq}, with $w=1$, are:
\begin{flalign}
&N_{21111}^{(3,0)_1}=A_{21111}^{3} \xi_{2}^3+A_{21111}^{1} \xi_{2}, \quad N_{21111}^{(3,0)_2}=B_{21111}^{3} \xi_{2}^3+B_{21111}^{1} \xi_{2},&
\end{flalign}
\begin{flalign}
&N_{11222}^{(3,0)_1}=A_{11222}^{3} \xi_{2}^3+A_{11222}^{1} \xi_{2}, \quad N_{11222}^{(3,0)_2}=B_{11222}^{3} \xi_{2}^3+B_{11222}^{1} \xi_{2},&
\end{flalign}
\begin{flalign}
&N_{12111}^{(3,0)_1}=A_{12111}^{3} \xi_{2}^3+A_{12111}^{1} \xi_{2}, \quad N_{12111}^{(3,0)_2}=B_{12111}^{3} \xi_{2}^3+B_{12111}^{1} \xi_{2},&
\end{flalign}
\begin{flalign}
&N_{22222}^{(3,0)_1}=A_{22222}^{3} \xi_{2}^3+A_{22222}^{1} \xi_{2}, \quad N_{22222}^{(3,0)_2}=B_{22222}^{3} \xi_{2}^3+B_{22222}^{1} \xi_{2},&
\end{flalign}
where the constants $A_{21111}^{3}$, $A_{21111}^{1}$, $B_{21111}^{3}$, $B_{21111}^{1}$, $A_{11222}^{3}$, $A_{11222}^{1}$, $B_{11222}^{3}$, $B_{11222}^{1}$, $A_{12111}^{3}$, $A_{12111}^{1}$, $B_{12111}^{3}$,\\ \\ $B_{12111}^{1}$, $A_{22222}^{3}$, $A_{22222}^{1}$, $B_{22222}^{3}$ and $B_{22222}^{1}$ are
\begin{flalign}
&A_{21111}^{3}=\frac{A_{21111}^{3,1}\eta+A_{21111}^{3,0}}{18\, \left( { \hat{C}_{2222}^{{2}}}\,\eta+{ \hat{C}_{2222}^{{1}}} \right)  \left( \eta+1 \right) { \hat{C}_{1212}^{{1}}}\,{ \hat{C}_{2222}^{{1}}}},&
\end{flalign}
\begin{flalign}
&A_{21111}^{1}=\frac{\eta(A_{21111}^{1,4}\eta^{4}+A_{21111}^{1,3}\eta^{3}+A_{21111}^{1,2}\eta^{2}+A_{21111}^{1,1} \eta+A_{21111}^{1,0})}{72\, \left( { \hat{C}_{2222}^{{2}}}\,\eta+{ \hat{C}_{2222}^{{1}}} \right) ^{2} \left(\eta+1 \right) ^{4}{ \hat{C}_{1212}^{{1}}}\,{ \hat{C}_{2222}^{{1}}}\,{ \hat{C}_{1212}^{{2}}}},&
\end{flalign}
\begin{flalign}
&B_{21111}^{3}=\frac{\eta (B_{21111}^{3,2}\eta+B_{21111}^{3,1})}{18\, \left( { \hat{C}_{2222}^{{2}}}\,\eta+{ \hat{C}_{2222}^{{1}}} \right)  \left( \eta+1 \right) { \hat{C}_{1212}^{{2}}}\,{ \hat{C}_{2222}^{{2}}}},&
\end{flalign}
\begin{flalign}
&B_{21111}^{1}=\frac{\eta(B_{21111}^{1,4}\eta^{4}+B_{21111}^{1,3}\eta^{3}+B_{21111}^{1,2}\eta^{2}+B_{21111}^{1,1} \eta+B_{21111}^{1,0})}{72\, \left( { \hat{C}_{2222}^{{2}}}\,\eta+{ \hat{C}_{2222}^{{1}}} \right) ^{2} \left(\eta+1 \right) ^{4}{ \hat{C}_{1212}^{{2}}}\,{ \hat{C}_{2222}^{{2}}}\,{ \hat{C}_{1212}^{{1}}}},&
\end{flalign}
\begin{flalign}
&A_{11222}^{3}=-\frac{1}{18}\,{\frac { \left( { \hat{C}_{1212}^{{2}}}\,{\eta}^{3}+ \left( { 
			\hat{C}_{1212}^{{1}}}+2\,{ \hat{C}_{1212}^{{2}}} \right) {\eta}^{2}+ \left( 2\,{ 
			\hat{C}_{1212}^{{1}}}+{ \hat{C}_{1212}^{{2}}} \right) \eta+{ \hat{C}_{1212}^{{1}}} \right)  \left( {
			\hat{C}_{1212}^{{1}}}-{ \hat{C}_{1212}^{{2}}} \right) }{ \left( { \hat{C}_{1212}^{{2}}}\,\eta+{
			\hat{C}_{1212}^{{1}}} \right) ^{2} \left( \eta+1 \right) ^{2}}},&
\end{flalign}
\begin{flalign}
&A_{11222}^{1}=-\frac{1}{18}\,{\frac { \left( -\frac{1}{4}\,{ \hat{C}_{1212}^{{2}}}\,{\eta}^{3}+ \left( \frac{3}{4}\,{
			\hat{C}_{1212}^{{1}}}-\frac{3}{2}\,{ \hat{C}_{1212}^{{2}}} \right) {\eta}^{2}+ \left( -\frac{3}{2}\,{
			\hat{C}_{1212}^{{1}}}+{ \hat{C}_{1212}^{{2}}} \right) \eta \right)  \left( { 
			\hat{C}_{1212}^{{1}}}-{ \hat{C}_{1212}^{{2}}} \right) }{ \left( { \hat{C}_{1212}^{{2}}}\,\eta+{ 
			\hat{C}_{1212}^{{1}}} \right) ^{2} \left( \eta+1 \right) ^{2}}},&
\end{flalign}
\begin{flalign}
&B_{11222}^{3}={\frac {\eta\, \left( { \hat{C}_{1212}^{{1}}}-{ \hat{C}_{1212}^{{2}}} \right) }{18\,{
			\hat{C}_{1212}^{{2}}}\,\eta+18\,{ \hat{C}_{1212}^{{1}}}}},&
\end{flalign}
\begin{flalign}
&B_{11222}^{1}={\frac {\eta\, \left( { \hat{C}_{1212}^{{1}}}-{ \hat{C}_{1212}^{{2}}} \right)  \left( 
		\left( 4\,{ \hat{C}_{1212}^{{1}}}-6\,{ \hat{C}_{1212}^{{2}}} \right) {\eta}^{2}+
		\left( -6\,{ \hat{C}_{1212}^{{1}}}+3\,{ \hat{C}_{1212}^{{2}}} \right) \eta-{ 
			\hat{C}_{1212}^{{1}}} \right) }{72\, \left( \eta+1 \right) ^{2} \left( { 
			\hat{C}_{1212}^{{2}}}\,\eta+{ \hat{C}_{1212}^{{1}}} \right) ^{2}}},&
\end{flalign}
\begin{flalign}
&A_{12111}^{3}=\frac{\eta(A_{12111}^{3,3}\eta^{3}+A_{12111}^{3,2}\eta^{2}+A_{12111}^{3,1}\eta^{1}+A_{12111}^{3,0})}{18\, \left( { \hat{C}_{1212}^{{2}}}\,\eta+{ \hat{C}_{1212}^{{1}}} \right) ^{2}{ 
		\hat{C}_{2222}^{{1}}}\,{ \hat{C}_{1212}^{{1}}}\,{ \hat{C}_{2222}^{{2}}}\, \left( \eta+1 \right) ^{3}
},&
\end{flalign}
\begin{flalign}
&A_{12111}^{1}=\frac{\eta(A_{12111}^{1,3}\eta^{3}+A_{12111}^{1,2}\eta^{2}+A_{12111}^{1,1}\eta^{1}+A_{12111}^{1,0})}{72\, \left( { \hat{C}_{1212}^{{2}}}\,\eta+{ \hat{C}_{1212}^{{1}}} \right) ^{2}{ 
		\hat{C}_{2222}^{{1}}}\,{ \hat{C}_{1212}^{{1}}}\,{ \hat{C}_{2222}^{{2}}}\, \left( \eta+1 \right) ^{3}
},&
\end{flalign}
\begin{flalign}
&B_{12111}^{3}=-\frac{1}{18}\,{\frac {\eta\, \left( { \hat{C}_{1212}^{{1}}}-{ \hat{C}_{1212}^{{2}}} \right) 
		\left( { C_{1111}^{{2}}}\,{ \hat{C}_{2222}^{{2}}}-{{ C_{1122}^{{2}}}}^{2}-{ 
			\hat{C}_{1122}^{{2}}}\,{ \hat{C}_{1212}^{{2}}} \right) }{ \left( { \hat{C}_{1212}^{{2}}}\,\eta+{ 
			\hat{C}_{1212}^{{1}}} \right) { \hat{C}_{2222}^{{2}}}\,{ \hat{C}_{1212}^{{2}}}}},&
\end{flalign}
\begin{flalign}
&B_{12111}^{1}=\frac{\eta(B_{12111}^{1,3}\eta^{3}+B_{12111}^{1,2}\eta^{2}+B_{12111}^{1,1}\eta^{1}+B_{12111}^{1,0})}{72\, \left( { \hat{C}_{1212}^{{2}}}\,\eta+{ \hat{C}_{1212}^{{1}}} \right) ^{2} \left( 
	\eta+1 \right) ^{3}{ \hat{C}_{2222}^{{2}}}\,{ \hat{C}_{1212}^{{2}}}\,{ \hat{C}_{2222}^{{1}}}
},&
\end{flalign}
\begin{flalign}
&A_{22222}^{3}=-\frac{1}{18}\,{\frac { \left( { \hat{C}_{2222}^{{1}}}-{ \hat{C}_{2222}^{{2}}} \right)  \left( {
			\hat{C}_{2222}^{{2}}}\,{\eta}^{3}+ \left( { \hat{C}_{2222}^{{1}}}+2\,{ \hat{C}_{2222}^{{2}}}
		\right) {\eta}^{2}+ \left( 2\,{ \hat{C}_{2222}^{{1}}}+{ \hat{C}_{2222}^{{2}}} \right) 
		\eta+{ \hat{C}_{2222}^{{1}}} \right) }{ \left( { \hat{C}_{2222}^{{2}}}\,\eta+{ 
			\hat{C}_{2222}^{{1}}} \right) ^{2} \left( \eta+1 \right) ^{2}}},&
\end{flalign}
\begin{flalign}
&A_{22222}^{1}=-\frac{1}{18}\,{\frac { \left( { \hat{C}_{2222}^{{1}}}-{ \hat{C}_{2222}^{{2}}} \right)  \left( -
		\frac{1}{4}\,{ \hat{C}_{2222}^{{2}}}\,{\eta}^{3}+ \left( \frac{3}{4}\,{ \hat{C}_{2222}^{{1}}}-\frac{3}{2}\,{ 
			\hat{C}_{2222}^{{2}}} \right) {\eta}^{2}+ \left( -\frac{3}{2}\,{ \hat{C}_{2222}^{{1}}}+{ 
			\hat{C}_{2222}^{{2}}} \right) \eta \right) }{ \left( { \hat{C}_{2222}^{{2}}}\,\eta+{ 
			\hat{C}_{2222}^{{1}}} \right) ^{2} \left( \eta+1 \right) ^{2}}},&
\end{flalign}
\begin{flalign}
&B_{22222}^{3}=\frac{1}{18}\,{\frac { \left( { \hat{C}_{2222}^{{1}}}-{ \hat{C}_{2222}^{{2}}} \right)  \left( {
			\eta}^{3}{ \hat{C}_{2222}^{{2}}}+ \left( { \hat{C}_{2222}^{{1}}}+2\,{ \hat{C}_{2222}^{{2}}}
		\right) {\eta}^{2}+ \left( 2\,{ \hat{C}_{2222}^{{1}}}+{ \hat{C}_{2222}^{{2}}} \right) 
		\eta+{ \hat{C}_{2222}^{{1}}} \right) \eta}{ \left( \eta+1 \right) ^{2} \left( {
			\hat{C}_{2222}^{{2}}}\,\eta+{ \hat{C}_{2222}^{{1}}} \right) ^{2}}},&
\end{flalign}
\begin{flalign}
&B_{22222}^{1}=\frac{1}{18}\,{\frac { \left( { \hat{C}_{2222}^{{1}}}-{ \hat{C}_{2222}^{{2}}} \right)  \left( 
		\left( { \hat{C}_{2222}^{{1}}}-\frac{3}{2}\,{ \hat{C}_{2222}^{{2}}} \right) {\eta}^{2}+ \left( 
		-\frac{3}{2}\,{ \hat{C}_{2222}^{{1}}}+\frac{3}{4}\,{ \hat{C}_{2222}^{{2}}} \right) \eta-{ \frac{\hat{C}_{2222}^{{1}}}{4}}
		\right) \eta}{ \left( \eta+1 \right) ^{2} \left( { \hat{C}_{2222}^{{2}}}\,\eta
		+{ \hat{C}_{2222}^{{1}}} \right) ^{2}}}.&
\end{flalign}   
\normalsize \emph{Appendix C.5. Perturbation functions $N_{hpq}^{(3,2)}$}  \\ \\   
The perturbation functions $N_{hpq}^{(3,2)_{i}}$, $i=\{1,2\}$, obtained by performing the cell problem  \eqref{32N} are:
\begin{flalign}
&N_{211}^{(3,2)_1}=A_{211}^{3} \xi_{2}^3+A_{211}^{1} \xi_{2}, \quad N_{211}^{(3,2)_2}=B_{211}^{3} \xi_{2}^3+B_{211}^{1} \xi_{2},&
\end{flalign}	
\begin{flalign}
&N_{121}^{(3,2)_1}=A_{121}^{3} \xi_{2}^3+A_{121}^{1} \xi_{2}, \quad N_{121}^{(3,2)_2}=B_{121}^{3} \xi_{2}^3+B_{121}^{1} \xi_{2},&
\end{flalign}
\begin{flalign}
&N_{112}^{(3,2)_1}=A_{112}^{3} \xi_{2}^3+A_{112}^{1} \xi_{2}, \quad N_{112}^{(3,2)_2}=B_{112}^{3} \xi_{2}^3+B_{112}^{1} \xi_{2},&
\end{flalign}
\begin{flalign}
\label{sta}
&N_{222}^{(3,2)_1}=A_{222}^{3} \xi_{2}^3+A_{222}^{1} \xi_{2}, \quad N_{222}^{(3,2)_2}=B_{222}^{3} \xi_{2}^3+B_{222}^{1} \xi_{2},&
\end{flalign}
where the constants $A_{211}^{3}$, $A_{211}^{1}$, $B_{211}^{3}$, $B_{211}^{1}$, $A_{121}^{3}$, $A_{121}^{1}$, $B_{121}^{3}$, $B_{121}^{1}$, $A_{112}^{3}$, $A_{112}^{1}$, $B_{112}^{3}$, $B_{112}^{1}$, $B_{122}^{1}$, $A_{222}^{3}$,\\ \\ $A_{222}^{1}$, $B_{222}^{3}$ and $B_{222}^{1}$ are 
\begin{flalign}
\label{3.51}
&A_{211}^{3}=\frac{1}{6}\,{\frac {A_{211}^{3,1}\eta+A_{211}^{3,0}}{\left( \eta+1 \right)  \left( { \hat{C}_{2222}^{{2}}}\,\eta+{ \hat{C}_{2222}^{{1}}}
		\right) { \hat{C}_{1212}^{{1}}}\,{ \hat{C}_{2222}^{{1}}}}},&
\end{flalign}
\begin{flalign}
&A_{211}^{1}=\frac{1}{24}\frac{\eta(A_{211}^{1,4}\eta^{4}+A_{211}^{1,3}\eta^{3}+A_{211}^{1,2}\eta^{2}+A_{211}^{1,1} \eta+A_{211}^{1,0})}{\left( \eta+1 \right) ^{4} \left( { \hat{C}_{2222}^{{2}}}\,\eta+{ 
		\hat{C}_{2222}^{{1}}} \right) ^{2}{ \hat{C}_{1212}^{{1}}}\,{ \hat{C}_{2222}^{{1}}}\,{ \hat{C}_{1212}^{{2}}}},&
\end{flalign}
\begin{flalign}
&B_{211}^{3}=\frac{1}{6}\,{\frac {\eta(B_{211}^{3,2}\eta+B_{211}^{3,1})}{\left( \eta+1 \right)  \left( { \hat{C}_{2222}^{{2}}}\,\eta+{ \hat{C}_{2222}^{{1}}}
		\right) { \hat{C}_{1212}^{{2}}}\,{ \hat{C}_{2222}^{{2}}}}},&
\end{flalign}
\begin{flalign}
&B_{211}^{1}=\frac{1}{24}\frac{\eta(A_{211}^{1,4}\eta^{4}+A_{211}^{1,3}\eta^{3}+A_{211}^{1,2}\eta^{2}+A_{211}^{1,1} \eta+A_{211}^{1,0})}{\left( \eta+1 \right) ^{4} \left( { \hat{C}_{2222}^{{2}}}\,\eta+{ 
		\hat{C}_{2222}^{{1}}} \right) ^{2}{ \hat{C}_{1212}^{2}}\,{ \hat{C}_{2222}^{{2}}}\,{ \hat{C}_{1212}^{{1}}}},&
\end{flalign}
\begin{flalign}
&A_{121}^{3}=\frac{\eta A_{121}^{3,1}+A_{121}^{3,0} }{6\, \left( \eta+1 \right)  \left( { \hat{C}_{1212}^{{2}}}\,\eta+{ \hat{C}_{1212}^{{1}}}
	\right) {{ C_{1212}^{{1}}}}^{2}},&
\end{flalign}
\begin{flalign}
&A_{121}^{1}=\frac{\eta(A_{121}^{1,4} \eta^{4}+A_{121}^{1,3} \eta^{3}+A_{121}^{1,2} \eta^{2}+A_{121}^{1,1} \eta+A_{121}^{1,0})}{24\, \left( \eta+1 \right) ^{4} \left( { \hat{C}_{1212}^{{2}}}\,\eta+{ 
		C_{1212}^{{1}}} \right) ^{2}{ \hat{C}_{1212}^{{2}}}\,{{ C_{1212}^{{1}}}}^{2}},&
\end{flalign}
\begin{flalign}
&B_{121}^{3}=\frac{\eta(B_{121}^{3,2}\eta+B_{121}^{3,1})}{6\, \left( \eta+1 \right) {{ C_{1212}^{{2}}}}^{2} \left( { \hat{C}_{1212}^{{2}}}\,
	\eta+{ \hat{C}_{1212}^{{1}}} \right)},&
\end{flalign}
\begin{flalign}
&B_{121}^{1}=\frac{\eta(B_{121}^{1,4} \eta^{4}+B_{121}^{1,3} \eta^{3}+B_{121}^{1,2} \eta^{2}+B_{121}^{1,1} \eta+B_{121}^{1,0})}{24\, \left( \eta+1 \right) ^{4} \left( { \hat{C}_{1212}^{{2}}}\,\eta+{ 
		\hat{C}_{1212}^{{1}}} \right) ^{2}{{ C_{1212}^{{2}}}}^{2}{ \hat{C}_{1212}^{{1}}}},&
\end{flalign}
\begin{flalign}
&A_{112}^{3}=\frac{1}{6}\,{\frac { \left( - \left( \eta+3 \right) { 
			\rho_1}+2 \,{ \rho_{2}} \right) { \hat{C}_{1212}^{{1}}}}{
		\left( { \hat{C}_{1212}^{{2}}}\,\eta+{ \hat{C}_{1212}^{{1}}} \right)  \left( \eta+1
		\right) { \hat{C}_{1212}^{{1}}}}},&
\end{flalign}
\begin{flalign}
&A_{112}^{1}=\frac{\eta(A_{112}^{1,4} \eta^{4}+A_{112}^{1,3} \eta^{3}+A_{112}^{1,2} \eta^{2}+A_{112}^{1,1} \eta+A_{112}^{1,0})}{24\, \left( { \hat{C}_{1212}^{{2}}}\,\eta+{ \hat{C}_{1212}^{{1}}} \right) ^{2} \left( 
	\eta+1 \right) ^{4}{ \hat{C}_{1212}^{{2}}}\,{ \hat{C}_{1212}^{{1}}}},&
\end{flalign}
\begin{flalign}
&B_{112}^{3}=\frac{1}{3}\,{\frac { \left(  \left(  \left( -  \frac{3}{2}
		{ \rho_2}+{ \rho_1} \right) { \hat{C}_{1212}^{{2}}
		}+\frac{1}{2}\,{ \rho_2}\,{ \hat{C}_{1212}^{{1}}} \right) \eta-\frac{1}{2}\,{ \rho_2}\,{
			\hat{C}_{1212}^{{2}}}+{ \hat{C}_{1212}^{{1}}}\, \left(  -\frac{1}{2}
		{ \rho_2}+{ \rho_1} \right)  \right) \eta
	}{ \left( { \hat{C}_{1212}^{{2}}}\,\eta+{ \hat{C}_{1212}^{{1}}} \right)  \left( \eta+1
		\right) { \hat{C}_{1212}^{{2}}}}},&
\end{flalign}
\begin{flalign}
&B_{112}^{1}=\frac{\eta(B_{112}^{1,4} \eta^{4}+B_{112}^{1,3} \eta^{3}+B_{112}^{1,2} \eta^{2}+B_{112}^{1,1} \eta+B_{112}^{1,0})}{24\, \left( \eta+1 \right) ^{4} \left( { \hat{C}_{1212}^{{2}}}\,\eta+{ 
		\hat{C}_{1212}^{{1}}} \right) ^{2}{ \hat{C}_{1212}^{{2}}}\,{ \hat{C}_{1212}^{{1}}}
},&
\end{flalign}
\begin{flalign}
&A_{222}^{3}=\frac{1}{6}\,{\frac { \left( - \left( \eta+3 \right) { 
			\rho_{1}}+2\,{ \rho_{2}} \right) { \hat{C}_{2222}^{{1}}}+2\,{ 
			\hat{C}_{2222}^{{2}}}\, \eta\,{ \ \rho_{2}} }{
		\left( { \hat{C}_{2222}^{{2}}}\,\eta+{ \hat{C}_{2222}^{{1}}} \right)  \left( \eta+1
		\right) { \hat{C}_{2222}^{{1}}}}},&
\end{flalign}
\begin{flalign}
&A_{222}^{1}=\frac{\eta(A_{222}^{1,4} \eta^{4}+A_{222}^{1,3} \eta^{3}+A_{222}^{1,2} \eta^{2}+A_{222}^{1,1} \eta+A_{222}^{1,0})}{24\, \left( \eta+1 \right) ^{4} \left( { \hat{C}_{2222}^{{2}}}\,\eta+{ 
		\hat{C}_{2222}^{{1}}} \right) ^{2}{ \hat{C}_{2222}^{{1}}}\,{ \hat{C}_{2222}^{{2}}}
},&
\end{flalign}
\begin{flalign}
&B_{222}^{3}=\frac{1}{3}\,{\frac { \left(  \left(  \left(   \frac{3}{2}
		{ \ \rho_{2}}+\,{ \rho_{1}} \right) { \hat{C}_{2222}^{{2}}
		}+\frac{1}{2}\,{ \ \rho_{2}}\,{ \hat{C}_{2222}^{{1}}} \right) \eta-\frac{1}{2}\,{ \ \rho_{2}}\,{
			\hat{C}_{2222}^{{2}}}+ \left(   -\frac{1}{2} { \ \rho_{2}}+{
			\delta_{22}}\,{ \rho_{1}} \right) { \hat{C}_{2222}^{{1}}} \right) \eta}{
		\left( { \hat{C}_{2222}^{{2}}}\,\eta+{ \hat{C}_{2222}^{{1}}} \right)  \left( \eta+1
		\right) { \hat{C}_{2222}^{{2}}}}},&
\end{flalign}
\begin{flalign}
\label{3.62}
&B_{222}^{1}=\frac{\eta(B_{222}^{1,4} \eta^{4}+B_{222}^{1,3} \eta^{3}+B_{222}^{1,2} \eta^{2}+B_{222}^{1,1} \eta+B_{222}^{1,0})}{24\, \left( \eta+1 \right) ^{4} \left( { \hat{C}_{2222}^{{2}}}\,\eta+{ 
		\hat{C}_{2222}^{{1}}} \right) ^{2}{ \hat{C}_{2222}^{{1}}}\,{ \hat{C}_{2222}^{{2}}}
}.&
\end{flalign}
The constants that appear in the perturbation functions Eq. \eqref{3.51}-Eq. \eqref{3.62} depend on the geometric and mechanical properties of the phases $1$ and $2$ and for sake of simplcity are not reported here.  
\section*{\small Appendix D. Constants appearing in the perturbation functions}
The constants that characterize the perturbation functions in appendix D.2, $A_{1111}^{2,0}$, $A_{1111}^{2,1}$, $A_{1111}^{0,3}$, $A_{1111}^{0,2}$,\\ \\ $A_{1111}^{0,1}$, $A_{1111}^{0,0}$, $B_{1111}^{2,1}$, $B_{1111}^{2,2}$, $B_{2222}^{0,2}$ and $B_{2222}^{0,1}$, assume the form 
\begin{flalign}
&A_{1111}^{2,1}=\left( { {\hat{C}_{1122}^{2}}}+{ {\hat{C}_{1212}^{1}}} \right) { {\hat{C}_{1122}^{1}}}-({\hat{C}_{1122}^{2}})^{2}-{ {\hat{C}_{1122}^{2}}}\,{ {\hat{C}_{1212}^{1}}}-{ {\hat{C}_{2222}^{2}}}\, \left( 
{ \hat{C}_{1111}^1}-{ {\hat{C}_{1111}^{2}}} \right),&
\end{flalign}
\begin{flalign}
&A_{1111}^{2,0}={{ {C_{1122}^{1}}}}^{2}+ \left( { {\hat{C}_{1212}^{1}}}-{ {\hat{C}_{1122}^{2}}} \right) {
	{\hat{C}_{1122}^{1}}}-{ {\hat{C}_{1122}^{2}}}\,{ {\hat{C}_{1212}^{1}}}-{ {\hat{C}_{2222}^{1}}}\, \left( {
	\hat{C}_{1111}^{1}}-{ {\hat{C}_{1111}^{2}}} \right),&
\end{flalign}
\begin{flalign}
&A_{1111}^{0,3}={ {\hat{C}_{1212}^{2}}}\, \left(  \left({ {\hat{C}_{1122}^{2}}} -{ {\hat{C}_{1122}^{1}}}
\right) { {\hat{C}_{1212}^{1}}}-{ {\hat{C}_{1122}^{1}}}\,{ {\hat{C}_{1122}^{2}}}+{{ ({\hat{C}_{1122}^{2}})}
}^{2}+{ {\hat{C}_{2222}^{2}}}\, \left( { {\hat{C}_{1111}^{1}}}-{ {\hat{C}_{1111}^{2}}} \right) 
\right),&
\end{flalign}
\begin{flalign}
&A_{1111}^{0,2}=  \left( 4\,{ {\hat{C}_{1122}^{2}}}-4\,{ {\hat{C}_{1122}^{1}}} \right) { 
	{\hat{C}_{1212}^{1}}}-{{ {C_{1122}^{1}}}}^{2}-2\,{ {\hat{C}_{1122}^{1}}}\,{ {\hat{C}_{1122}^{2}}}+3\,{{
		({\hat{C}_{1122}^{2}})}}^{2}+&
\end{flalign}	
\begin{flalign*}
&+( {\hat{C}_{2222}^{1}}+3 {\hat{C}_{2222}^{2}})( {\hat{C}_{1111}^{1}}- {\hat{C}_{1111}^{2}}) {\hat{C}_{1212}^{2}},&
\end{flalign*}
\begin{flalign}
&A_{1111}^{0,1}=\left(  \left(5\,{ {\hat{C}_{1122}^{2}}}-5\,{ {\hat{C}_{1122}^{1}}} \right) { 
	{\hat{C}_{1212}^{1}}}-3\,{{ {C_{1122}^{1}}}}^{2}+3\,{ {\hat{C}_{1122}^{1}}}\,{ {\hat{C}_{1122}^{2}}}+3
\,{ {\hat{C}_{2222}^{1}}}\, \left( { {\hat{C}_{1111}^{1}}}-{ {\hat{C}_{1111}^{2}}} \right)\right){ {\hat{C}_{1212}^{2}}} \,&
\end{flalign}
\begin{flalign*}
&+2\,{ {\hat{C}_{1212}^{1}}}\, \left( {{ ({\hat{C}_{1122}^{2}})}}^{2}-{ {\hat{C}_{1122}^{1}}}\,{ {\hat{C}_{1122}^{2}}}+{ {\hat{C}_{2222}^{2}}}\, \left( { {\hat{C}_{1111}^{1}}}-{ {\hat{C}_{1111}^{2}}} \right)  \right),&
\end{flalign*}
\begin{flalign}
&A_{1111}^{0,0}=2\,{ {\hat{C}_{1212}^{1}}}\, \left(  \left( { {\hat{C}_{1122}^{2}}}-{ {\hat{C}_{1122}^{1}}}
\right) { {\hat{C}_{1212}^{2}}}-{{ {C_{1122}^{1}}}}^{2}+{ {\hat{C}_{1122}^{1}}}\,{ 
	{\hat{C}_{1122}^{2}}}+{ {\hat{C}_{2222}^{1}}}\, \left( { {\hat{C}_{1111}^{1}}}-{ {\hat{C}_{1111}^{2}}}
\right)  \right),&
\end{flalign}
\begin{flalign}
&B_{1111}^{2,2}={{({\hat{C}_{1122}^{2}})}}^{2}+ \left( { {\hat{C}_{1212}^{2}}}-{ {\hat{C}_{1122}^{1}}} \right) {
	{\hat{C}_{1122}^{2}}}-{ {\hat{C}_{1212}^{2}}}\,{ {\hat{C}_{1122}^{1}}}+{ {\hat{C}_{2222}^{2}}}\, \left( {
	{\hat{C}_{1111}^{1}}}-{ {\hat{C}_{1111}^{2}}} \right),&
\end{flalign}
\begin{flalign}
&B_{1111}^{2,1}=-{{ {C_{1122}^{1}}}}^{2}+ \left( { {\hat{C}_{1122}^{2}}}-{ {\hat{C}_{1212}^{2}}} \right) {
	{\hat{C}_{1122}^{1}}}+{ {\hat{C}_{1122}^{2}}}\,{ {\hat{C}_{1212}^{2}}}+{ {\hat{C}_{2222}^{1}}}\, \left( {
	{\hat{C}_{1111}^{1}}}-{ {\hat{C}_{1111}^{2}}} \right),&
\end{flalign}
\begin{flalign}
&B_{1111}^{0,2}=\left(  \left( 5\,{ {\hat{C}_{1122}^{1}}}-5\,{ {\hat{C}_{1122}^{2}}} \right) { 
	{\hat{C}_{1212}^{2}}}+3\,{ {\hat{C}_{1122}^{1}}}\,{ {\hat{C}_{1122}^{2}}}-3\,{{ ({\hat{C}_{1122}^{2}})}}^{2}-3
\,{ {\hat{C}_{2222}^{2}}}\, \left( { {\hat{C}_{1111}^{1}}}-{ {\hat{C}_{1111}^{2}}} \right) 
\right){ {\hat{C}_{1212}^{1}}}+&
\end{flalign}
\begin{flalign*}
&-2\, \left( -{{ {C_{1122}^{1}}}}^{2}+{ {\hat{C}_{1122}^{1}}
}\,{ {\hat{C}_{1122}^{2}}}+{ {\hat{C}_{2222}^{1}}}\, \left( { {\hat{C}_{1111}^{1}}}-{ {\hat{C}_{1111}^{2}}
} \right)  \right) { {\hat{C}_{1212}^{2}}},&
\end{flalign*}
\begin{flalign}
&B_{1111}^{0,1}=-3\, \left( -{{ {C_{1122}^{1}}}}^{2}+ \left( \frac{2}{3}\,{ ({\hat{C}_{1122}^{2}})}-\frac{4}{3}\,{
	{\hat{C}_{1212}^{2}}} \right) { {\hat{C}_{1122}^{1}}}+\frac{1}{3}\,{{ ({\hat{C}_{1122}^{2}})}}^{2}+\frac{4}{3}\,{
	({\hat{C}_{1122}^{2}})}\,{ {\hat{C}_{1212}^{2}}}\right )+&
\end{flalign}
\begin{flalign*}
& -3( \hat{C}_{1111}^{1}- \hat{C}_{1111}^{2})\Big (\frac{ \hat{C}_{2222}^{2}}{3}+{ \hat{C}_{2222}^{1}}\Big ) \hat{C}_{1212}^{1}.&
\end{flalign*}	
\section*{ \small Appendix E. Overall inertial terms and overall consitutive tensors}
By taking into account the method I, the transformed inertial tensors appearing in Eq. $\eqref{latrasdi}$ assume the form:

\begin{flalign}
\label{primo}
&\hat{I}_{2121}=\varepsilon^{2}\langle \rho^{m} (N_{121}^{(1,0)}N_{121}^{(1,0)})-2\rho^{m}  N_{2211}^{(2,0)}\rangle \frac{1}{\rho},&
\end{flalign}

\begin{flalign}
&\hat{I}_{2222}=\varepsilon^{2}\langle \rho^{m} (N_{222}^{(1,0)}N_{222}^{(1,0)})-2\rho^{m} N_{2222}^{(2,0)}\rangle \frac{1}{\rho},&
\end{flalign}
\begin{flalign}
&\hat{I}_{22}=\varepsilon^{2}\langle \rho^{m} (N_{22}^{(2,2)}+N_{22}^{(2,2)})\rangle \frac{1}{\rho},&
\end{flalign}

\begin{flalign}
&\hat{I}_{1212}=\varepsilon^{2}\langle \rho^{m} (N_{112}^{(1,0)}N_{112}^{(1,0)})-2\rho^{m}  N_{1122}^{(2,0)}\rangle \frac{1}{\rho},&
\end{flalign}

\begin{flalign}
&\hat{I}_{1111}=\varepsilon^{2}\langle \rho^{m} (N_{211}^{(1,0)}N_{211}^{(1,0)})-2\rho^{m} N_{1111}^{(2,0)}\rangle \frac{1}{\rho},&
\end{flalign}
\begin{flalign}
&\hat{I}_{11}=\varepsilon^{2}\langle \rho^{m} (N_{11}^{(2,2)}+N_{11}^{(2,2)})\rangle \frac{1}{\rho}.&
\end{flalign}
Referred to the Eq. $\eqref{latrasdi}$, the constitutive tensors in the Laplace space are
\begin{flalign}
&\hat{J}_{11}=\varepsilon^{2}\langle \hat{G}_{1212}B_{121}^{(2,2)}B_{121}^{(2,2)} \rangle,&
\end{flalign}
\begin{flalign}
&\hat{J}_{22}=\varepsilon^{2}\langle \hat{G}_{2222}B_{222}^{(2,2)}B_{222}^{(2,2)} \rangle,&
\end{flalign}
\begin{flalign}
&\hat{G}_{1111}=\langle \hat{G}_{1111}B_{1111}^{(1,0)}B_{1111}^{(1,0)}+ \hat{G}_{2222}B_{2211}^{(1,0)}B_{2211}^{(1,0)}+2\hat{G}_{1122}B_{2211}^{(1,0)}B_{1111}^{(1,0)} \rangle,&
\end{flalign}
\begin{flalign}
&\hat{G}_{2121}=\langle \hat{G}_{2222}B_{2221}^{(1,0)}B_{2221}^{(1,0)} \rangle=\hat{G}_{1212},&
\end{flalign}
\begin{flalign}
\label{ultimo}
&\hat{G}_{2222}=\langle \hat{G}_{2222}B_{2222}^{(1,0)}B_{2222}^{(1,0)} \rangle,&
\end{flalign}

%\begin{flalign}
%&\hat{J}_{1111}^{I}=\varepsilon^{2}\langle 2\hat{G}_{1111}B_{1111}^{(1,0)}B_{1111}^{(3,2)}+ %2\hat{G}_{2222}B_{2211}^{(1,0)}B_{2211}^{(3,2)}-2\hat{G}_{1212}B_{12111}^{(2,0)}B_{121}^{(2,2)}%+&
%\end{flalign}
%\begin{flalign*}
%&-2\hat{G}_{1212}B_{21111}^{(2,0)}B_{121}^{(2,2)}+2\hat{G}_{1122}B_{1111}^{(1,0)}B_{2211}^{(3,2%)} +2\hat{G}_{1122}B_{2211}^{(1,0)}B_{1111}^{(3,2)}\rangle,&
%\end{flalign*}
\begin{flalign*}
\label{secondo}
&\hat{J}_{2121}^{1}=-\varepsilon^{2}\langle 2\hat{G}_{1122}B_{11211}^{(2,0)}B_{222}^{(2,2)}- 2\hat{G}_{1212}B_{1221}^{(1,0)}B_{1221}^{(3,2)}+2\hat{G}_{2222}B_{22211}^{(2,0)}B_{222}^{(2,2)}-2\hat{G}_{1212}B_{2121}^{(1,0)}B_{2121}^{(3,2)}+&
\end{flalign*}
\begin{flalign}
&-2\hat{G}_{1212}B_{2121}^{(1,0)}B_{1221}^{(3,2)}-2\hat{G}_{1212}B_{2121}^{(1,0)}B_{2121}^{(3,2)}\rangle,&
\end{flalign}
\begin{flalign}
&\hat{J}_{1212}^{1}=-\varepsilon^{2}\langle -2\hat{G}_{1212}B_{1212}^{(1,0)}B_{1212}^{(3,2)}+2\hat{G}_{1212}B_{12122}^{(2,0)}B_{121}^{(2,2)} \rangle,&
\end{flalign}
\begin{flalign}
&\hat{J}_{2222}^{1}=\varepsilon^{2}\langle -2\hat{G}_{2222}B_{2222}^{(1,0)}B_{2222}^{(3,2)}+2\hat{G}_{2222}B_{22222}^{(2,0)}B_{222}^{(2,2)} \rangle,&
\end{flalign}
\begin{flalign}
\label{S111111_1}
&\hat{S}_{111111}^{1}=\varepsilon^{2}\langle -2\hat{G}_{1111}B_{1111}^{(1,0)}B_{111111}^{(3,0)}+ \hat{G}_{1212}B_{12111}^{(2,0)}B_{12111}^{(2,0)}+\hat{G}_{1212}B_{21111}^{(2,0)}B_{21111}^{(2,0)}+&
\end{flalign}
\begin{flalign*}
&+2\hat{G}_{1212}B_{12111}^{(2,0)}B_{21111}^{(2,0)}-2\hat{G}_{2222}B_{2211}^{(1,0)}B_{221111}^{(3,0)}-2\hat{G}_{1122}B_{2211}^{(1,0)}B_{111111}^{(3,0)} -2\hat{G}_{2211}B_{1111}^{(1,0)}B_{221111}^{(3,0)}\rangle,&
\end{flalign*}
\begin{flalign}
\label{penultimo}
&\hat{S}_{211211}^{1}=\varepsilon^{2}\langle \hat{G}_{1111}B_{11211}^{(2,0)}B_{11211}^{(2,0)}+2\hat{G}_{1122}B_{11211}^{(2,0)}B_{22211}^{(2,0)}+\hat{G}_{2222}B_{22211}^{(2,0)}B_{22211}^{(2,0)}-2\hat{G}_{1212}B_{1212}^{(1,0)}B_{12111}^{(3,0)}+&
\end{flalign}
\begin{flalign*}
&-2\hat{G}_{1212}B_{1221}^{(1,0)}B_{122111}^{(3,0)}-2\hat{G}_{1212}B_{2121}^{(1,0)}B_{122111}^{(3,0)}-2\hat{G}_{1212}B_{1221}^{(1,0)}B_{212111}^{(3,0)}-2\hat{G}_{1212}B_{2121}^{(1,0)}B_{212111}^{(3,0)}   \rangle.&
\end{flalign*}
The inertial and constitutive tensors derived from the method $2$ coincide with those ones from \eqref{primo} to \eqref{ultimo}, whereas the ones from \eqref{secondo} to \eqref{penultimo} are slightly different because the localization tensors $B^{(3,0)}_{hkpq_{1}q_{2}q_{3}}$ and $B^{(3,2)}_{hkpq_{1}}$ disappear in the formulation and they assume the form: 
\begin{flalign}
&\hat{J}_{1111}^{2}=-\varepsilon^{2}\langle 2\hat{G}_{1212}B_{12111}^{(2,0)}B_{121}^{(2,2)}+2\hat{G}_{1212}B_{21111}^{(2,0)}B_{121}^{(2,2)}\rangle &
\end{flalign}\begin{flalign}
&\hat{J}_{2121}^{2}=-\varepsilon^{2}\langle 2\hat{G}_{1122}B_{11211}^{(2,0)}B_{222}^{(2,2)}+2\hat{G}_{2222}B_{22211}^{(2,0)}B_{222}^{(2,2)}\rangle&
\end{flalign}

\begin{flalign}
&\hat{J}_{1212}^{2}=-\varepsilon^{2}\langle 2\hat{G}_{1212}B_{12122}^{(2,0)}B_{121}^{(2,2)} \rangle,&
\end{flalign}
\begin{flalign}
&\hat{J}_{2222}^{2}=-\varepsilon^{2}\langle 2\hat{G}_{2222}B_{22222}^{(2,0)}B_{222}^{(2,2)} \rangle,&
\end{flalign}
\begin{flalign}
\label{S211211_2}
&\hat{S}_{111111}^{2}=\varepsilon^{2}\langle  \hat{G}_{1212}B_{12111}^{(2,0)}B_{12111}^{(2,0)}+\hat{G}_{1212}B_{21111}^{(2,0)}B_{21111}^{(2,0)}+2\hat{G}_{1212}B_{12111}^{(2,0)}B_{21111}^{(2,0)}\rangle, &
\end{flalign}
\begin{flalign}
&\hat{S}_{211211}^{2}=\varepsilon^{2}\langle \hat{G}_{1111}B_{11211}^{(2,0)}B_{11211}^{(2,0)}+2\hat{G}_{1122}B_{11211}^{(2,0)}B_{22211}^{(2,0)}+\hat{G}_{2222}B_{22211}^{(2,0)}B_{22211}^{(2,0)} \rangle.&
\end{flalign}

\end{document}